\documentclass[amsmath,amssymb,reprint,aps,pra,floatfix,superscriptaddress]{revtex4-2}
\usepackage{bm}
\usepackage{amsmath}
\usepackage{graphicx}
\usepackage{braket}
\usepackage{bbm}
\usepackage[hidelinks]{hyperref}
\usepackage{pifont}
\usepackage{bbold}
\usepackage[dvipsnames]{xcolor}
\usepackage{amssymb}
\usepackage{slashed}
\usepackage{comment}
\usepackage{pifont}

\begin{document}

\title{Gain-induced group delay in spontaneous parametric down-conversion}

\author{Guillaume Thekkadath}
\email{guillaume.thekkadath@nrc.ca}
\affiliation{National Research Council of Canada, 100 Sussex Drive, Ottawa, Ontario K1N 5A2, Canada}

\author{Martin Houde}
\affiliation{Department of Engineering Physics, École Polytechnique de Montréal, 2500 Chem. de Polytechnique, Montréal, Quebec H3T 1J4,
Canada}

\author{Duncan England}
\affiliation{National Research Council of Canada, 100 Sussex Drive, Ottawa, Ontario K1N 5A2, Canada}

\author{Philip~Bustard}
\affiliation{National Research Council of Canada, 100 Sussex Drive, Ottawa, Ontario K1N 5A2, Canada}

\author{Frédéric Bouchard}
\affiliation{National Research Council of Canada, 100 Sussex Drive, Ottawa, Ontario K1N 5A2, Canada}

\author{Nicolás Quesada}
\email{nicolas.quesada@polymtl.ca}
\affiliation{Department of Engineering Physics, École Polytechnique de Montréal, 2500 Chem. de Polytechnique, Montréal, Quebec H3T 1J4,
Canada}

\author{Ben Sussman}
\affiliation{National Research Council of Canada, 100 Sussex Drive, Ottawa, Ontario K1N 5A2, Canada}

\begin{abstract}
Strongly-driven nonlinear optical processes such as spontaneous parametric down-conversion and spontaneous four-wave mixing can produce multiphoton nonclassical beams of light which have applications in quantum information processing and sensing.
In contrast to the low-gain regime, new physical effects arise in a high-gain regime due to the interactions between the nonclassical light and the strong pump driving the nonlinear process.
Here, we describe and experimentally observe a gain-induced group delay between the multiphoton pulses generated in a high-gain type-II spontaneous parametric down-conversion source.
Since the group delay introduces distinguishability between the generated photons, it will be important to compensate for it when designing quantum interference devices in which strong optical nonlinearities are required.  
\end{abstract}

\maketitle
Nonlinear optical processes like spontaneous parametric down-conversion (SPDC) and four-wave mixing have long been used as nonclassical light sources for quantum optics experiments~\cite{louisell1961quantum, harris1967observation, burnham1970observation, hong1987measurement,kwiat1995new,li2004all}.
In these processes, photons from a pump laser interact with a nonlinear medium and scatter into pairs of photons with different energies.
Over the years, advances in the ability to manipulate the spatiotemporal properties of the photon pairs~\cite{yamada1993first,grice2001eliminating,branczyk2011engineered} have led to improvements in their purity~\cite{mosley2008heralded}, correlation strength~\cite{nasr2008ultra, devaux2020imaging}, and collection efficiency~\cite{bennink2010optimal}.
Photon pair sources have enabled landmark experiments such as the first demonstration of quantum teleportation~\cite{bouwmeester1997experimental} and the loophole-free violation of Bell's inequalities~\cite{shalm2015strong,giustina2015significant}.
There have also been efforts to increase the strength of the nonlinear interaction beyond the photon pair regime by using pulsed lasers~\cite{slusher1987pulsed,kim1994quadrature, zhong2021phase}, optical cavities~\cite{wu1986generation,vahlbruch2016detection, zhang2021squeezed, arrazola2021quantum}, and waveguides~\cite{eckstein2011highly,harder2016single, paesani2020near}.
When the nonlinear interaction is strong, multiple pump photons scatter coherently into superpositions of many photon pairs, thereby generating squeezed light or twin beams.
These light sources have nonclassical properties which can be used for metrology~\cite{ganapathy2023broadband} and information processing~\cite{menicucci2006universal,bourassa2021blueprint,zhong2021phase} applications.

\begin{figure}[t]
\centering
\includegraphics[width=1\columnwidth]{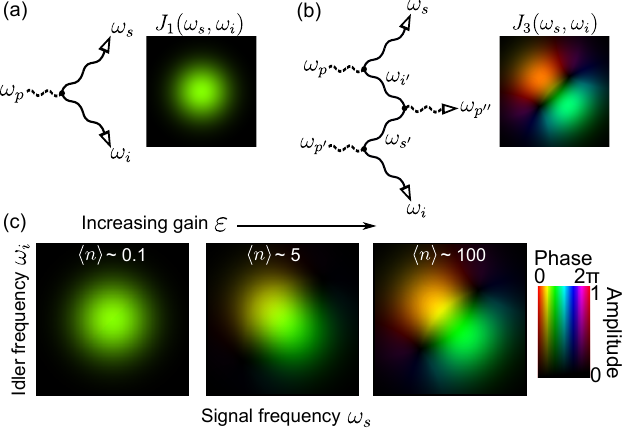}
\caption{(a) SPDC scattering process to first-order in perturbation theory.
Assuming a Gaussian pump and phase-matching function, photon pairs are generated in the joint spectral mode $J_1(\omega_s, \omega_i)$ with probability amplitude $
\sim \varepsilon$.
(b) Leading-order correction in the perturbation series describes a more complex scattering process which generates photon pairs in an orthogonal mode $J_3(\omega_s, \omega_i)$ with probability amplitude $\sim \varepsilon ^3$.
(c) Total joint spectral amplitude is a superposition of both scattering processes as well as higher-order ones [Eq.~\eqref{eqn:jsa_timeCorrected}].
As the gain $\varepsilon$ increases, a linear spectral phase gradient develops along the anti-diagonal axis, leading to a group delay between the generated photons.
This delay becomes appreciable when $\braket{n} \gtrsim 1$, where $\braket{n}$ is the average number of photon pairs generated per pump pulse.}
\label{fig:fig1}
\end{figure}

Given the potential of increasing the strength of nonclassical features such as squeezing, there has been a significant amount of work trying to understand the rich physics that emerges in a strong nonlinear interaction regime, also known as the high-gain regime~\cite{laporta1991squeezing,wasilewski2006pulsed,dayan2007theory,cassemiro2007quest,leung2009spectral,iskhakov2009generation,brambilla2010spatiotemporal,branczyk2011time,spasibko2012spectral,iskhakov2012superbunched,christ2013theory,allevi2014coherence,quesada2014effects,quesada2015time,sharapova2015schmidt,finger2015raman,guo2015complete,chekhova2015bright,bell2015effects,pevrina2016internal,pevrina2016spatial,pevrina2016coherent,liu2016approaching,allevi2017nonlinear,sharapova2020properties,florez2020pump,triginer2020understanding,chen2021mode,quesada2022beyond,yanagimoto2022onset,kulkarni2022classical,houde2023waveguided,kalash2023wigner,chinni2024beyond}.
For example, pump depletion can lead to correlations between the quantum fields and the classical pump~\cite{cassemiro2007quest,allevi2014coherence,pevrina2016internal,florez2020pump,chinni2024beyond}.
Moreover, the spatiotemporal mode structure of photon pairs can be modified by so-called time-ordering corrections~\cite{leung2009spectral,branczyk2011time,christ2013theory,quesada2014effects,quesada2015time} and higher-order nonlinearities~\cite{bell2015effects,quesada2020theory}.
These gain-dependent mode modifications play an important role in determining a source's squeezing strength and interference visibility at large pump intensities. 
They have previously been measured experimentally, either directly with the spontaneously generated photons~\cite{spasibko2012spectral,allevi2014coherence,sharapova2015schmidt,pevrina2016coherent,liu2016approaching,sharapova2020properties,kulkarni2022classical}, or by seeding the nonlinear medium with classical beams~\cite{triginer2020understanding,huo2020direct}.

\begin{figure}
\centering
\includegraphics[width=1\columnwidth]{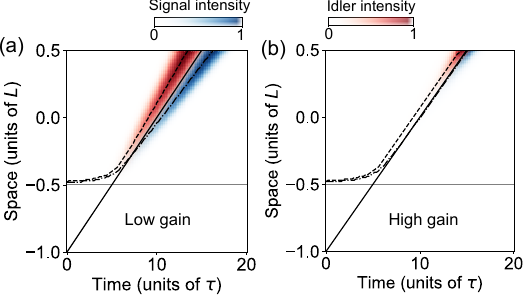}
\caption{Numerically calculated pulse dynamics in (a) low-gain $\braket{n}\sim0.1$ and (b) high-gain regime $\braket{n}\sim 100$ using parameters that model the ppKTP source used in the experiment.
Black line shows pulse peak for signal (dot dashed), idler (dashed), and pump (continuous).
Horizontal grey line indicates the crystal entrance facet.
Colormap transparency scales linearly with the pulse energy.
Pulses are normalized to their maximal value.
$L$ is the crystal length and $\tau$ is the pump pulse duration.
}
\label{fig:fig2}
\end{figure}

While these previous works focused on gain-induced modifications in the intensity of the mode structure, in this Letter, we focus on the phase.
We show that the photon pair spectral phase is modified even at modest gains (e.g. a single pair generated per pump pulse), which leads to a group delay between the generated photons.
We experimentally measure this effect using a type-II SPDC source pumped by ultrashort pulses.
We also provide a physical explanation both in an interaction picture in terms of scattering processes [Fig.~\ref{fig:fig1}] and in a Heisenberg picture in terms of pulse dynamics [Fig.~\ref{fig:fig2}].

Under the parametric (i.e. undepleted pump) and rotating-wave approximations, the interaction Hamiltonian describing SPDC in a single spatial mode is~\cite{SM}:
\begin{equation}
\begin{split}
\hat{\mathcal{H}}(t) &= -\frac{\hbar \varepsilon}{2\pi} \int d\omega_p d\omega_s d\omega_i e^{i\Delta t} \Phi(\omega_s, \omega_i, \omega_p) \\
& \qquad \times \beta(\omega_p) \hat{a}^{\dagger}(\omega_s) \hat{b}^{\dagger}(\omega_i) + \mathrm{h.c.},
\end{split}
\label{eqn:hamiltonian}
\end{equation}
where $\varepsilon$ is the interaction strength or parametric gain, $\Delta = \omega_s + \omega_i - \omega_p$, and $\hat{a}^{\dagger}(\omega_s)$ [$\hat{b}^{\dagger}(\omega_i)$] is the photon creation operator for the signal [idler] mode.
The quantities $\Phi(\omega_s, \omega_i, \omega_p)$ and $\beta(\omega_p)$ are the phase-matching and pump spectral mode functions, respectively.
The dynamics of the fields under Eq.~\eqref{eqn:hamiltonian} are described by the unitary time-evolution operator $\hat{\mathcal{U}}(t, t_0) = \hat{\mathcal{T}} \exp{[-\frac{i}{\hbar} \int_{t_0}^t dt' \hat{\mathcal{H}}(t')]}$ where $\hat{\mathcal{T}}$ is the time-ordering operator.
We are interested in obtaining the state of the fields at the end of their interaction, i.e. $\hat{\mathcal{U}}(t \rightarrow \infty, t_0\rightarrow -\infty) \equiv \hat{\mathcal{U}}$. 
If the gain is small ($\varepsilon \ll 1$), the unitary can be expanded to first-order which describes the creation of single photon pairs in the initially vacuum states of the signal and idler modes $\ket{0}_a\ket{0}_b$:
\begin{align}
\ket{\Psi}_{ab} &= \hat{\mathcal{U}}\ket{0}_a\ket{0}_b \label{eqn:spdc_lowGain} \\
&\approx \left(\hat{\mathbbm{1}} - \frac{i}{\hbar} \int_{-\infty}^{\infty} dt \hat{\mathcal{H}}(t)\right)\ket{0}_a\ket{0}_b \nonumber \\
&\approx \ket{0}_a\ket{0}_b \nonumber \\
&\qquad  +  i \varepsilon \int d\omega_s d\omega_i J_1(\omega_s, \omega_i) \hat{a}^{\dagger}(\omega_s) \hat{b}^{\dagger}(\omega_i)  \ket{0}_a\ket{0}_b \nonumber,
\end{align}
where $J_1(\omega_s, \omega_i) =   \beta(\omega_s+\omega_i) \Phi(\omega_s, \omega_i, \omega_s+\omega_i)$ is the normalized joint spectral mode of the photon pairs in a low-gain regime.
For many applications, it is desirable to produce photons in a single spectral-temporal mode, in which case the joint mode should be uncorrelated, $J_1(\omega_s, \omega_i) = f(\omega_s)g(\omega_i)$.
This can be achieved by satisfying the group velocity condition $v_s \leq v_p \leq v_i$~\cite{grice2001eliminating,graffitti2018design} and employing crystals with apodized periodic poling~\cite{branczyk2011engineered}.
The average number of photon pairs generated per pump pulse in the single mode is $\braket{n}=\mathrm{sinh}^2(\varepsilon)$.

With increasing $\varepsilon$, one must include higher-order terms in the expansion of $\hat{\mathcal{U}}$.
Since
$[\hat{\mathcal{H}}(t), \hat{\mathcal{H}}(t')]\neq 0$, the Dyson or Magnus series should be used to maintain proper time ordering~\cite{branczyk2011time}.
As the Dyson expansion does not preserve the correct photon statistics when truncated, we turn to the Magnus expansion~\cite{quesada2014effects,quesada2015time}:
\begin{equation}
\hat{\mathcal{U}} = \exp{\left(\hat{\Omega}_1 + \hat{\Omega}_2 + \hat{\Omega}_3 +...\right)}.
\label{eqn:magnus}
\end{equation}
In particular, $\hat{\Omega}_1 = -\frac{i}{\hbar} \int_{-\infty}^{\infty} dt \hat{\mathcal{H}}(t)$ while higher-order terms depend on commutators of the Hamiltonian at different times, e.g. $\hat{\Omega}_2 = \frac{(-i)^2}{2\hbar^2} \int_{-\infty}^{\infty} dt \int_{-\infty}^{t} dt' [\hat{\mathcal{H}}(t), \hat{\mathcal{H}}(t')]$.
Each term $\hat{\Omega}_i$ can be associated with a different type of scattering process which can occur multiple times throughout the unitary evolution.
The first-order term $\hat{\Omega}_1$ corresponds to the usual picture of one pump photon scattering into a pair of lower energy photons [Fig.~\ref{fig:fig1}(a)].
The second-order term $\hat{\Omega}_2$ describes a frequency-conversion process which requires seeding one of the down-converted modes.
We assume the signal and idler fields are initially in the vacuum state, and so this term does not contribute, i.e. $\exp{\left(\hat{\Omega}_2\right)}\ket{0}_a\ket{0}_b = \ket{0}_a\ket{0}_b$.
The third order term $\hat{\Omega}_3$ describes a scattering process in which two pump photons each produce pairs of down-converted photons followed by the up-conversion of two of the latter photons [Fig.~\ref{fig:fig1}(b)].
Since $\hat{\Omega}_3$ scales with $\varepsilon ^3$, the probability of this process occurring is similar to the probability of three photon pairs being generated due to $\hat{\Omega}_1$.
Thus, $\hat{\Omega}_3$ and higher-order terms become appreciable only when multiple photon pairs are generated per pump pulse.

The higher-order terms in the Magnus expansion [Eq.~\eqref{eqn:magnus}] modify the spectral-temporal properties of the photon pairs, i.e. $\ket{\Psi}_{ab} = \exp{ \left( i \int d\omega_s d\omega_i J(\omega_s,\omega_i) \hat{a}^{\dagger}(\omega_s) \hat{b}^{\dagger}(\omega_i) \right)}\ket{0}_a\ket{0}_b$ with:
\begin{equation}
J(\omega_s, \omega_i) = \varepsilon J_1(\omega_s, \omega_i) + \frac{\varepsilon^3}{\sqrt{18}} J_3(\omega_s, \omega_i) + ...,
\label{eqn:jsa_timeCorrected}
\end{equation}
where $J(\omega_s, \omega_i)$ is the joint spectral amplitude (JSA) of generating a photon pair.
Two of the present authors have developed a Python library, \texttt{NeedALight}, to compute $J(\omega_s, \omega_i)$ for arbitrary phase-matching and pump mode shapes~\cite{houde2023waveguided,NeedALight}.
When these functions are Gaussian, we can derive analytic expressions for $J_1(\omega_s, \omega_i)$ and $J_3(\omega_s, \omega_i)$ (see the Supplementary Materials [SM]~\cite{SM}).
We find that, even when $J_1(\omega_s, \omega_i)$ is uncorrelated, the total JSA is no longer uncorrelated due to the presence of a new orthogonal mode $J_3(\omega_s, \omega_i)$, as has previously been discussed~\cite{quesada2020theory,houde2023waveguided}.
Another notable feature is that $J_3(\omega_s, \omega_i)$ contains an imaginary component even when the pump pulse has a uniform spectral phase (i.e. no chirp).
This results in a non-trivial joint spectral phase [Fig.~\ref{fig:fig1}(c)] that is well approximated by a linear gradient along the anti-diagonal axis, i.e. $\mathrm{arg}[J(\omega_s, \omega_i)] \sim \frac{\beta}{2} (\omega_s-\omega_i)$ where $\beta =  24\tau \varepsilon^2 /\sqrt{3\pi}(12+ \varepsilon^2)$ and $\tau$ is the pump pulse duration.
As we show in the SM, this causes a temporal group delay of $T=\beta_0 - \beta$ between the signal and idler photons, where $\beta_0= L(v_s^{-1} - v_i^{-1})/2$ is a gain-independent group delay caused by group-velocity walk-off.
Thus, the down-converted photons emerge from the crystal closer in time with increasing $\varepsilon$.
Since their coherence time and the group delay both scale with the pump duration $\tau$, the relative size of this effect depends only on the gain $\varepsilon$.
In fact, a gain-induced group delay was also discussed in Ref.~\cite{brambilla2010spatiotemporal} for the case of a monochromatic pump.
Below we provide a physical explanation for its origin, and present an experiment which measures it.

One intuitive but qualitative way to understand the group delay is from the fact that down-converted photons can stimulate further scattering processes.
As a result, most photons are created towards the end of the crystal in a high-gain regime~\cite{brambilla2010spatiotemporal,spasibko2012spectral}.
Since they propagate a shorter distance in the crystal, they should suffer less group velocity walk-off.
We can connect this explanation to the Magnus expansion.
The number of photons generated in the mode $J_1(\omega_s, \omega_i)$ grows exponentially with $\varepsilon$, as one might expect from stimulated processes.
However, there are also photons generated in new modes such as $J_3(\omega_s, \omega_i)$ which arise due to interactions between the pump and the down-converted photons.
For example, the scattering process leading to $J_3(\omega_s, \omega_i)$ involves the up-conversion of a signal photon $\omega_{s'}$ and idler photon $\omega_{i'}$ into a pump photon $\omega_{p''}$  [Fig.~\ref{fig:fig1}(b)].
Due to their different group velocities ($v_s \leq v_p \leq v_i$), for $\omega_{s'}$ and $\omega_{i'}$ to interact, the production $\omega_{p'} \rightarrow (\omega_{s'}, \omega_{i})$ is more likely to occur before $\omega_p \rightarrow (\omega_s, \omega_{i'})$.
Thus, while Fig.~\ref{fig:fig1}(a) describes a scattering process in which $(\omega_s, \omega_i)$ are produced simultaneously with a probability $\sim \varepsilon$, the process in Fig.~\ref{fig:fig1}(b) produces $\omega_i$ before $\omega_s$ with a probability $\sim \varepsilon^3$.
Since the SPDC interaction is a superposition of both processes (as well as higher order ones), the group delay scales with $\varepsilon^2$ to leading order.
Finally, we note that the group velocity matching condition leading to a separable $J_1(\omega_s, \omega_i) = f(\omega_s)g(\omega_i)$ maximizes the time both down-converted photons spend overlapped with the pump.
This enhances their interaction and thus the relative contribution of higher-order Magnus terms is much larger than in a source with a highly-correlated $J_1(\omega_s, \omega_i)$ and the same $\varepsilon$~\cite{quesada2014effects}.

The interaction between the pump and down-converted light can also be visualized in the Heisenberg picture.
By numerically solving the equations of motion~\cite{SM}, we can plot the dynamics of the signal and idler pulses.
In a low-gain regime [Fig.~\ref{fig:fig2}(a)], these are uniformly amplified throughout the crystal as they travel at their respective group velocities.
On average, photons are produced half-way through the crystal~\cite{kwiat1995new}, leading to a group delay of $\beta_0= L(v_s^{-1} - v_i^{-1})/2$.
In contrast, in a high-gain regime [Fig.~\ref{fig:fig2}(b)], most amplification occurs towards the end of the crystal.
Moreover, there is a larger amplification for temporal components closest to the peak of the pump pulse.
As a result, the down-converted pulses tend to become temporally narrower and ``stick" to the pump pulse, i.e. their effective group velocities converge to that of the pump with increasing gain.
This effect also occurs in classical amplifiers with pulsed gain profiles~\cite{longhi2002dispersive}.

We can measure the group delay using intensity interferometry.
As in Hong-Ou-Mandel interferometry~\cite{hong1987measurement}, the signal $\hat{a}(\omega_s)$ and idler $\hat{b}(\omega_i)$ modes are combined on a balanced beam splitter, $\hat{a}(\omega_s)\rightarrow(\hat{c}(\omega_s)+\hat{d}(\omega_s))/\sqrt{2}$ and $\hat{b}(\omega_i)\rightarrow(\hat{c}(\omega_i)-\hat{d}(\omega_i))/\sqrt{2}$.
If we consider the contributions to the output state having photons simultaneously in both modes, we find: 
\begin{align}
\ket{\tilde{\Psi}}_{cd} = \mathrm{exp} \Bigg[ i\int & d\omega_s d\omega_i J(\omega_s, \omega_i) \times \\
& \frac{1}{2}[\hat{c}^\dagger(\omega_i) \hat{d}^\dagger(\omega_s) - \hat{c}^\dagger(\omega_s) \hat{d}^\dagger(\omega_i)] \Bigg] \ket{0},
\label{eqn:pdc_output_BS} \nonumber
\end{align}
where $J(\omega_s, \omega_i)$ is the complete JSA, i.e. including all terms in the Magnus expansion.
Consider the probability density of measuring a coincidence between a photon with frequency $\omega_1$ in mode $c$ and $\omega_2$ in mode $d$:
\begin{equation}
\begin{split}
\mathrm{pr}(\omega_1, \omega_2) &= \left|\braket{0|\hat{c}(\omega_1)\hat{d}(\omega_2)|\tilde{\Psi}}_{cd}\right|^2 \\
&= \frac{1}{4}|J(\omega_1, \omega_2) - J(\omega_2,\omega_1)|^2,
\end{split}
\label{eqn:spectrally_resolved_coincidence}
\end{equation}
which is a projection onto the anti-symmetric component of the JSA.
If the signal and idler photons are symmetric (i.e. indistinguishable) in amplitude, $|J(\omega_1, \omega_2)|=|J(\omega_2, \omega_1)|$, then we find:
\begin{equation}
\mathrm{pr}(\omega_1, \omega_2) = \frac{1}{2}|J(\omega_1, \omega_2)|^2 \left( 1 - \cos{[\tfrac{T}{2}(\omega_1-\omega_2)]} \right)
\label{eqn:interferogram}
\end{equation}
where $\mathrm{arg}\left[J(\omega_1, \omega_2)\right] = \tfrac{T}{2}(\omega_1-\omega_2)$ is a linear spectral phase gradient due to a group delay of $T$ between the photons.
Thus, the single-pair coincidence probability is an interferogram whose fringe pattern can be related to the group delay $T$~\cite{gerrits2015spectral,jin2015spectrally,Orre2019interference,thekkadath2022measuring}.

\begin{figure*}
\centering
\includegraphics[width=0.9\textwidth]{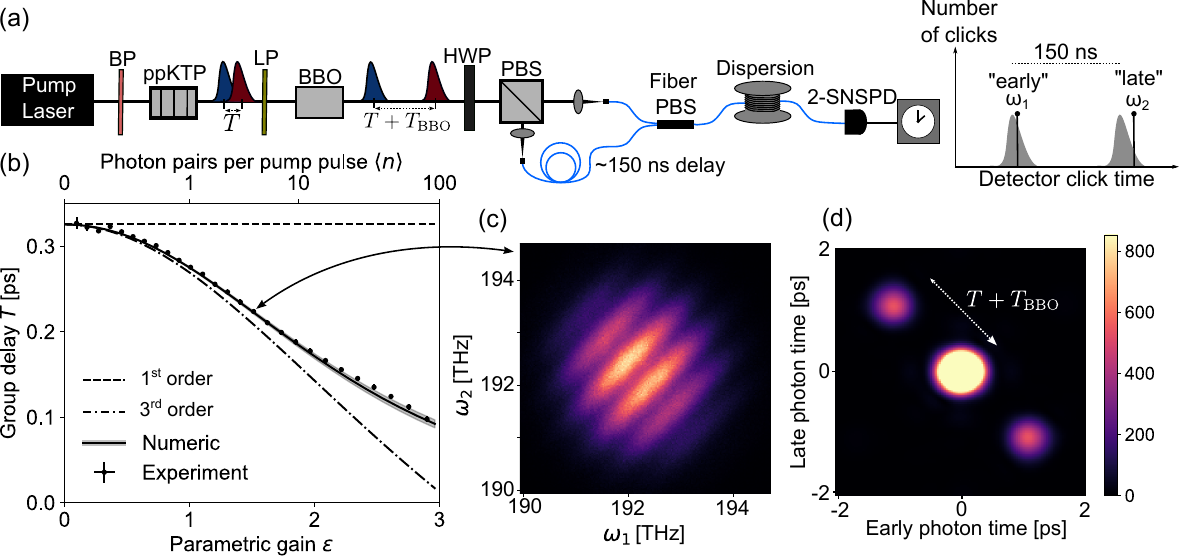}
\caption{(a) Experimental setup. BP: bandpass filter, LP: longpass filter, HWP: half-wave plate, PBS: polarizing beam splitter, 2-SNSPD: two-element superconducting nanowire single photon detector.
BBO crystal is used as a linear optical element to increase the group delay between the signal and idler photons produced by the ppKTP.
Detection time of the photons is related to their frequency due to the dispersive fiber.
(b) Group delay $T$ measured as a function of parametric gain $\varepsilon$. 
Horizontal dashed line is an analytic model limited to the first-order Magnus term $\hat{\Omega}_1$, in which case the expected group delay is simply the group velocity walk-off $\beta_0= L(v_s^{-1} - v_i^{-1})/2$.
Dotted-dashed line is $T=\beta_0 - 24\tau \varepsilon^2 /\sqrt{3\pi}(12+ \varepsilon^2)$ based on Magnus expansion up to $\hat{\Omega}_3$.
Shaded black curve is the numerical model \texttt{NeedALight}~\cite{NeedALight}, where shaded region represents 95\% confidence interval.
(c) Example of a spectrally-resolved coincidence histogram $N(\omega_1, \omega_2)$ measured when $\varepsilon =1.407(3)$.
(d) Fourier transform of $N(\omega_1, \omega_2)$.
Colorbar shows number of events per bin collected in five minutes.
}
\label{fig:fig3}
\end{figure*}

In practice, it is challenging to measure Eq.~\eqref{eqn:interferogram} in a high-gain regime due to optical losses and a lack of spectrally-resolving photon-number-resolving detectors.
If $\ket{\tilde{\Psi}}_{cd}$ is attenuated to the single-photon level such that it can be measured with click detectors, the correlations between photon pairs become muddled.
However, spectrally-resolved coincidences still reveal an interference pattern which can be used to determine the group delay between the two incident pulses~\cite{thekkadath2022measuring}.
In the SM, we show that after attenuation:
\begin{equation}
\mathrm{pr}(\omega_1, \omega_2) \propto [j(\omega_1)j(\omega_2)]^2(1-\mathcal{V}\cos{[\tfrac{T}{2}(\omega_1-\omega_2)]})
\label{eqn:high_gain_coincidences}
\end{equation}
where $j(\omega)=\int d\omega'|J(\omega,\omega')|=\int d\omega'|J(\omega',\omega)|$ is the marginal spectral amplitude of the signal and idler photons.
The interference visibility $\mathcal{V}\leq 1/3$ is a phenomenological parameter that accounts for the signal and idler thermal statistics, dark counts, and any other effect which reduces the fringe contrast.

The experimental setup is shown in Fig~\ref{fig:fig3}(a).
The pump laser is an optical parametric amplifier that generates 180 fs pulses (center wavelength 779\,nm) at a 200\,kHz repetition rate.
The pump bandwidth is set to 5.37(5)\,nm full-width-at-half-maximum using a pair of bandpass filters.
The SPDC source uses a group-velocity matching condition as in Refs.~\cite{gerrits2011generation,weston2016efficient,zhong2020quantum}.
We weakly focus the pump (beam waist $\sim 125\,\mu$m) into a 2-mm-long periodically poled potassium titanyl phosphate (ppKTP) crystal to produce collinear and orthogonally-polarized degenerate photon pairs with center wavelength of 1558\,nm (192\,THz).
The photon pairs are mixed by a half-wave plate and spatially-separated by a polarizing beam splitter (PBS).
The vertical output of the PBS is delayed by about 150 ns and recombined with the horizontal output such that both modes can be detected by the same detector.
To achieve spectrally-resolved detection, we use a dispersive fiber (group delay dispersion 1.033\,ns/nm, insertion loss 3.3\,dB) and a superconducting nanowire detector (jitter 0.1\,ns) to implement a time-of-flight spectrometer~\cite{avenhaus2009fiber}.
This detector is two interleaved but independent nanowires (i.e. a two-element detector) which helps reduce the effect of detector saturation.
We estimate the parametric gain $\varepsilon$ from the detector threshold statistics using a procedure described in the Supplemental Materials~\cite{SM}.
Similarly, using a Klyshko measurement~\cite{klyshko1980use}, we estimate the total (including detection) efficiency of the setup to be $\eta=7(1)$\,\% in both outputs of the PBS.
This efficiency provides sufficient attenuation to measure $\mathrm{pr}(\omega_1, \omega_2)$ [Eq.~\eqref{eqn:high_gain_coincidences}] without saturating the detectors even at the highest gain reached in our experiment ($\varepsilon \sim 3$, $\braket{n} \sim 100$).
Single-mode fibers are used to ensure that we collect photons from the same spatial mode for all gain values.

We scan the pump power from 0 to 60\,mW and record timetags of coincidence events where at least one photon was detected in both outputs of the PBS.
The timetags are converted to a frequency and then placed into a histogram with 30\,GHz wide bins, $N(\omega_1, \omega_2) \propto \mathrm{pr}(\omega_1, \omega_2)$ [Fig.~\ref{fig:fig3}(c)].
The group delay $T$ is measured by taking a Fourier transform of $N(\omega_1, \omega_2)$ and determining the distance between the center peak and the sidebands [Fig.~\ref{fig:fig3}(d)].
We increase this distance beyond the bandwidth of the center peak by inserting a 2-mm-long $\alpha$-phase barium borate (BBO) crystal to introduce a group delay offset $T_{\mathrm{BBO}}=0.833$\,ps.
This offset is then subtracted from the distance to obtain $T$.
The values of $T$ measured at different gains $\varepsilon$ are plotted in Fig.~\ref{fig:fig3}(b).
This is the main result of the paper.
At low gain, we find that $T = 0.327(7)$\,ps which is in agreement with the expected value of $\beta_0= L(v_s^{-1} - v_i^{-1})/2 = 0.325$\,ps given by the birefringent walk-off in the ppKTP~\cite{kato2002sellmeier}.
With increasing gain, the signal and idler photons exit the crystal closer in time, as expected.
The group delay agrees with the analytic third-order Magnus expansion until $\varepsilon \sim 1$.
Beyond this point, we find good agreement with a numerical model which solves the Heisenberg equations of motion (including $\chi^{(3)}$ effects) using \texttt{NeedALight}~\cite{NeedALight}.
All its parameters are determined from independent measurements which are described in the Supplemental Materials~\cite{SM}.
The raw data and code to reproduce the presented results are provided in Ref.~\cite{SPDC_GroupDelay}.

In summary, we demonstrated that photon pairs generated by spontaneous parametric down-conversion exhibit a gain-induced group delay.
While our discussion focused on a bulk crystal pumped by ultrashort pulses, we expect the group delay to affect a broad range of photon pair and squeezed light sources.
For instance, the Hamiltonian in Eq.~\eqref{eqn:hamiltonian} also describes spontaneous four-wave mixing~\cite{quesada2014effects} which is often used in integrated circuits or optical fibers.
Furthermore, the group delay will also affect sources pumped by longer pulses or even continuous-wave light~\cite{brambilla2010spatiotemporal}.
Our results show that the joint spectral amplitude is not simply determined by the product of the phase-matching and pump mode functions in a high-gain regime~\cite{triginer2020understanding,sharapova2020properties,huo2020direct}.
This regime is becoming increasingly relevant for applications like photonic quantum computing~\cite{bourassa2021blueprint}, Gaussian boson sampling~\cite{hamilton2017gaussian,zhong2020quantum,grier2022complexity,zhong2021phase}, interferometry~\cite{chekhova2016nonlinear,thekkadath2020quantum,machado2020optical,qin2023unconditional}, and quantum frequency conversion~\cite{eckstein2011quantum,manurkar2016multidimensional,reddy2017engineering}.
In some of these applications, the group delay introduces distinguishability between the generated photons which can severely reduce the performance of the device.
For example, in a nonlinear interferometer, the signal and idler pulses generated by the first parametric amplifier must be temporally overlapped in the second one to maximize the interference visibility~\cite{machado2020optical}.
With bulk optics, it is relatively straightforward to compensate for the group delay using translation stages.
However, the delay may pose complications when designing quantum interference circuits integrated into chips~\cite{silverstone2014chip,luo2019nonlinear,paesani2020near,arrazola2021quantum,bao2023very}.

\begin{acknowledgments}
We thank Denis Guay and Rune Lausten for their technical support.
N.Q. and M.H. acknowledge support from the MEI du Québec and EU's Horizon under agreement 101070700 project MIRAQLS.
All authors acknowledge financial support from NSERC.
\end{acknowledgments}

\nocite{bulmer2022threshold,jeong2000dynamics,quesada2019broadband,agarwal2013nonlinear,killoran2019strawberry,boyd2003nonlinear,quesada2018gaussian,helt2020degenerate,quesada2015thesis}

\bibliographystyle{apsrev4-2}
\bibliography{refs}

\begin{thebibliography}{101}%
\makeatletter
\providecommand \@ifxundefined [1]{%
 \@ifx{#1\undefined}
}%
\providecommand \@ifnum [1]{%
 \ifnum #1\expandafter \@firstoftwo
 \else \expandafter \@secondoftwo
 \fi
}%
\providecommand \@ifx [1]{%
 \ifx #1\expandafter \@firstoftwo
 \else \expandafter \@secondoftwo
 \fi
}%
\providecommand \natexlab [1]{#1}%
\providecommand \enquote  [1]{``#1''}%
\providecommand \bibnamefont  [1]{#1}%
\providecommand \bibfnamefont [1]{#1}%
\providecommand \citenamefont [1]{#1}%
\providecommand \href@noop [0]{\@secondoftwo}%
\providecommand \href [0]{\begingroup \@sanitize@url \@href}%
\providecommand \@href[1]{\@@startlink{#1}\@@href}%
\providecommand \@@href[1]{\endgroup#1\@@endlink}%
\providecommand \@sanitize@url [0]{\catcode `\\12\catcode `\$12\catcode `\&12\catcode `\#12\catcode `\^12\catcode `\_12\catcode `\%12\relax}%
\providecommand \@@startlink[1]{}%
\providecommand \@@endlink[0]{}%
\providecommand \url  [0]{\begingroup\@sanitize@url \@url }%
\providecommand \@url [1]{\endgroup\@href {#1}{\urlprefix }}%
\providecommand \urlprefix  [0]{URL }%
\providecommand \Eprint [0]{\href }%
\providecommand \doibase [0]{https://doi.org/}%
\providecommand \selectlanguage [0]{\@gobble}%
\providecommand \bibinfo  [0]{\@secondoftwo}%
\providecommand \bibfield  [0]{\@secondoftwo}%
\providecommand \translation [1]{[#1]}%
\providecommand \BibitemOpen [0]{}%
\providecommand \bibitemStop [0]{}%
\providecommand \bibitemNoStop [0]{.\EOS\space}%
\providecommand \EOS [0]{\spacefactor3000\relax}%
\providecommand \BibitemShut  [1]{\csname bibitem#1\endcsname}%
\let\auto@bib@innerbib\@empty
\bibitem [{\citenamefont {Louisell}\ \emph {et~al.}(1961)\citenamefont {Louisell}, \citenamefont {Yariv},\ and\ \citenamefont {Siegman}}]{louisell1961quantum}%
  \BibitemOpen
  \bibfield  {author} {\bibinfo {author} {\bibfnamefont {W.~H.}\ \bibnamefont {Louisell}}, \bibinfo {author} {\bibfnamefont {A.}~\bibnamefont {Yariv}},\ and\ \bibinfo {author} {\bibfnamefont {A.~E.}\ \bibnamefont {Siegman}},\ }\href {https://doi.org/10.1103/PhysRev.124.1646} {\bibfield  {journal} {\bibinfo  {journal} {Phys. Rev.}\ }\textbf {\bibinfo {volume} {124}},\ \bibinfo {pages} {1646} (\bibinfo {year} {1961})}\BibitemShut {NoStop}%
\bibitem [{\citenamefont {Harris}\ \emph {et~al.}(1967)\citenamefont {Harris}, \citenamefont {Oshman},\ and\ \citenamefont {Byer}}]{harris1967observation}%
  \BibitemOpen
  \bibfield  {author} {\bibinfo {author} {\bibfnamefont {S.~E.}\ \bibnamefont {Harris}}, \bibinfo {author} {\bibfnamefont {M.~K.}\ \bibnamefont {Oshman}},\ and\ \bibinfo {author} {\bibfnamefont {R.~L.}\ \bibnamefont {Byer}},\ }\href {https://doi.org/10.1103/PhysRevLett.18.732} {\bibfield  {journal} {\bibinfo  {journal} {Phys. Rev. Lett.}\ }\textbf {\bibinfo {volume} {18}},\ \bibinfo {pages} {732} (\bibinfo {year} {1967})}\BibitemShut {NoStop}%
\bibitem [{\citenamefont {Burnham}\ and\ \citenamefont {Weinberg}(1970)}]{burnham1970observation}%
  \BibitemOpen
  \bibfield  {author} {\bibinfo {author} {\bibfnamefont {D.~C.}\ \bibnamefont {Burnham}}\ and\ \bibinfo {author} {\bibfnamefont {D.~L.}\ \bibnamefont {Weinberg}},\ }\href {https://doi.org/10.1103/PhysRevLett.25.84} {\bibfield  {journal} {\bibinfo  {journal} {Phys. Rev. Lett.}\ }\textbf {\bibinfo {volume} {25}},\ \bibinfo {pages} {84} (\bibinfo {year} {1970})}\BibitemShut {NoStop}%
\bibitem [{\citenamefont {Hong}\ \emph {et~al.}(1987)\citenamefont {Hong}, \citenamefont {Ou},\ and\ \citenamefont {Mandel}}]{hong1987measurement}%
  \BibitemOpen
  \bibfield  {author} {\bibinfo {author} {\bibfnamefont {C.~K.}\ \bibnamefont {Hong}}, \bibinfo {author} {\bibfnamefont {Z.~Y.}\ \bibnamefont {Ou}},\ and\ \bibinfo {author} {\bibfnamefont {L.}~\bibnamefont {Mandel}},\ }\href {https://doi.org/10.1103/PhysRevLett.59.2044} {\bibfield  {journal} {\bibinfo  {journal} {Phys. Rev. Lett.}\ }\textbf {\bibinfo {volume} {59}},\ \bibinfo {pages} {2044} (\bibinfo {year} {1987})}\BibitemShut {NoStop}%
\bibitem [{\citenamefont {Kwiat}\ \emph {et~al.}(1995)\citenamefont {Kwiat}, \citenamefont {Mattle}, \citenamefont {Weinfurter}, \citenamefont {Zeilinger}, \citenamefont {Sergienko},\ and\ \citenamefont {Shih}}]{kwiat1995new}%
  \BibitemOpen
  \bibfield  {author} {\bibinfo {author} {\bibfnamefont {P.~G.}\ \bibnamefont {Kwiat}}, \bibinfo {author} {\bibfnamefont {K.}~\bibnamefont {Mattle}}, \bibinfo {author} {\bibfnamefont {H.}~\bibnamefont {Weinfurter}}, \bibinfo {author} {\bibfnamefont {A.}~\bibnamefont {Zeilinger}}, \bibinfo {author} {\bibfnamefont {A.~V.}\ \bibnamefont {Sergienko}},\ and\ \bibinfo {author} {\bibfnamefont {Y.}~\bibnamefont {Shih}},\ }\href {https://doi.org/10.1103/PhysRevLett.75.4337} {\bibfield  {journal} {\bibinfo  {journal} {Phys. Rev. Lett.}\ }\textbf {\bibinfo {volume} {75}},\ \bibinfo {pages} {4337} (\bibinfo {year} {1995})}\BibitemShut {NoStop}%
\bibitem [{\citenamefont {Li}\ \emph {et~al.}(2004)\citenamefont {Li}, \citenamefont {Chen}, \citenamefont {Voss}, \citenamefont {Sharping},\ and\ \citenamefont {Kumar}}]{li2004all}%
  \BibitemOpen
  \bibfield  {author} {\bibinfo {author} {\bibfnamefont {X.}~\bibnamefont {Li}}, \bibinfo {author} {\bibfnamefont {J.}~\bibnamefont {Chen}}, \bibinfo {author} {\bibfnamefont {P.}~\bibnamefont {Voss}}, \bibinfo {author} {\bibfnamefont {J.}~\bibnamefont {Sharping}},\ and\ \bibinfo {author} {\bibfnamefont {P.}~\bibnamefont {Kumar}},\ }\href {https://doi.org/10.1364/OPEX.12.003737} {\bibfield  {journal} {\bibinfo  {journal} {Opt. Exp.}\ }\textbf {\bibinfo {volume} {12}},\ \bibinfo {pages} {3737} (\bibinfo {year} {2004})}\BibitemShut {NoStop}%
\bibitem [{\citenamefont {Yamada}\ \emph {et~al.}(1993)\citenamefont {Yamada}, \citenamefont {Nada}, \citenamefont {Saitoh},\ and\ \citenamefont {Watanabe}}]{yamada1993first}%
  \BibitemOpen
  \bibfield  {author} {\bibinfo {author} {\bibfnamefont {M.}~\bibnamefont {Yamada}}, \bibinfo {author} {\bibfnamefont {N.}~\bibnamefont {Nada}}, \bibinfo {author} {\bibfnamefont {M.}~\bibnamefont {Saitoh}},\ and\ \bibinfo {author} {\bibfnamefont {K.}~\bibnamefont {Watanabe}},\ }\href {https://doi.org/10.1063/1.108925} {\bibfield  {journal} {\bibinfo  {journal} {App. Phys. Lett.}\ }\textbf {\bibinfo {volume} {62}},\ \bibinfo {pages} {435} (\bibinfo {year} {1993})}\BibitemShut {NoStop}%
\bibitem [{\citenamefont {Grice}\ \emph {et~al.}(2001)\citenamefont {Grice}, \citenamefont {U'Ren},\ and\ \citenamefont {Walmsley}}]{grice2001eliminating}%
  \BibitemOpen
  \bibfield  {author} {\bibinfo {author} {\bibfnamefont {W.~P.}\ \bibnamefont {Grice}}, \bibinfo {author} {\bibfnamefont {A.~B.}\ \bibnamefont {U'Ren}},\ and\ \bibinfo {author} {\bibfnamefont {I.~A.}\ \bibnamefont {Walmsley}},\ }\href {https://doi.org/10.1103/PhysRevA.64.063815} {\bibfield  {journal} {\bibinfo  {journal} {Phys. Rev. A}\ }\textbf {\bibinfo {volume} {64}},\ \bibinfo {pages} {063815} (\bibinfo {year} {2001})}\BibitemShut {NoStop}%
\bibitem [{\citenamefont {Bra{\'n}czyk}\ \emph {et~al.}(2011{\natexlab{a}})\citenamefont {Bra{\'n}czyk}, \citenamefont {Fedrizzi}, \citenamefont {Stace}, \citenamefont {Ralph},\ and\ \citenamefont {White}}]{branczyk2011engineered}%
  \BibitemOpen
  \bibfield  {author} {\bibinfo {author} {\bibfnamefont {A.~M.}\ \bibnamefont {Bra{\'n}czyk}}, \bibinfo {author} {\bibfnamefont {A.}~\bibnamefont {Fedrizzi}}, \bibinfo {author} {\bibfnamefont {T.~M.}\ \bibnamefont {Stace}}, \bibinfo {author} {\bibfnamefont {T.~C.}\ \bibnamefont {Ralph}},\ and\ \bibinfo {author} {\bibfnamefont {A.~G.}\ \bibnamefont {White}},\ }\href {https://doi.org/10.1364/OE.19.000055} {\bibfield  {journal} {\bibinfo  {journal} {Opt. Exp.}\ }\textbf {\bibinfo {volume} {19}},\ \bibinfo {pages} {55} (\bibinfo {year} {2011}{\natexlab{a}})}\BibitemShut {NoStop}%
\bibitem [{\citenamefont {Mosley}\ \emph {et~al.}(2008)\citenamefont {Mosley}, \citenamefont {Lundeen}, \citenamefont {Smith}, \citenamefont {Wasylczyk}, \citenamefont {U'Ren}, \citenamefont {Silberhorn},\ and\ \citenamefont {Walmsley}}]{mosley2008heralded}%
  \BibitemOpen
  \bibfield  {author} {\bibinfo {author} {\bibfnamefont {P.~J.}\ \bibnamefont {Mosley}}, \bibinfo {author} {\bibfnamefont {J.~S.}\ \bibnamefont {Lundeen}}, \bibinfo {author} {\bibfnamefont {B.~J.}\ \bibnamefont {Smith}}, \bibinfo {author} {\bibfnamefont {P.}~\bibnamefont {Wasylczyk}}, \bibinfo {author} {\bibfnamefont {A.~B.}\ \bibnamefont {U'Ren}}, \bibinfo {author} {\bibfnamefont {C.}~\bibnamefont {Silberhorn}},\ and\ \bibinfo {author} {\bibfnamefont {I.~A.}\ \bibnamefont {Walmsley}},\ }\href {https://doi.org/10.1103/PhysRevLett.100.133601} {\bibfield  {journal} {\bibinfo  {journal} {Phys. Rev. Lett.}\ }\textbf {\bibinfo {volume} {100}},\ \bibinfo {pages} {133601} (\bibinfo {year} {2008})}\BibitemShut {NoStop}%
\bibitem [{\citenamefont {Nasr}\ \emph {et~al.}(2008)\citenamefont {Nasr}, \citenamefont {Carrasco}, \citenamefont {Saleh}, \citenamefont {Sergienko}, \citenamefont {Teich}, \citenamefont {Torres}, \citenamefont {Torner}, \citenamefont {Hum},\ and\ \citenamefont {Fejer}}]{nasr2008ultra}%
  \BibitemOpen
  \bibfield  {author} {\bibinfo {author} {\bibfnamefont {M.~B.}\ \bibnamefont {Nasr}}, \bibinfo {author} {\bibfnamefont {S.}~\bibnamefont {Carrasco}}, \bibinfo {author} {\bibfnamefont {B.~E.~A.}\ \bibnamefont {Saleh}}, \bibinfo {author} {\bibfnamefont {A.~V.}\ \bibnamefont {Sergienko}}, \bibinfo {author} {\bibfnamefont {M.~C.}\ \bibnamefont {Teich}}, \bibinfo {author} {\bibfnamefont {J.~P.}\ \bibnamefont {Torres}}, \bibinfo {author} {\bibfnamefont {L.}~\bibnamefont {Torner}}, \bibinfo {author} {\bibfnamefont {D.~S.}\ \bibnamefont {Hum}},\ and\ \bibinfo {author} {\bibfnamefont {M.~M.}\ \bibnamefont {Fejer}},\ }\href {https://doi.org/10.1103/PhysRevLett.100.183601} {\bibfield  {journal} {\bibinfo  {journal} {Phys. Rev. Lett.}\ }\textbf {\bibinfo {volume} {100}},\ \bibinfo {pages} {183601} (\bibinfo {year} {2008})}\BibitemShut {NoStop}%
\bibitem [{\citenamefont {Devaux}\ \emph {et~al.}(2020)\citenamefont {Devaux}, \citenamefont {Mosset}, \citenamefont {Moreau},\ and\ \citenamefont {Lantz}}]{devaux2020imaging}%
  \BibitemOpen
  \bibfield  {author} {\bibinfo {author} {\bibfnamefont {F.}~\bibnamefont {Devaux}}, \bibinfo {author} {\bibfnamefont {A.}~\bibnamefont {Mosset}}, \bibinfo {author} {\bibfnamefont {P.-A.}\ \bibnamefont {Moreau}},\ and\ \bibinfo {author} {\bibfnamefont {E.}~\bibnamefont {Lantz}},\ }\href {https://doi.org/10.1103/PhysRevX.10.031031} {\bibfield  {journal} {\bibinfo  {journal} {Phys. Rev. X}\ }\textbf {\bibinfo {volume} {10}},\ \bibinfo {pages} {031031} (\bibinfo {year} {2020})}\BibitemShut {NoStop}%
\bibitem [{\citenamefont {Bennink}(2010)}]{bennink2010optimal}%
  \BibitemOpen
  \bibfield  {author} {\bibinfo {author} {\bibfnamefont {R.~S.}\ \bibnamefont {Bennink}},\ }\href {https://doi.org/10.1103/PhysRevA.81.053805} {\bibfield  {journal} {\bibinfo  {journal} {Phys. Rev. A}\ }\textbf {\bibinfo {volume} {81}},\ \bibinfo {pages} {053805} (\bibinfo {year} {2010})}\BibitemShut {NoStop}%
\bibitem [{\citenamefont {Bouwmeester}\ \emph {et~al.}(1997)\citenamefont {Bouwmeester}, \citenamefont {Pan}, \citenamefont {Mattle}, \citenamefont {Eibl}, \citenamefont {Weinfurter},\ and\ \citenamefont {Zeilinger}}]{bouwmeester1997experimental}%
  \BibitemOpen
  \bibfield  {author} {\bibinfo {author} {\bibfnamefont {D.}~\bibnamefont {Bouwmeester}}, \bibinfo {author} {\bibfnamefont {J.-W.}\ \bibnamefont {Pan}}, \bibinfo {author} {\bibfnamefont {K.}~\bibnamefont {Mattle}}, \bibinfo {author} {\bibfnamefont {M.}~\bibnamefont {Eibl}}, \bibinfo {author} {\bibfnamefont {H.}~\bibnamefont {Weinfurter}},\ and\ \bibinfo {author} {\bibfnamefont {A.}~\bibnamefont {Zeilinger}},\ }\href {https://doi.org/10.1038/37539} {\bibfield  {journal} {\bibinfo  {journal} {Nature}\ }\textbf {\bibinfo {volume} {390}},\ \bibinfo {pages} {575} (\bibinfo {year} {1997})}\BibitemShut {NoStop}%
\bibitem [{\citenamefont {Shalm}\ \emph {et~al.}(2015)\citenamefont {Shalm}, \citenamefont {Meyer-Scott}, \citenamefont {Christensen}, \citenamefont {Bierhorst}, \citenamefont {Wayne}, \citenamefont {Stevens}, \citenamefont {Gerrits}, \citenamefont {Glancy}, \citenamefont {Hamel}, \citenamefont {Allman} \emph {et~al.}}]{shalm2015strong}%
  \BibitemOpen
  \bibfield  {author} {\bibinfo {author} {\bibfnamefont {L.~K.}\ \bibnamefont {Shalm}}, \bibinfo {author} {\bibfnamefont {E.}~\bibnamefont {Meyer-Scott}}, \bibinfo {author} {\bibfnamefont {B.~G.}\ \bibnamefont {Christensen}}, \bibinfo {author} {\bibfnamefont {P.}~\bibnamefont {Bierhorst}}, \bibinfo {author} {\bibfnamefont {M.~A.}\ \bibnamefont {Wayne}}, \bibinfo {author} {\bibfnamefont {M.~J.}\ \bibnamefont {Stevens}}, \bibinfo {author} {\bibfnamefont {T.}~\bibnamefont {Gerrits}}, \bibinfo {author} {\bibfnamefont {S.}~\bibnamefont {Glancy}}, \bibinfo {author} {\bibfnamefont {D.~R.}\ \bibnamefont {Hamel}}, \bibinfo {author} {\bibfnamefont {M.~S.}\ \bibnamefont {Allman}}, \emph {et~al.},\ }\href {https://doi.org/10.1103/PhysRevLett.115.250402} {\bibfield  {journal} {\bibinfo  {journal} {Phys. Rev. Lett.}\ }\textbf {\bibinfo {volume} {115}},\ \bibinfo {pages} {250402} (\bibinfo {year} {2015})}\BibitemShut {NoStop}%
\bibitem [{\citenamefont {Giustina}\ \emph {et~al.}(2015)\citenamefont {Giustina}, \citenamefont {Versteegh}, \citenamefont {Wengerowsky}, \citenamefont {Handsteiner}, \citenamefont {Hochrainer}, \citenamefont {Phelan}, \citenamefont {Steinlechner}, \citenamefont {Kofler}, \citenamefont {Larsson}, \citenamefont {Abell\'an} \emph {et~al.}}]{giustina2015significant}%
  \BibitemOpen
  \bibfield  {author} {\bibinfo {author} {\bibfnamefont {M.}~\bibnamefont {Giustina}}, \bibinfo {author} {\bibfnamefont {M.~A.~M.}\ \bibnamefont {Versteegh}}, \bibinfo {author} {\bibfnamefont {S.}~\bibnamefont {Wengerowsky}}, \bibinfo {author} {\bibfnamefont {J.}~\bibnamefont {Handsteiner}}, \bibinfo {author} {\bibfnamefont {A.}~\bibnamefont {Hochrainer}}, \bibinfo {author} {\bibfnamefont {K.}~\bibnamefont {Phelan}}, \bibinfo {author} {\bibfnamefont {F.}~\bibnamefont {Steinlechner}}, \bibinfo {author} {\bibfnamefont {J.}~\bibnamefont {Kofler}}, \bibinfo {author} {\bibfnamefont {J.-A.}\ \bibnamefont {Larsson}}, \bibinfo {author} {\bibfnamefont {C.}~\bibnamefont {Abell\'an}}, \emph {et~al.},\ }\href {https://doi.org/10.1103/PhysRevLett.115.250401} {\bibfield  {journal} {\bibinfo  {journal} {Phys. Rev. Lett.}\ }\textbf {\bibinfo {volume} {115}},\ \bibinfo {pages} {250401} (\bibinfo {year} {2015})}\BibitemShut {NoStop}%
\bibitem [{\citenamefont {Slusher}\ \emph {et~al.}(1987)\citenamefont {Slusher}, \citenamefont {Grangier}, \citenamefont {LaPorta}, \citenamefont {Yurke},\ and\ \citenamefont {Potasek}}]{slusher1987pulsed}%
  \BibitemOpen
  \bibfield  {author} {\bibinfo {author} {\bibfnamefont {R.~E.}\ \bibnamefont {Slusher}}, \bibinfo {author} {\bibfnamefont {P.}~\bibnamefont {Grangier}}, \bibinfo {author} {\bibfnamefont {A.}~\bibnamefont {LaPorta}}, \bibinfo {author} {\bibfnamefont {B.}~\bibnamefont {Yurke}},\ and\ \bibinfo {author} {\bibfnamefont {M.~J.}\ \bibnamefont {Potasek}},\ }\href {https://doi.org/10.1103/PhysRevLett.59.2566} {\bibfield  {journal} {\bibinfo  {journal} {Phys. Rev. Lett.}\ }\textbf {\bibinfo {volume} {59}},\ \bibinfo {pages} {2566} (\bibinfo {year} {1987})}\BibitemShut {NoStop}%
\bibitem [{\citenamefont {Kim}\ and\ \citenamefont {Kumar}(1994)}]{kim1994quadrature}%
  \BibitemOpen
  \bibfield  {author} {\bibinfo {author} {\bibfnamefont {C.}~\bibnamefont {Kim}}\ and\ \bibinfo {author} {\bibfnamefont {P.}~\bibnamefont {Kumar}},\ }\href {https://doi.org/10.1103/PhysRevLett.73.1605} {\bibfield  {journal} {\bibinfo  {journal} {Phys. Rev. Lett.}\ }\textbf {\bibinfo {volume} {73}},\ \bibinfo {pages} {1605} (\bibinfo {year} {1994})}\BibitemShut {NoStop}%
\bibitem [{\citenamefont {Zhong}\ \emph {et~al.}(2021)\citenamefont {Zhong}, \citenamefont {Deng}, \citenamefont {Qin}, \citenamefont {Wang}, \citenamefont {Chen}, \citenamefont {Peng}, \citenamefont {Luo}, \citenamefont {Wu}, \citenamefont {Gong}, \citenamefont {Su} \emph {et~al.}}]{zhong2021phase}%
  \BibitemOpen
  \bibfield  {author} {\bibinfo {author} {\bibfnamefont {H.-S.}\ \bibnamefont {Zhong}}, \bibinfo {author} {\bibfnamefont {Y.-H.}\ \bibnamefont {Deng}}, \bibinfo {author} {\bibfnamefont {J.}~\bibnamefont {Qin}}, \bibinfo {author} {\bibfnamefont {H.}~\bibnamefont {Wang}}, \bibinfo {author} {\bibfnamefont {M.-C.}\ \bibnamefont {Chen}}, \bibinfo {author} {\bibfnamefont {L.-C.}\ \bibnamefont {Peng}}, \bibinfo {author} {\bibfnamefont {Y.-H.}\ \bibnamefont {Luo}}, \bibinfo {author} {\bibfnamefont {D.}~\bibnamefont {Wu}}, \bibinfo {author} {\bibfnamefont {S.-Q.}\ \bibnamefont {Gong}}, \bibinfo {author} {\bibfnamefont {H.}~\bibnamefont {Su}}, \emph {et~al.},\ }\href {https://doi.org/10.1103/PhysRevLett.127.180502} {\bibfield  {journal} {\bibinfo  {journal} {Phys. Rev. Lett.}\ }\textbf {\bibinfo {volume} {127}},\ \bibinfo {pages} {180502} (\bibinfo {year} {2021})}\BibitemShut {NoStop}%
\bibitem [{\citenamefont {Wu}\ \emph {et~al.}(1986)\citenamefont {Wu}, \citenamefont {Kimble}, \citenamefont {Hall},\ and\ \citenamefont {Wu}}]{wu1986generation}%
  \BibitemOpen
  \bibfield  {author} {\bibinfo {author} {\bibfnamefont {L.-A.}\ \bibnamefont {Wu}}, \bibinfo {author} {\bibfnamefont {H.~J.}\ \bibnamefont {Kimble}}, \bibinfo {author} {\bibfnamefont {J.~L.}\ \bibnamefont {Hall}},\ and\ \bibinfo {author} {\bibfnamefont {H.}~\bibnamefont {Wu}},\ }\href {https://doi.org/10.1103/PhysRevLett.57.2520} {\bibfield  {journal} {\bibinfo  {journal} {Phys. Rev. Lett.}\ }\textbf {\bibinfo {volume} {57}},\ \bibinfo {pages} {2520} (\bibinfo {year} {1986})}\BibitemShut {NoStop}%
\bibitem [{\citenamefont {Vahlbruch}\ \emph {et~al.}(2016)\citenamefont {Vahlbruch}, \citenamefont {Mehmet}, \citenamefont {Danzmann},\ and\ \citenamefont {Schnabel}}]{vahlbruch2016detection}%
  \BibitemOpen
  \bibfield  {author} {\bibinfo {author} {\bibfnamefont {H.}~\bibnamefont {Vahlbruch}}, \bibinfo {author} {\bibfnamefont {M.}~\bibnamefont {Mehmet}}, \bibinfo {author} {\bibfnamefont {K.}~\bibnamefont {Danzmann}},\ and\ \bibinfo {author} {\bibfnamefont {R.}~\bibnamefont {Schnabel}},\ }\href {https://doi.org/10.1103/PhysRevLett.117.110801} {\bibfield  {journal} {\bibinfo  {journal} {Phys. Rev. Lett.}\ }\textbf {\bibinfo {volume} {117}},\ \bibinfo {pages} {110801} (\bibinfo {year} {2016})}\BibitemShut {NoStop}%
\bibitem [{\citenamefont {Zhang}\ \emph {et~al.}(2021)\citenamefont {Zhang}, \citenamefont {Menotti}, \citenamefont {Tan}, \citenamefont {Vaidya}, \citenamefont {Mahler}, \citenamefont {Helt}, \citenamefont {Zatti}, \citenamefont {Liscidini}, \citenamefont {Morrison},\ and\ \citenamefont {Vernon}}]{zhang2021squeezed}%
  \BibitemOpen
  \bibfield  {author} {\bibinfo {author} {\bibfnamefont {Y.}~\bibnamefont {Zhang}}, \bibinfo {author} {\bibfnamefont {M.}~\bibnamefont {Menotti}}, \bibinfo {author} {\bibfnamefont {K.}~\bibnamefont {Tan}}, \bibinfo {author} {\bibfnamefont {V.}~\bibnamefont {Vaidya}}, \bibinfo {author} {\bibfnamefont {D.}~\bibnamefont {Mahler}}, \bibinfo {author} {\bibfnamefont {L.}~\bibnamefont {Helt}}, \bibinfo {author} {\bibfnamefont {L.}~\bibnamefont {Zatti}}, \bibinfo {author} {\bibfnamefont {M.}~\bibnamefont {Liscidini}}, \bibinfo {author} {\bibfnamefont {B.}~\bibnamefont {Morrison}},\ and\ \bibinfo {author} {\bibfnamefont {Z.}~\bibnamefont {Vernon}},\ }\href {https://doi.org/10.1038/s41467-021-22540-2} {\bibfield  {journal} {\bibinfo  {journal} {Nat. Commun.}\ }\textbf {\bibinfo {volume} {12}},\ \bibinfo {pages} {2233} (\bibinfo {year} {2021})}\BibitemShut {NoStop}%
\bibitem [{\citenamefont {Arrazola}\ \emph {et~al.}(2021)\citenamefont {Arrazola}, \citenamefont {Bergholm}, \citenamefont {Br{\'a}dler}, \citenamefont {Bromley}, \citenamefont {Collins}, \citenamefont {Dhand}, \citenamefont {Fumagalli}, \citenamefont {Gerrits}, \citenamefont {Goussev}, \citenamefont {Helt} \emph {et~al.}}]{arrazola2021quantum}%
  \BibitemOpen
  \bibfield  {author} {\bibinfo {author} {\bibfnamefont {J.~M.}\ \bibnamefont {Arrazola}}, \bibinfo {author} {\bibfnamefont {V.}~\bibnamefont {Bergholm}}, \bibinfo {author} {\bibfnamefont {K.}~\bibnamefont {Br{\'a}dler}}, \bibinfo {author} {\bibfnamefont {T.~R.}\ \bibnamefont {Bromley}}, \bibinfo {author} {\bibfnamefont {M.~J.}\ \bibnamefont {Collins}}, \bibinfo {author} {\bibfnamefont {I.}~\bibnamefont {Dhand}}, \bibinfo {author} {\bibfnamefont {A.}~\bibnamefont {Fumagalli}}, \bibinfo {author} {\bibfnamefont {T.}~\bibnamefont {Gerrits}}, \bibinfo {author} {\bibfnamefont {A.}~\bibnamefont {Goussev}}, \bibinfo {author} {\bibfnamefont {L.~G.}\ \bibnamefont {Helt}}, \emph {et~al.},\ }\href {https://doi.org/10.1038/s41586-021-03202-1} {\bibfield  {journal} {\bibinfo  {journal} {Nature}\ }\textbf {\bibinfo {volume} {591}},\ \bibinfo {pages} {54} (\bibinfo {year} {2021})}\BibitemShut {NoStop}%
\bibitem [{\citenamefont {Eckstein}\ \emph {et~al.}(2011{\natexlab{a}})\citenamefont {Eckstein}, \citenamefont {Christ}, \citenamefont {Mosley},\ and\ \citenamefont {Silberhorn}}]{eckstein2011highly}%
  \BibitemOpen
  \bibfield  {author} {\bibinfo {author} {\bibfnamefont {A.}~\bibnamefont {Eckstein}}, \bibinfo {author} {\bibfnamefont {A.}~\bibnamefont {Christ}}, \bibinfo {author} {\bibfnamefont {P.~J.}\ \bibnamefont {Mosley}},\ and\ \bibinfo {author} {\bibfnamefont {C.}~\bibnamefont {Silberhorn}},\ }\href {https://doi.org/10.1103/PhysRevLett.106.013603} {\bibfield  {journal} {\bibinfo  {journal} {Phys. Rev. Lett.}\ }\textbf {\bibinfo {volume} {106}},\ \bibinfo {pages} {013603} (\bibinfo {year} {2011}{\natexlab{a}})}\BibitemShut {NoStop}%
\bibitem [{\citenamefont {Harder}\ \emph {et~al.}(2016)\citenamefont {Harder}, \citenamefont {Bartley}, \citenamefont {Lita}, \citenamefont {Nam}, \citenamefont {Gerrits},\ and\ \citenamefont {Silberhorn}}]{harder2016single}%
  \BibitemOpen
  \bibfield  {author} {\bibinfo {author} {\bibfnamefont {G.}~\bibnamefont {Harder}}, \bibinfo {author} {\bibfnamefont {T.~J.}\ \bibnamefont {Bartley}}, \bibinfo {author} {\bibfnamefont {A.~E.}\ \bibnamefont {Lita}}, \bibinfo {author} {\bibfnamefont {S.~W.}\ \bibnamefont {Nam}}, \bibinfo {author} {\bibfnamefont {T.}~\bibnamefont {Gerrits}},\ and\ \bibinfo {author} {\bibfnamefont {C.}~\bibnamefont {Silberhorn}},\ }\href {https://doi.org/10.1103/PhysRevLett.116.143601} {\bibfield  {journal} {\bibinfo  {journal} {Phys. Rev. Lett.}\ }\textbf {\bibinfo {volume} {116}},\ \bibinfo {pages} {143601} (\bibinfo {year} {2016})}\BibitemShut {NoStop}%
\bibitem [{\citenamefont {Paesani}\ \emph {et~al.}(2020)\citenamefont {Paesani}, \citenamefont {Borghi}, \citenamefont {Signorini}, \citenamefont {Ma{\"\i}nos}, \citenamefont {Pavesi},\ and\ \citenamefont {Laing}}]{paesani2020near}%
  \BibitemOpen
  \bibfield  {author} {\bibinfo {author} {\bibfnamefont {S.}~\bibnamefont {Paesani}}, \bibinfo {author} {\bibfnamefont {M.}~\bibnamefont {Borghi}}, \bibinfo {author} {\bibfnamefont {S.}~\bibnamefont {Signorini}}, \bibinfo {author} {\bibfnamefont {A.}~\bibnamefont {Ma{\"\i}nos}}, \bibinfo {author} {\bibfnamefont {L.}~\bibnamefont {Pavesi}},\ and\ \bibinfo {author} {\bibfnamefont {A.}~\bibnamefont {Laing}},\ }\href {https://doi.org/10.1038/s41467-020-16187-8} {\bibfield  {journal} {\bibinfo  {journal} {Nat. Commun.}\ }\textbf {\bibinfo {volume} {11}},\ \bibinfo {pages} {2505} (\bibinfo {year} {2020})}\BibitemShut {NoStop}%
\bibitem [{\citenamefont {Ganapathy}\ \emph {et~al.}(2023)\citenamefont {Ganapathy}, \citenamefont {Jia}, \citenamefont {Nakano}, \citenamefont {Xu}, \citenamefont {Aritomi}, \citenamefont {Cullen}, \citenamefont {Kijbunchoo}, \citenamefont {Dwyer}, \citenamefont {Mullavey}, \citenamefont {McCuller} \emph {et~al.}}]{ganapathy2023broadband}%
  \BibitemOpen
  \bibfield  {author} {\bibinfo {author} {\bibfnamefont {D.}~\bibnamefont {Ganapathy}}, \bibinfo {author} {\bibfnamefont {W.}~\bibnamefont {Jia}}, \bibinfo {author} {\bibfnamefont {M.}~\bibnamefont {Nakano}}, \bibinfo {author} {\bibfnamefont {V.}~\bibnamefont {Xu}}, \bibinfo {author} {\bibfnamefont {N.}~\bibnamefont {Aritomi}}, \bibinfo {author} {\bibfnamefont {T.}~\bibnamefont {Cullen}}, \bibinfo {author} {\bibfnamefont {N.}~\bibnamefont {Kijbunchoo}}, \bibinfo {author} {\bibfnamefont {S.~E.}\ \bibnamefont {Dwyer}}, \bibinfo {author} {\bibfnamefont {A.}~\bibnamefont {Mullavey}}, \bibinfo {author} {\bibfnamefont {L.}~\bibnamefont {McCuller}}, \emph {et~al.} (\bibinfo {collaboration} {LIGO O4 Detector Collaboration}),\ }\href {https://doi.org/10.1103/PhysRevX.13.041021} {\bibfield  {journal} {\bibinfo  {journal} {Phys. Rev. X}\ }\textbf {\bibinfo {volume} {13}},\ \bibinfo {pages} {041021} (\bibinfo {year} {2023})}\BibitemShut {NoStop}%
\bibitem [{\citenamefont {Menicucci}\ \emph {et~al.}(2006)\citenamefont {Menicucci}, \citenamefont {van Loock}, \citenamefont {Gu}, \citenamefont {Weedbrook}, \citenamefont {Ralph},\ and\ \citenamefont {Nielsen}}]{menicucci2006universal}%
  \BibitemOpen
  \bibfield  {author} {\bibinfo {author} {\bibfnamefont {N.~C.}\ \bibnamefont {Menicucci}}, \bibinfo {author} {\bibfnamefont {P.}~\bibnamefont {van Loock}}, \bibinfo {author} {\bibfnamefont {M.}~\bibnamefont {Gu}}, \bibinfo {author} {\bibfnamefont {C.}~\bibnamefont {Weedbrook}}, \bibinfo {author} {\bibfnamefont {T.~C.}\ \bibnamefont {Ralph}},\ and\ \bibinfo {author} {\bibfnamefont {M.~A.}\ \bibnamefont {Nielsen}},\ }\href {https://doi.org/10.1103/PhysRevLett.97.110501} {\bibfield  {journal} {\bibinfo  {journal} {Phys. Rev. Lett.}\ }\textbf {\bibinfo {volume} {97}},\ \bibinfo {pages} {110501} (\bibinfo {year} {2006})}\BibitemShut {NoStop}%
\bibitem [{\citenamefont {Bourassa}\ \emph {et~al.}(2021)\citenamefont {Bourassa}, \citenamefont {Alexander}, \citenamefont {Vasmer}, \citenamefont {Patil}, \citenamefont {Tzitrin}, \citenamefont {Matsuura}, \citenamefont {Su}, \citenamefont {Baragiola}, \citenamefont {Guha}, \citenamefont {Dauphinais} \emph {et~al.}}]{bourassa2021blueprint}%
  \BibitemOpen
  \bibfield  {author} {\bibinfo {author} {\bibfnamefont {J.~E.}\ \bibnamefont {Bourassa}}, \bibinfo {author} {\bibfnamefont {R.~N.}\ \bibnamefont {Alexander}}, \bibinfo {author} {\bibfnamefont {M.}~\bibnamefont {Vasmer}}, \bibinfo {author} {\bibfnamefont {A.}~\bibnamefont {Patil}}, \bibinfo {author} {\bibfnamefont {I.}~\bibnamefont {Tzitrin}}, \bibinfo {author} {\bibfnamefont {T.}~\bibnamefont {Matsuura}}, \bibinfo {author} {\bibfnamefont {D.}~\bibnamefont {Su}}, \bibinfo {author} {\bibfnamefont {B.~Q.}\ \bibnamefont {Baragiola}}, \bibinfo {author} {\bibfnamefont {S.}~\bibnamefont {Guha}}, \bibinfo {author} {\bibfnamefont {G.}~\bibnamefont {Dauphinais}}, \emph {et~al.},\ }\href {https://doi.org/10.22331/q-2021-02-04-392} {\bibfield  {journal} {\bibinfo  {journal} {Quantum}\ }\textbf {\bibinfo {volume} {5}},\ \bibinfo {pages} {392} (\bibinfo {year} {2021})}\BibitemShut {NoStop}%
\bibitem [{\citenamefont {La~Porta}\ and\ \citenamefont {Slusher}(1991)}]{laporta1991squeezing}%
  \BibitemOpen
  \bibfield  {author} {\bibinfo {author} {\bibfnamefont {A.}~\bibnamefont {La~Porta}}\ and\ \bibinfo {author} {\bibfnamefont {R.~E.}\ \bibnamefont {Slusher}},\ }\href {https://doi.org/10.1103/PhysRevA.44.2013} {\bibfield  {journal} {\bibinfo  {journal} {Phys. Rev. A}\ }\textbf {\bibinfo {volume} {44}},\ \bibinfo {pages} {2013} (\bibinfo {year} {1991})}\BibitemShut {NoStop}%
\bibitem [{\citenamefont {Wasilewski}\ \emph {et~al.}(2006)\citenamefont {Wasilewski}, \citenamefont {Lvovsky}, \citenamefont {Banaszek},\ and\ \citenamefont {Radzewicz}}]{wasilewski2006pulsed}%
  \BibitemOpen
  \bibfield  {author} {\bibinfo {author} {\bibfnamefont {W.}~\bibnamefont {Wasilewski}}, \bibinfo {author} {\bibfnamefont {A.~I.}\ \bibnamefont {Lvovsky}}, \bibinfo {author} {\bibfnamefont {K.}~\bibnamefont {Banaszek}},\ and\ \bibinfo {author} {\bibfnamefont {C.}~\bibnamefont {Radzewicz}},\ }\href {https://doi.org/10.1103/PhysRevA.73.063819} {\bibfield  {journal} {\bibinfo  {journal} {Phys. Rev. A}\ }\textbf {\bibinfo {volume} {73}},\ \bibinfo {pages} {063819} (\bibinfo {year} {2006})}\BibitemShut {NoStop}%
\bibitem [{\citenamefont {Dayan}(2007)}]{dayan2007theory}%
  \BibitemOpen
  \bibfield  {author} {\bibinfo {author} {\bibfnamefont {B.}~\bibnamefont {Dayan}},\ }\href {https://doi.org/10.1103/PhysRevA.76.043813} {\bibfield  {journal} {\bibinfo  {journal} {Phys. Rev. A}\ }\textbf {\bibinfo {volume} {76}},\ \bibinfo {pages} {043813} (\bibinfo {year} {2007})}\BibitemShut {NoStop}%
\bibitem [{\citenamefont {Cassemiro}\ \emph {et~al.}(2007)\citenamefont {Cassemiro}, \citenamefont {Villar}, \citenamefont {Martinelli},\ and\ \citenamefont {Nussenzveig}}]{cassemiro2007quest}%
  \BibitemOpen
  \bibfield  {author} {\bibinfo {author} {\bibfnamefont {K.~N.}\ \bibnamefont {Cassemiro}}, \bibinfo {author} {\bibfnamefont {A.}~\bibnamefont {Villar}}, \bibinfo {author} {\bibfnamefont {M.}~\bibnamefont {Martinelli}},\ and\ \bibinfo {author} {\bibfnamefont {P.}~\bibnamefont {Nussenzveig}},\ }\href {https://doi.org/10.1364/OE.15.018236} {\bibfield  {journal} {\bibinfo  {journal} {Opt. Exp.}\ }\textbf {\bibinfo {volume} {15}},\ \bibinfo {pages} {18236} (\bibinfo {year} {2007})}\BibitemShut {NoStop}%
\bibitem [{\citenamefont {Leung}\ \emph {et~al.}(2009)\citenamefont {Leung}, \citenamefont {Munro}, \citenamefont {Nemoto},\ and\ \citenamefont {Ralph}}]{leung2009spectral}%
  \BibitemOpen
  \bibfield  {author} {\bibinfo {author} {\bibfnamefont {P.~M.}\ \bibnamefont {Leung}}, \bibinfo {author} {\bibfnamefont {W.~J.}\ \bibnamefont {Munro}}, \bibinfo {author} {\bibfnamefont {K.}~\bibnamefont {Nemoto}},\ and\ \bibinfo {author} {\bibfnamefont {T.~C.}\ \bibnamefont {Ralph}},\ }\href {https://doi.org/10.1103/PhysRevA.79.042307} {\bibfield  {journal} {\bibinfo  {journal} {Phys. Rev. A}\ }\textbf {\bibinfo {volume} {79}},\ \bibinfo {pages} {042307} (\bibinfo {year} {2009})}\BibitemShut {NoStop}%
\bibitem [{\citenamefont {Iskhakov}\ \emph {et~al.}(2009)\citenamefont {Iskhakov}, \citenamefont {Chekhova},\ and\ \citenamefont {Leuchs}}]{iskhakov2009generation}%
  \BibitemOpen
  \bibfield  {author} {\bibinfo {author} {\bibfnamefont {T.}~\bibnamefont {Iskhakov}}, \bibinfo {author} {\bibfnamefont {M.~V.}\ \bibnamefont {Chekhova}},\ and\ \bibinfo {author} {\bibfnamefont {G.}~\bibnamefont {Leuchs}},\ }\href {https://doi.org/10.1103/PhysRevLett.102.183602} {\bibfield  {journal} {\bibinfo  {journal} {Phys. Rev. Lett.}\ }\textbf {\bibinfo {volume} {102}},\ \bibinfo {pages} {183602} (\bibinfo {year} {2009})}\BibitemShut {NoStop}%
\bibitem [{\citenamefont {Brambilla}\ \emph {et~al.}(2010)\citenamefont {Brambilla}, \citenamefont {Caspani}, \citenamefont {Lugiato},\ and\ \citenamefont {Gatti}}]{brambilla2010spatiotemporal}%
  \BibitemOpen
  \bibfield  {author} {\bibinfo {author} {\bibfnamefont {E.}~\bibnamefont {Brambilla}}, \bibinfo {author} {\bibfnamefont {L.}~\bibnamefont {Caspani}}, \bibinfo {author} {\bibfnamefont {L.~A.}\ \bibnamefont {Lugiato}},\ and\ \bibinfo {author} {\bibfnamefont {A.}~\bibnamefont {Gatti}},\ }\href {https://doi.org/10.1103/PhysRevA.82.013835} {\bibfield  {journal} {\bibinfo  {journal} {Phys. Rev. A}\ }\textbf {\bibinfo {volume} {82}},\ \bibinfo {pages} {013835} (\bibinfo {year} {2010})}\BibitemShut {NoStop}%
\bibitem [{\citenamefont {Bra{\'n}czyk}\ \emph {et~al.}(2011{\natexlab{b}})\citenamefont {Bra{\'n}czyk}, \citenamefont {Stace},\ and\ \citenamefont {Ralph}}]{branczyk2011time}%
  \BibitemOpen
  \bibfield  {author} {\bibinfo {author} {\bibfnamefont {A.~M.}\ \bibnamefont {Bra{\'n}czyk}}, \bibinfo {author} {\bibfnamefont {T.~M.}\ \bibnamefont {Stace}},\ and\ \bibinfo {author} {\bibfnamefont {T.}~\bibnamefont {Ralph}},\ }in\ \href {https://doi.org/10.1063/1.3630207} {\emph {\bibinfo {booktitle} {AIP Conference Proceedings}}},\ Vol.\ \bibinfo {volume} {1363}\ (\bibinfo {organization} {American Institute of Physics},\ \bibinfo {year} {2011})\ pp.\ \bibinfo {pages} {335--338}\BibitemShut {NoStop}%
\bibitem [{\citenamefont {Spasibko}\ \emph {et~al.}(2012)\citenamefont {Spasibko}, \citenamefont {Iskhakov},\ and\ \citenamefont {Chekhova}}]{spasibko2012spectral}%
  \BibitemOpen
  \bibfield  {author} {\bibinfo {author} {\bibfnamefont {K.~Y.}\ \bibnamefont {Spasibko}}, \bibinfo {author} {\bibfnamefont {T.~S.}\ \bibnamefont {Iskhakov}},\ and\ \bibinfo {author} {\bibfnamefont {M.~V.}\ \bibnamefont {Chekhova}},\ }\href {https://doi.org/10.1364/OE.20.007507} {\bibfield  {journal} {\bibinfo  {journal} {Opt. Exp.}\ }\textbf {\bibinfo {volume} {20}},\ \bibinfo {pages} {7507} (\bibinfo {year} {2012})}\BibitemShut {NoStop}%
\bibitem [{\citenamefont {Iskhakov}\ \emph {et~al.}(2012)\citenamefont {Iskhakov}, \citenamefont {P{\'e}rez}, \citenamefont {Spasibko}, \citenamefont {Chekhova},\ and\ \citenamefont {Leuchs}}]{iskhakov2012superbunched}%
  \BibitemOpen
  \bibfield  {author} {\bibinfo {author} {\bibfnamefont {T.~S.}\ \bibnamefont {Iskhakov}}, \bibinfo {author} {\bibfnamefont {A.}~\bibnamefont {P{\'e}rez}}, \bibinfo {author} {\bibfnamefont {K.~Y.}\ \bibnamefont {Spasibko}}, \bibinfo {author} {\bibfnamefont {M.}~\bibnamefont {Chekhova}},\ and\ \bibinfo {author} {\bibfnamefont {G.}~\bibnamefont {Leuchs}},\ }\href {https://doi.org/10.1364/OL.37.001919} {\bibfield  {journal} {\bibinfo  {journal} {Opt. Lett.}\ }\textbf {\bibinfo {volume} {37}},\ \bibinfo {pages} {1919} (\bibinfo {year} {2012})}\BibitemShut {NoStop}%
\bibitem [{\citenamefont {Christ}\ \emph {et~al.}(2013)\citenamefont {Christ}, \citenamefont {Brecht}, \citenamefont {Mauerer},\ and\ \citenamefont {Silberhorn}}]{christ2013theory}%
  \BibitemOpen
  \bibfield  {author} {\bibinfo {author} {\bibfnamefont {A.}~\bibnamefont {Christ}}, \bibinfo {author} {\bibfnamefont {B.}~\bibnamefont {Brecht}}, \bibinfo {author} {\bibfnamefont {W.}~\bibnamefont {Mauerer}},\ and\ \bibinfo {author} {\bibfnamefont {C.}~\bibnamefont {Silberhorn}},\ }\href {https://doi.org/10.1088/1367-2630/15/5/053038} {\bibfield  {journal} {\bibinfo  {journal} {New J. Phys.}\ }\textbf {\bibinfo {volume} {15}},\ \bibinfo {pages} {053038} (\bibinfo {year} {2013})}\BibitemShut {NoStop}%
\bibitem [{\citenamefont {Allevi}\ \emph {et~al.}(2014)\citenamefont {Allevi}, \citenamefont {Jedrkiewicz}, \citenamefont {Brambilla}, \citenamefont {Gatti}, \citenamefont {Pe\ifmmode~\check{r}\else \v{r}\fi{}ina}, \citenamefont {Haderka},\ and\ \citenamefont {Bondani}}]{allevi2014coherence}%
  \BibitemOpen
  \bibfield  {author} {\bibinfo {author} {\bibfnamefont {A.}~\bibnamefont {Allevi}}, \bibinfo {author} {\bibfnamefont {O.}~\bibnamefont {Jedrkiewicz}}, \bibinfo {author} {\bibfnamefont {E.}~\bibnamefont {Brambilla}}, \bibinfo {author} {\bibfnamefont {A.}~\bibnamefont {Gatti}}, \bibinfo {author} {\bibfnamefont {J.}~\bibnamefont {Pe\ifmmode~\check{r}\else \v{r}\fi{}ina}}, \bibinfo {author} {\bibfnamefont {O.}~\bibnamefont {Haderka}},\ and\ \bibinfo {author} {\bibfnamefont {M.}~\bibnamefont {Bondani}},\ }\href {https://doi.org/10.1103/PhysRevA.90.063812} {\bibfield  {journal} {\bibinfo  {journal} {Phys. Rev. A}\ }\textbf {\bibinfo {volume} {90}},\ \bibinfo {pages} {063812} (\bibinfo {year} {2014})}\BibitemShut {NoStop}%
\bibitem [{\citenamefont {Quesada}\ and\ \citenamefont {Sipe}(2014)}]{quesada2014effects}%
  \BibitemOpen
  \bibfield  {author} {\bibinfo {author} {\bibfnamefont {N.}~\bibnamefont {Quesada}}\ and\ \bibinfo {author} {\bibfnamefont {J.~E.}\ \bibnamefont {Sipe}},\ }\href {https://doi.org/10.1103/PhysRevA.90.063840} {\bibfield  {journal} {\bibinfo  {journal} {Phys. Rev. A}\ }\textbf {\bibinfo {volume} {90}},\ \bibinfo {pages} {063840} (\bibinfo {year} {2014})}\BibitemShut {NoStop}%
\bibitem [{\citenamefont {Quesada}\ and\ \citenamefont {Sipe}(2015)}]{quesada2015time}%
  \BibitemOpen
  \bibfield  {author} {\bibinfo {author} {\bibfnamefont {N.}~\bibnamefont {Quesada}}\ and\ \bibinfo {author} {\bibfnamefont {J.~E.}\ \bibnamefont {Sipe}},\ }\href {https://doi.org/10.1103/PhysRevLett.114.093903} {\bibfield  {journal} {\bibinfo  {journal} {Phys. Rev. Lett.}\ }\textbf {\bibinfo {volume} {114}},\ \bibinfo {pages} {093903} (\bibinfo {year} {2015})}\BibitemShut {NoStop}%
\bibitem [{\citenamefont {Sharapova}\ \emph {et~al.}(2015)\citenamefont {Sharapova}, \citenamefont {P\'erez}, \citenamefont {Tikhonova},\ and\ \citenamefont {Chekhova}}]{sharapova2015schmidt}%
  \BibitemOpen
  \bibfield  {author} {\bibinfo {author} {\bibfnamefont {P.}~\bibnamefont {Sharapova}}, \bibinfo {author} {\bibfnamefont {A.~M.}\ \bibnamefont {P\'erez}}, \bibinfo {author} {\bibfnamefont {O.~V.}\ \bibnamefont {Tikhonova}},\ and\ \bibinfo {author} {\bibfnamefont {M.~V.}\ \bibnamefont {Chekhova}},\ }\href {https://doi.org/10.1103/PhysRevA.91.043816} {\bibfield  {journal} {\bibinfo  {journal} {Phys. Rev. A}\ }\textbf {\bibinfo {volume} {91}},\ \bibinfo {pages} {043816} (\bibinfo {year} {2015})}\BibitemShut {NoStop}%
\bibitem [{\citenamefont {Finger}\ \emph {et~al.}(2015)\citenamefont {Finger}, \citenamefont {Iskhakov}, \citenamefont {Joly}, \citenamefont {Chekhova},\ and\ \citenamefont {Russell}}]{finger2015raman}%
  \BibitemOpen
  \bibfield  {author} {\bibinfo {author} {\bibfnamefont {M.~A.}\ \bibnamefont {Finger}}, \bibinfo {author} {\bibfnamefont {T.~S.}\ \bibnamefont {Iskhakov}}, \bibinfo {author} {\bibfnamefont {N.~Y.}\ \bibnamefont {Joly}}, \bibinfo {author} {\bibfnamefont {M.~V.}\ \bibnamefont {Chekhova}},\ and\ \bibinfo {author} {\bibfnamefont {P.~S.~J.}\ \bibnamefont {Russell}},\ }\href {https://doi.org/10.1103/PhysRevLett.115.143602} {\bibfield  {journal} {\bibinfo  {journal} {Phys. Rev. Lett.}\ }\textbf {\bibinfo {volume} {115}},\ \bibinfo {pages} {143602} (\bibinfo {year} {2015})}\BibitemShut {NoStop}%
\bibitem [{\citenamefont {Guo}\ \emph {et~al.}(2015)\citenamefont {Guo}, \citenamefont {Liu}, \citenamefont {Li},\ and\ \citenamefont {Ou}}]{guo2015complete}%
  \BibitemOpen
  \bibfield  {author} {\bibinfo {author} {\bibfnamefont {X.}~\bibnamefont {Guo}}, \bibinfo {author} {\bibfnamefont {N.}~\bibnamefont {Liu}}, \bibinfo {author} {\bibfnamefont {X.}~\bibnamefont {Li}},\ and\ \bibinfo {author} {\bibfnamefont {Z.}~\bibnamefont {Ou}},\ }\href {https://doi.org/10.1364/OE.23.029369} {\bibfield  {journal} {\bibinfo  {journal} {Opt. Express}\ }\textbf {\bibinfo {volume} {23}},\ \bibinfo {pages} {29369} (\bibinfo {year} {2015})}\BibitemShut {NoStop}%
\bibitem [{\citenamefont {Chekhova}\ \emph {et~al.}(2015)\citenamefont {Chekhova}, \citenamefont {Leuchs},\ and\ \citenamefont {{\.Z}ukowski}}]{chekhova2015bright}%
  \BibitemOpen
  \bibfield  {author} {\bibinfo {author} {\bibfnamefont {M.}~\bibnamefont {Chekhova}}, \bibinfo {author} {\bibfnamefont {G.}~\bibnamefont {Leuchs}},\ and\ \bibinfo {author} {\bibfnamefont {M.}~\bibnamefont {{\.Z}ukowski}},\ }\href {https://doi.org/10.1016/j.optcom.2014.07.050} {\bibfield  {journal} {\bibinfo  {journal} {Opt. Commun.}\ }\textbf {\bibinfo {volume} {337}},\ \bibinfo {pages} {27} (\bibinfo {year} {2015})}\BibitemShut {NoStop}%
\bibitem [{\citenamefont {Bell}\ \emph {et~al.}(2015)\citenamefont {Bell}, \citenamefont {McMillan}, \citenamefont {McCutcheon},\ and\ \citenamefont {Rarity}}]{bell2015effects}%
  \BibitemOpen
  \bibfield  {author} {\bibinfo {author} {\bibfnamefont {B.}~\bibnamefont {Bell}}, \bibinfo {author} {\bibfnamefont {A.}~\bibnamefont {McMillan}}, \bibinfo {author} {\bibfnamefont {W.}~\bibnamefont {McCutcheon}},\ and\ \bibinfo {author} {\bibfnamefont {J.}~\bibnamefont {Rarity}},\ }\href {https://doi.org/10.1103/PhysRevA.92.053849} {\bibfield  {journal} {\bibinfo  {journal} {Phys. Rev. A}\ }\textbf {\bibinfo {volume} {92}},\ \bibinfo {pages} {053849} (\bibinfo {year} {2015})}\BibitemShut {NoStop}%
\bibitem [{\citenamefont {Pe{\v{r}}ina~Jr}\ \emph {et~al.}(2016)\citenamefont {Pe{\v{r}}ina~Jr}, \citenamefont {Haderka}, \citenamefont {Allevi},\ and\ \citenamefont {Bondani}}]{pevrina2016internal}%
  \BibitemOpen
  \bibfield  {author} {\bibinfo {author} {\bibfnamefont {J.}~\bibnamefont {Pe{\v{r}}ina~Jr}}, \bibinfo {author} {\bibfnamefont {O.}~\bibnamefont {Haderka}}, \bibinfo {author} {\bibfnamefont {A.}~\bibnamefont {Allevi}},\ and\ \bibinfo {author} {\bibfnamefont {M.}~\bibnamefont {Bondani}},\ }\href {https://doi.org/10.1038/srep22320} {\bibfield  {journal} {\bibinfo  {journal} {Sci. Rep.}\ }\textbf {\bibinfo {volume} {6}},\ \bibinfo {pages} {22320} (\bibinfo {year} {2016})}\BibitemShut {NoStop}%
\bibitem [{\citenamefont {Pe\ifmmode~\check{r}\else \v{r}\fi{}ina}(2016{\natexlab{a}})}]{pevrina2016spatial}%
  \BibitemOpen
  \bibfield  {author} {\bibinfo {author} {\bibfnamefont {J.}~\bibnamefont {Pe\ifmmode~\check{r}\else \v{r}\fi{}ina}},\ }\href {https://doi.org/10.1103/PhysRevA.93.013852} {\bibfield  {journal} {\bibinfo  {journal} {Phys. Rev. A}\ }\textbf {\bibinfo {volume} {93}},\ \bibinfo {pages} {013852} (\bibinfo {year} {2016}{\natexlab{a}})}\BibitemShut {NoStop}%
\bibitem [{\citenamefont {Pe\ifmmode~\check{r}\else \v{r}\fi{}ina}(2016{\natexlab{b}})}]{pevrina2016coherent}%
  \BibitemOpen
  \bibfield  {author} {\bibinfo {author} {\bibfnamefont {J.}~\bibnamefont {Pe\ifmmode~\check{r}\else \v{r}\fi{}ina}},\ }\href {https://doi.org/10.1103/PhysRevA.93.063857} {\bibfield  {journal} {\bibinfo  {journal} {Phys. Rev. A}\ }\textbf {\bibinfo {volume} {93}},\ \bibinfo {pages} {063857} (\bibinfo {year} {2016}{\natexlab{b}})}\BibitemShut {NoStop}%
\bibitem [{\citenamefont {Liu}\ \emph {et~al.}(2016)\citenamefont {Liu}, \citenamefont {Liu}, \citenamefont {Guo}, \citenamefont {Yang}, \citenamefont {Li},\ and\ \citenamefont {Ou}}]{liu2016approaching}%
  \BibitemOpen
  \bibfield  {author} {\bibinfo {author} {\bibfnamefont {N.}~\bibnamefont {Liu}}, \bibinfo {author} {\bibfnamefont {Y.}~\bibnamefont {Liu}}, \bibinfo {author} {\bibfnamefont {X.}~\bibnamefont {Guo}}, \bibinfo {author} {\bibfnamefont {L.}~\bibnamefont {Yang}}, \bibinfo {author} {\bibfnamefont {X.}~\bibnamefont {Li}},\ and\ \bibinfo {author} {\bibfnamefont {Z.}~\bibnamefont {Ou}},\ }\href {https://doi.org/10.1364/OE.24.001096} {\bibfield  {journal} {\bibinfo  {journal} {Opt. Express}\ }\textbf {\bibinfo {volume} {24}},\ \bibinfo {pages} {1096} (\bibinfo {year} {2016})}\BibitemShut {NoStop}%
\bibitem [{\citenamefont {Allevi}\ and\ \citenamefont {Bondani}(2017)}]{allevi2017nonlinear}%
  \BibitemOpen
  \bibfield  {author} {\bibinfo {author} {\bibfnamefont {A.}~\bibnamefont {Allevi}}\ and\ \bibinfo {author} {\bibfnamefont {M.}~\bibnamefont {Bondani}},\ }\href {https://doi.org/10.1016/bs.aamop.2017.02.001} {\bibfield  {journal} {\bibinfo  {journal} {Adv. At. Mol. Opt. Phys.}\ }\textbf {\bibinfo {volume} {66}},\ \bibinfo {pages} {49} (\bibinfo {year} {2017})}\BibitemShut {NoStop}%
\bibitem [{\citenamefont {Sharapova}\ \emph {et~al.}(2020)\citenamefont {Sharapova}, \citenamefont {Frascella}, \citenamefont {Riabinin}, \citenamefont {P\'erez}, \citenamefont {Tikhonova}, \citenamefont {Lemieux}, \citenamefont {Boyd}, \citenamefont {Leuchs},\ and\ \citenamefont {Chekhova}}]{sharapova2020properties}%
  \BibitemOpen
  \bibfield  {author} {\bibinfo {author} {\bibfnamefont {P.~R.}\ \bibnamefont {Sharapova}}, \bibinfo {author} {\bibfnamefont {G.}~\bibnamefont {Frascella}}, \bibinfo {author} {\bibfnamefont {M.}~\bibnamefont {Riabinin}}, \bibinfo {author} {\bibfnamefont {A.~M.}\ \bibnamefont {P\'erez}}, \bibinfo {author} {\bibfnamefont {O.~V.}\ \bibnamefont {Tikhonova}}, \bibinfo {author} {\bibfnamefont {S.}~\bibnamefont {Lemieux}}, \bibinfo {author} {\bibfnamefont {R.~W.}\ \bibnamefont {Boyd}}, \bibinfo {author} {\bibfnamefont {G.}~\bibnamefont {Leuchs}},\ and\ \bibinfo {author} {\bibfnamefont {M.~V.}\ \bibnamefont {Chekhova}},\ }\href {https://doi.org/10.1103/PhysRevResearch.2.013371} {\bibfield  {journal} {\bibinfo  {journal} {Phys. Rev. Res.}\ }\textbf {\bibinfo {volume} {2}},\ \bibinfo {pages} {013371} (\bibinfo {year} {2020})}\BibitemShut {NoStop}%
\bibitem [{\citenamefont {Fl{\'o}rez}\ \emph {et~al.}(2020)\citenamefont {Fl{\'o}rez}, \citenamefont {Lundeen},\ and\ \citenamefont {Chekhova}}]{florez2020pump}%
  \BibitemOpen
  \bibfield  {author} {\bibinfo {author} {\bibfnamefont {J.}~\bibnamefont {Fl{\'o}rez}}, \bibinfo {author} {\bibfnamefont {J.~S.}\ \bibnamefont {Lundeen}},\ and\ \bibinfo {author} {\bibfnamefont {M.~V.}\ \bibnamefont {Chekhova}},\ }\href {https://doi.org/10.1364/OL.394925} {\bibfield  {journal} {\bibinfo  {journal} {Opt. Lett.}\ }\textbf {\bibinfo {volume} {45}},\ \bibinfo {pages} {4264} (\bibinfo {year} {2020})}\BibitemShut {NoStop}%
\bibitem [{\citenamefont {Triginer}\ \emph {et~al.}(2020)\citenamefont {Triginer}, \citenamefont {Vidrighin}, \citenamefont {Quesada}, \citenamefont {Eckstein}, \citenamefont {Moore}, \citenamefont {Kolthammer}, \citenamefont {Sipe},\ and\ \citenamefont {Walmsley}}]{triginer2020understanding}%
  \BibitemOpen
  \bibfield  {author} {\bibinfo {author} {\bibfnamefont {G.}~\bibnamefont {Triginer}}, \bibinfo {author} {\bibfnamefont {M.~D.}\ \bibnamefont {Vidrighin}}, \bibinfo {author} {\bibfnamefont {N.}~\bibnamefont {Quesada}}, \bibinfo {author} {\bibfnamefont {A.}~\bibnamefont {Eckstein}}, \bibinfo {author} {\bibfnamefont {M.}~\bibnamefont {Moore}}, \bibinfo {author} {\bibfnamefont {W.~S.}\ \bibnamefont {Kolthammer}}, \bibinfo {author} {\bibfnamefont {J.~E.}\ \bibnamefont {Sipe}},\ and\ \bibinfo {author} {\bibfnamefont {I.~A.}\ \bibnamefont {Walmsley}},\ }\href {https://doi.org/10.1103/PhysRevX.10.031063} {\bibfield  {journal} {\bibinfo  {journal} {Phys. Rev. X}\ }\textbf {\bibinfo {volume} {10}},\ \bibinfo {pages} {031063} (\bibinfo {year} {2020})}\BibitemShut {NoStop}%
\bibitem [{\citenamefont {Chen}\ \emph {et~al.}(2021)\citenamefont {Chen}, \citenamefont {Zhang},\ and\ \citenamefont {Ou}}]{chen2021mode}%
  \BibitemOpen
  \bibfield  {author} {\bibinfo {author} {\bibfnamefont {X.}~\bibnamefont {Chen}}, \bibinfo {author} {\bibfnamefont {J.}~\bibnamefont {Zhang}},\ and\ \bibinfo {author} {\bibfnamefont {Z.~Y.}\ \bibnamefont {Ou}},\ }\href {https://doi.org/10.1103/PhysRevResearch.3.023186} {\bibfield  {journal} {\bibinfo  {journal} {Phys. Rev. Res.}\ }\textbf {\bibinfo {volume} {3}},\ \bibinfo {pages} {023186} (\bibinfo {year} {2021})}\BibitemShut {NoStop}%
\bibitem [{\citenamefont {Quesada}\ \emph {et~al.}(2022)\citenamefont {Quesada}, \citenamefont {Helt}, \citenamefont {Menotti}, \citenamefont {Liscidini},\ and\ \citenamefont {Sipe}}]{quesada2022beyond}%
  \BibitemOpen
  \bibfield  {author} {\bibinfo {author} {\bibfnamefont {N.}~\bibnamefont {Quesada}}, \bibinfo {author} {\bibfnamefont {L.}~\bibnamefont {Helt}}, \bibinfo {author} {\bibfnamefont {M.}~\bibnamefont {Menotti}}, \bibinfo {author} {\bibfnamefont {M.}~\bibnamefont {Liscidini}},\ and\ \bibinfo {author} {\bibfnamefont {J.}~\bibnamefont {Sipe}},\ }\href {https://doi.org/10.1364/AOP.445496} {\bibfield  {journal} {\bibinfo  {journal} {Adv. Opt. Photonics}\ }\textbf {\bibinfo {volume} {14}},\ \bibinfo {pages} {291} (\bibinfo {year} {2022})}\BibitemShut {NoStop}%
\bibitem [{\citenamefont {Yanagimoto}\ \emph {et~al.}(2022)\citenamefont {Yanagimoto}, \citenamefont {Ng}, \citenamefont {Yamamura}, \citenamefont {Onodera}, \citenamefont {Wright}, \citenamefont {Jankowski}, \citenamefont {Fejer}, \citenamefont {McMahon},\ and\ \citenamefont {Mabuchi}}]{yanagimoto2022onset}%
  \BibitemOpen
  \bibfield  {author} {\bibinfo {author} {\bibfnamefont {R.}~\bibnamefont {Yanagimoto}}, \bibinfo {author} {\bibfnamefont {E.}~\bibnamefont {Ng}}, \bibinfo {author} {\bibfnamefont {A.}~\bibnamefont {Yamamura}}, \bibinfo {author} {\bibfnamefont {T.}~\bibnamefont {Onodera}}, \bibinfo {author} {\bibfnamefont {L.~G.}\ \bibnamefont {Wright}}, \bibinfo {author} {\bibfnamefont {M.}~\bibnamefont {Jankowski}}, \bibinfo {author} {\bibfnamefont {M.}~\bibnamefont {Fejer}}, \bibinfo {author} {\bibfnamefont {P.~L.}\ \bibnamefont {McMahon}},\ and\ \bibinfo {author} {\bibfnamefont {H.}~\bibnamefont {Mabuchi}},\ }\href {https://doi.org/10.1364/OPTICA.447782} {\bibfield  {journal} {\bibinfo  {journal} {Optica}\ }\textbf {\bibinfo {volume} {9}},\ \bibinfo {pages} {379} (\bibinfo {year} {2022})}\BibitemShut {NoStop}%
\bibitem [{\citenamefont {Kulkarni}\ \emph {et~al.}(2022)\citenamefont {Kulkarni}, \citenamefont {Rioux}, \citenamefont {Braverman}, \citenamefont {Chekhova},\ and\ \citenamefont {Boyd}}]{kulkarni2022classical}%
  \BibitemOpen
  \bibfield  {author} {\bibinfo {author} {\bibfnamefont {G.}~\bibnamefont {Kulkarni}}, \bibinfo {author} {\bibfnamefont {J.}~\bibnamefont {Rioux}}, \bibinfo {author} {\bibfnamefont {B.}~\bibnamefont {Braverman}}, \bibinfo {author} {\bibfnamefont {M.~V.}\ \bibnamefont {Chekhova}},\ and\ \bibinfo {author} {\bibfnamefont {R.~W.}\ \bibnamefont {Boyd}},\ }\href {https://doi.org/10.1103/PhysRevResearch.4.033098} {\bibfield  {journal} {\bibinfo  {journal} {Phys. Rev. Res.}\ }\textbf {\bibinfo {volume} {4}},\ \bibinfo {pages} {033098} (\bibinfo {year} {2022})}\BibitemShut {NoStop}%
\bibitem [{\citenamefont {Houde}\ and\ \citenamefont {Quesada}(2023)}]{houde2023waveguided}%
  \BibitemOpen
  \bibfield  {author} {\bibinfo {author} {\bibfnamefont {M.}~\bibnamefont {Houde}}\ and\ \bibinfo {author} {\bibfnamefont {N.}~\bibnamefont {Quesada}},\ }\bibfield  {journal} {\bibinfo  {journal} {AVS Quantum Sci.}\ }\textbf {\bibinfo {volume} {5}},\ \href {https://doi.org/10.1116/5.0133009} {10.1116/5.0133009} (\bibinfo {year} {2023})\BibitemShut {NoStop}%
\bibitem [{\citenamefont {Kalash}\ and\ \citenamefont {Chekhova}(2023)}]{kalash2023wigner}%
  \BibitemOpen
  \bibfield  {author} {\bibinfo {author} {\bibfnamefont {M.}~\bibnamefont {Kalash}}\ and\ \bibinfo {author} {\bibfnamefont {M.~V.}\ \bibnamefont {Chekhova}},\ }\href {https://doi.org/10.1364/OPTICA.488697} {\bibfield  {journal} {\bibinfo  {journal} {Optica}\ }\textbf {\bibinfo {volume} {10}},\ \bibinfo {pages} {1142} (\bibinfo {year} {2023})}\BibitemShut {NoStop}%
\bibitem [{\citenamefont {Chinni}\ and\ \citenamefont {Quesada}(2024)}]{chinni2024beyond}%
  \BibitemOpen
  \bibfield  {author} {\bibinfo {author} {\bibfnamefont {K.}~\bibnamefont {Chinni}}\ and\ \bibinfo {author} {\bibfnamefont {N.}~\bibnamefont {Quesada}},\ }\href {https://doi.org/10.1103/PhysRevA.110.013712} {\bibfield  {journal} {\bibinfo  {journal} {Phys. Rev. A}\ }\textbf {\bibinfo {volume} {110}},\ \bibinfo {pages} {013712} (\bibinfo {year} {2024})}\BibitemShut {NoStop}%
\bibitem [{\citenamefont {Quesada}\ \emph {et~al.}(2020)\citenamefont {Quesada}, \citenamefont {Triginer}, \citenamefont {Vidrighin},\ and\ \citenamefont {Sipe}}]{quesada2020theory}%
  \BibitemOpen
  \bibfield  {author} {\bibinfo {author} {\bibfnamefont {N.}~\bibnamefont {Quesada}}, \bibinfo {author} {\bibfnamefont {G.}~\bibnamefont {Triginer}}, \bibinfo {author} {\bibfnamefont {M.~D.}\ \bibnamefont {Vidrighin}},\ and\ \bibinfo {author} {\bibfnamefont {J.~E.}\ \bibnamefont {Sipe}},\ }\href {https://doi.org/10.1103/PhysRevA.102.033519} {\bibfield  {journal} {\bibinfo  {journal} {Phys. Rev. A}\ }\textbf {\bibinfo {volume} {102}},\ \bibinfo {pages} {033519} (\bibinfo {year} {2020})}\BibitemShut {NoStop}%
\bibitem [{\citenamefont {Huo}\ \emph {et~al.}(2020)\citenamefont {Huo}, \citenamefont {Liu}, \citenamefont {Li}, \citenamefont {Cui}, \citenamefont {Chen}, \citenamefont {Palivela}, \citenamefont {Xie}, \citenamefont {Li},\ and\ \citenamefont {Ou}}]{huo2020direct}%
  \BibitemOpen
  \bibfield  {author} {\bibinfo {author} {\bibfnamefont {N.}~\bibnamefont {Huo}}, \bibinfo {author} {\bibfnamefont {Y.}~\bibnamefont {Liu}}, \bibinfo {author} {\bibfnamefont {J.}~\bibnamefont {Li}}, \bibinfo {author} {\bibfnamefont {L.}~\bibnamefont {Cui}}, \bibinfo {author} {\bibfnamefont {X.}~\bibnamefont {Chen}}, \bibinfo {author} {\bibfnamefont {R.}~\bibnamefont {Palivela}}, \bibinfo {author} {\bibfnamefont {T.}~\bibnamefont {Xie}}, \bibinfo {author} {\bibfnamefont {X.}~\bibnamefont {Li}},\ and\ \bibinfo {author} {\bibfnamefont {Z.~Y.}\ \bibnamefont {Ou}},\ }\href {https://doi.org/10.1103/PhysRevLett.124.213603} {\bibfield  {journal} {\bibinfo  {journal} {Phys. Rev. Lett.}\ }\textbf {\bibinfo {volume} {124}},\ \bibinfo {pages} {213603} (\bibinfo {year} {2020})}\BibitemShut {NoStop}%
\bibitem [{SM()}]{SM}%
  \BibitemOpen
  \href@noop {} {}\bibinfo {note} {See Supplemental Materials at [link], which includes Refs.~[93-101], for further details on the derivations and the experimental results.}\BibitemShut {Stop}%
\bibitem [{\citenamefont {Graffitti}\ \emph {et~al.}(2018)\citenamefont {Graffitti}, \citenamefont {Kelly-Massicotte}, \citenamefont {Fedrizzi},\ and\ \citenamefont {Bra\ifmmode~\acute{n}\else \'{n}\fi{}czyk}}]{graffitti2018design}%
  \BibitemOpen
  \bibfield  {author} {\bibinfo {author} {\bibfnamefont {F.}~\bibnamefont {Graffitti}}, \bibinfo {author} {\bibfnamefont {J.}~\bibnamefont {Kelly-Massicotte}}, \bibinfo {author} {\bibfnamefont {A.}~\bibnamefont {Fedrizzi}},\ and\ \bibinfo {author} {\bibfnamefont {A.~M.}\ \bibnamefont {Bra\ifmmode~\acute{n}\else \'{n}\fi{}czyk}},\ }\href {https://doi.org/10.1103/PhysRevA.98.053811} {\bibfield  {journal} {\bibinfo  {journal} {Phys. Rev. A}\ }\textbf {\bibinfo {volume} {98}},\ \bibinfo {pages} {053811} (\bibinfo {year} {2018})}\BibitemShut {NoStop}%
\bibitem [{\citenamefont {Houde}\ and\ \citenamefont {Quesada}(2024)}]{NeedALight}%
  \BibitemOpen
  \bibfield  {author} {\bibinfo {author} {\bibfnamefont {M.}~\bibnamefont {Houde}}\ and\ \bibinfo {author} {\bibfnamefont {N.}~\bibnamefont {Quesada}},\ }\href@noop {} {\bibinfo {title} {Needalight}},\ \bibinfo {howpublished} {\url{https://github.com/polyquantique/NeedALight}} (\bibinfo {year} {2024})\BibitemShut {NoStop}%
\bibitem [{\citenamefont {Longhi}\ \emph {et~al.}(2002)\citenamefont {Longhi}, \citenamefont {Marano},\ and\ \citenamefont {Laporta}}]{longhi2002dispersive}%
  \BibitemOpen
  \bibfield  {author} {\bibinfo {author} {\bibfnamefont {S.}~\bibnamefont {Longhi}}, \bibinfo {author} {\bibfnamefont {M.}~\bibnamefont {Marano}},\ and\ \bibinfo {author} {\bibfnamefont {P.}~\bibnamefont {Laporta}},\ }\href {https://doi.org/10.1103/PhysRevA.66.033803} {\bibfield  {journal} {\bibinfo  {journal} {Phys. Rev. A}\ }\textbf {\bibinfo {volume} {66}},\ \bibinfo {pages} {033803} (\bibinfo {year} {2002})}\BibitemShut {NoStop}%
\bibitem [{\citenamefont {Gerrits}\ \emph {et~al.}(2015)\citenamefont {Gerrits}, \citenamefont {Marsili}, \citenamefont {Verma}, \citenamefont {Shalm}, \citenamefont {Shaw}, \citenamefont {Mirin},\ and\ \citenamefont {Nam}}]{gerrits2015spectral}%
  \BibitemOpen
  \bibfield  {author} {\bibinfo {author} {\bibfnamefont {T.}~\bibnamefont {Gerrits}}, \bibinfo {author} {\bibfnamefont {F.}~\bibnamefont {Marsili}}, \bibinfo {author} {\bibfnamefont {V.~B.}\ \bibnamefont {Verma}}, \bibinfo {author} {\bibfnamefont {L.~K.}\ \bibnamefont {Shalm}}, \bibinfo {author} {\bibfnamefont {M.}~\bibnamefont {Shaw}}, \bibinfo {author} {\bibfnamefont {R.~P.}\ \bibnamefont {Mirin}},\ and\ \bibinfo {author} {\bibfnamefont {S.~W.}\ \bibnamefont {Nam}},\ }\href {https://doi.org/10.1103/PhysRevA.91.013830} {\bibfield  {journal} {\bibinfo  {journal} {Phys. Rev. A}\ }\textbf {\bibinfo {volume} {91}},\ \bibinfo {pages} {013830} (\bibinfo {year} {2015})}\BibitemShut {NoStop}%
\bibitem [{\citenamefont {Jin}\ \emph {et~al.}(2015)\citenamefont {Jin}, \citenamefont {Gerrits}, \citenamefont {Fujiwara}, \citenamefont {Wakabayashi}, \citenamefont {Yamashita}, \citenamefont {Miki}, \citenamefont {Terai}, \citenamefont {Shimizu}, \citenamefont {Takeoka},\ and\ \citenamefont {Sasaki}}]{jin2015spectrally}%
  \BibitemOpen
  \bibfield  {author} {\bibinfo {author} {\bibfnamefont {R.-B.}\ \bibnamefont {Jin}}, \bibinfo {author} {\bibfnamefont {T.}~\bibnamefont {Gerrits}}, \bibinfo {author} {\bibfnamefont {M.}~\bibnamefont {Fujiwara}}, \bibinfo {author} {\bibfnamefont {R.}~\bibnamefont {Wakabayashi}}, \bibinfo {author} {\bibfnamefont {T.}~\bibnamefont {Yamashita}}, \bibinfo {author} {\bibfnamefont {S.}~\bibnamefont {Miki}}, \bibinfo {author} {\bibfnamefont {H.}~\bibnamefont {Terai}}, \bibinfo {author} {\bibfnamefont {R.}~\bibnamefont {Shimizu}}, \bibinfo {author} {\bibfnamefont {M.}~\bibnamefont {Takeoka}},\ and\ \bibinfo {author} {\bibfnamefont {M.}~\bibnamefont {Sasaki}},\ }\href {https://doi.org/10.1364/OE.23.028836} {\bibfield  {journal} {\bibinfo  {journal} {Opt. Exp.}\ }\textbf {\bibinfo {volume} {23}},\ \bibinfo {pages} {28836} (\bibinfo {year} {2015})}\BibitemShut {NoStop}%
\bibitem [{\citenamefont {Orre}\ \emph {et~al.}(2019)\citenamefont {Orre}, \citenamefont {Goldschmidt}, \citenamefont {Deshpande}, \citenamefont {Gorshkov}, \citenamefont {Tamma}, \citenamefont {Hafezi},\ and\ \citenamefont {Mittal}}]{Orre2019interference}%
  \BibitemOpen
  \bibfield  {author} {\bibinfo {author} {\bibfnamefont {V.~V.}\ \bibnamefont {Orre}}, \bibinfo {author} {\bibfnamefont {E.~A.}\ \bibnamefont {Goldschmidt}}, \bibinfo {author} {\bibfnamefont {A.}~\bibnamefont {Deshpande}}, \bibinfo {author} {\bibfnamefont {A.~V.}\ \bibnamefont {Gorshkov}}, \bibinfo {author} {\bibfnamefont {V.}~\bibnamefont {Tamma}}, \bibinfo {author} {\bibfnamefont {M.}~\bibnamefont {Hafezi}},\ and\ \bibinfo {author} {\bibfnamefont {S.}~\bibnamefont {Mittal}},\ }\href {https://doi.org/10.1103/PhysRevLett.123.123603} {\bibfield  {journal} {\bibinfo  {journal} {Phys. Rev. Lett.}\ }\textbf {\bibinfo {volume} {123}},\ \bibinfo {pages} {123603} (\bibinfo {year} {2019})}\BibitemShut {NoStop}%
\bibitem [{\citenamefont {Thekkadath}\ \emph {et~al.}(2022)\citenamefont {Thekkadath}, \citenamefont {Bell}, \citenamefont {Patel}, \citenamefont {Kim},\ and\ \citenamefont {Walmsley}}]{thekkadath2022measuring}%
  \BibitemOpen
  \bibfield  {author} {\bibinfo {author} {\bibfnamefont {G.~S.}\ \bibnamefont {Thekkadath}}, \bibinfo {author} {\bibfnamefont {B.~A.}\ \bibnamefont {Bell}}, \bibinfo {author} {\bibfnamefont {R.~B.}\ \bibnamefont {Patel}}, \bibinfo {author} {\bibfnamefont {M.~S.}\ \bibnamefont {Kim}},\ and\ \bibinfo {author} {\bibfnamefont {I.~A.}\ \bibnamefont {Walmsley}},\ }\href {https://doi.org/10.1103/PhysRevLett.128.023601} {\bibfield  {journal} {\bibinfo  {journal} {Phys. Rev. Lett.}\ }\textbf {\bibinfo {volume} {128}},\ \bibinfo {pages} {023601} (\bibinfo {year} {2022})}\BibitemShut {NoStop}%
\bibitem [{\citenamefont {Gerrits}\ \emph {et~al.}(2011)\citenamefont {Gerrits}, \citenamefont {Stevens}, \citenamefont {Baek}, \citenamefont {Calkins}, \citenamefont {Lita}, \citenamefont {Glancy}, \citenamefont {Knill}, \citenamefont {Nam}, \citenamefont {Mirin}, \citenamefont {Hadfield} \emph {et~al.}}]{gerrits2011generation}%
  \BibitemOpen
  \bibfield  {author} {\bibinfo {author} {\bibfnamefont {T.}~\bibnamefont {Gerrits}}, \bibinfo {author} {\bibfnamefont {M.~J.}\ \bibnamefont {Stevens}}, \bibinfo {author} {\bibfnamefont {B.}~\bibnamefont {Baek}}, \bibinfo {author} {\bibfnamefont {B.}~\bibnamefont {Calkins}}, \bibinfo {author} {\bibfnamefont {A.}~\bibnamefont {Lita}}, \bibinfo {author} {\bibfnamefont {S.}~\bibnamefont {Glancy}}, \bibinfo {author} {\bibfnamefont {E.}~\bibnamefont {Knill}}, \bibinfo {author} {\bibfnamefont {S.~W.}\ \bibnamefont {Nam}}, \bibinfo {author} {\bibfnamefont {R.~P.}\ \bibnamefont {Mirin}}, \bibinfo {author} {\bibfnamefont {R.~H.}\ \bibnamefont {Hadfield}}, \emph {et~al.},\ }\href {https://doi.org/10.1364/OE.19.024434} {\bibfield  {journal} {\bibinfo  {journal} {Opt. Exp.}\ }\textbf {\bibinfo {volume} {19}},\ \bibinfo {pages} {24434} (\bibinfo {year} {2011})}\BibitemShut {NoStop}%
\bibitem [{\citenamefont {Weston}\ \emph {et~al.}(2016)\citenamefont {Weston}, \citenamefont {Chrzanowski}, \citenamefont {Wollmann}, \citenamefont {Boston}, \citenamefont {Ho}, \citenamefont {Shalm}, \citenamefont {Verma}, \citenamefont {Allman}, \citenamefont {Nam}, \citenamefont {Patel} \emph {et~al.}}]{weston2016efficient}%
  \BibitemOpen
  \bibfield  {author} {\bibinfo {author} {\bibfnamefont {M.~M.}\ \bibnamefont {Weston}}, \bibinfo {author} {\bibfnamefont {H.~M.}\ \bibnamefont {Chrzanowski}}, \bibinfo {author} {\bibfnamefont {S.}~\bibnamefont {Wollmann}}, \bibinfo {author} {\bibfnamefont {A.}~\bibnamefont {Boston}}, \bibinfo {author} {\bibfnamefont {J.}~\bibnamefont {Ho}}, \bibinfo {author} {\bibfnamefont {L.~K.}\ \bibnamefont {Shalm}}, \bibinfo {author} {\bibfnamefont {V.~B.}\ \bibnamefont {Verma}}, \bibinfo {author} {\bibfnamefont {M.~S.}\ \bibnamefont {Allman}}, \bibinfo {author} {\bibfnamefont {S.~W.}\ \bibnamefont {Nam}}, \bibinfo {author} {\bibfnamefont {R.~B.}\ \bibnamefont {Patel}}, \emph {et~al.},\ }\href {https://doi.org/10.1364/OE.24.010869} {\bibfield  {journal} {\bibinfo  {journal} {Opt. Exp.}\ }\textbf {\bibinfo {volume} {24}},\ \bibinfo {pages} {10869} (\bibinfo {year} {2016})}\BibitemShut {NoStop}%
\bibitem [{\citenamefont {Zhong}\ \emph {et~al.}(2020)\citenamefont {Zhong}, \citenamefont {Wang}, \citenamefont {Deng}, \citenamefont {Chen}, \citenamefont {Peng}, \citenamefont {Luo}, \citenamefont {Qin}, \citenamefont {Wu}, \citenamefont {Ding}, \citenamefont {Hu} \emph {et~al.}}]{zhong2020quantum}%
  \BibitemOpen
  \bibfield  {author} {\bibinfo {author} {\bibfnamefont {H.-S.}\ \bibnamefont {Zhong}}, \bibinfo {author} {\bibfnamefont {H.}~\bibnamefont {Wang}}, \bibinfo {author} {\bibfnamefont {Y.-H.}\ \bibnamefont {Deng}}, \bibinfo {author} {\bibfnamefont {M.-C.}\ \bibnamefont {Chen}}, \bibinfo {author} {\bibfnamefont {L.-C.}\ \bibnamefont {Peng}}, \bibinfo {author} {\bibfnamefont {Y.-H.}\ \bibnamefont {Luo}}, \bibinfo {author} {\bibfnamefont {J.}~\bibnamefont {Qin}}, \bibinfo {author} {\bibfnamefont {D.}~\bibnamefont {Wu}}, \bibinfo {author} {\bibfnamefont {X.}~\bibnamefont {Ding}}, \bibinfo {author} {\bibfnamefont {Y.}~\bibnamefont {Hu}}, \emph {et~al.},\ }\href {https://doi.org/10.1126/science.abe8770} {\bibfield  {journal} {\bibinfo  {journal} {Science}\ }\textbf {\bibinfo {volume} {370}},\ \bibinfo {pages} {1460} (\bibinfo {year} {2020})}\BibitemShut {NoStop}%
\bibitem [{\citenamefont {Avenhaus}\ \emph {et~al.}(2009)\citenamefont {Avenhaus}, \citenamefont {Eckstein}, \citenamefont {Mosley},\ and\ \citenamefont {Silberhorn}}]{avenhaus2009fiber}%
  \BibitemOpen
  \bibfield  {author} {\bibinfo {author} {\bibfnamefont {M.}~\bibnamefont {Avenhaus}}, \bibinfo {author} {\bibfnamefont {A.}~\bibnamefont {Eckstein}}, \bibinfo {author} {\bibfnamefont {P.~J.}\ \bibnamefont {Mosley}},\ and\ \bibinfo {author} {\bibfnamefont {C.}~\bibnamefont {Silberhorn}},\ }\href {https://doi.org/10.1364/OL.34.002873} {\bibfield  {journal} {\bibinfo  {journal} {Opt. Lett.}\ }\textbf {\bibinfo {volume} {34}},\ \bibinfo {pages} {2873} (\bibinfo {year} {2009})}\BibitemShut {NoStop}%
\bibitem [{\citenamefont {Klyshko}(1980)}]{klyshko1980use}%
  \BibitemOpen
  \bibfield  {author} {\bibinfo {author} {\bibfnamefont {D.}~\bibnamefont {Klyshko}},\ }\href {https://doi.org/10.1070/QE1980v010n09ABEH010660} {\bibfield  {journal} {\bibinfo  {journal} {Sov. J. Quantum Electron.}\ }\textbf {\bibinfo {volume} {10}},\ \bibinfo {pages} {1112} (\bibinfo {year} {1980})}\BibitemShut {NoStop}%
\bibitem [{\citenamefont {Kato}\ and\ \citenamefont {Takaoka}(2002)}]{kato2002sellmeier}%
  \BibitemOpen
  \bibfield  {author} {\bibinfo {author} {\bibfnamefont {K.}~\bibnamefont {Kato}}\ and\ \bibinfo {author} {\bibfnamefont {E.}~\bibnamefont {Takaoka}},\ }\href {https://doi.org/10.1364/AO.41.005040} {\bibfield  {journal} {\bibinfo  {journal} {App. Opt.}\ }\textbf {\bibinfo {volume} {41}},\ \bibinfo {pages} {5040} (\bibinfo {year} {2002})}\BibitemShut {NoStop}%
\bibitem [{\citenamefont {Thekkadath}\ \emph {et~al.}(2024)\citenamefont {Thekkadath}, \citenamefont {Houde}, \citenamefont {England}, \citenamefont {Bustard}, \citenamefont {Bouchard}, \citenamefont {Quesada},\ and\ \citenamefont {Sussman}}]{SPDC_GroupDelay}%
  \BibitemOpen
  \bibfield  {author} {\bibinfo {author} {\bibfnamefont {G.}~\bibnamefont {Thekkadath}}, \bibinfo {author} {\bibfnamefont {M.}~\bibnamefont {Houde}}, \bibinfo {author} {\bibfnamefont {D.}~\bibnamefont {England}}, \bibinfo {author} {\bibfnamefont {P.}~\bibnamefont {Bustard}}, \bibinfo {author} {\bibfnamefont {F.}~\bibnamefont {Bouchard}}, \bibinfo {author} {\bibfnamefont {N.}~\bibnamefont {Quesada}},\ and\ \bibinfo {author} {\bibfnamefont {B.}~\bibnamefont {Sussman}},\ }\href@noop {} {\bibinfo {title} {Spdc group delay}},\ \bibinfo {howpublished} {\url{https://github.com/UltrafastQO/SPDC_GroupDelay}} (\bibinfo {year} {2024})\BibitemShut {NoStop}%
\bibitem [{\citenamefont {Hamilton}\ \emph {et~al.}(2017)\citenamefont {Hamilton}, \citenamefont {Kruse}, \citenamefont {Sansoni}, \citenamefont {Barkhofen}, \citenamefont {Silberhorn},\ and\ \citenamefont {Jex}}]{hamilton2017gaussian}%
  \BibitemOpen
  \bibfield  {author} {\bibinfo {author} {\bibfnamefont {C.~S.}\ \bibnamefont {Hamilton}}, \bibinfo {author} {\bibfnamefont {R.}~\bibnamefont {Kruse}}, \bibinfo {author} {\bibfnamefont {L.}~\bibnamefont {Sansoni}}, \bibinfo {author} {\bibfnamefont {S.}~\bibnamefont {Barkhofen}}, \bibinfo {author} {\bibfnamefont {C.}~\bibnamefont {Silberhorn}},\ and\ \bibinfo {author} {\bibfnamefont {I.}~\bibnamefont {Jex}},\ }\href {https://doi.org/10.1103/PhysRevLett.119.170501} {\bibfield  {journal} {\bibinfo  {journal} {Phys. Rev. Lett.}\ }\textbf {\bibinfo {volume} {119}},\ \bibinfo {pages} {170501} (\bibinfo {year} {2017})}\BibitemShut {NoStop}%
\bibitem [{\citenamefont {Grier}\ \emph {et~al.}(2022)\citenamefont {Grier}, \citenamefont {Brod}, \citenamefont {Arrazola}, \citenamefont {de~Andrade~Alonso},\ and\ \citenamefont {Quesada}}]{grier2022complexity}%
  \BibitemOpen
  \bibfield  {author} {\bibinfo {author} {\bibfnamefont {D.}~\bibnamefont {Grier}}, \bibinfo {author} {\bibfnamefont {D.~J.}\ \bibnamefont {Brod}}, \bibinfo {author} {\bibfnamefont {J.~M.}\ \bibnamefont {Arrazola}}, \bibinfo {author} {\bibfnamefont {M.~B.}\ \bibnamefont {de~Andrade~Alonso}},\ and\ \bibinfo {author} {\bibfnamefont {N.}~\bibnamefont {Quesada}},\ }\href {https://doi.org/10.22331/q-2022-11-28-863} {\bibfield  {journal} {\bibinfo  {journal} {Quantum}\ }\textbf {\bibinfo {volume} {6}},\ \bibinfo {pages} {863} (\bibinfo {year} {2022})}\BibitemShut {NoStop}%
\bibitem [{\citenamefont {Chekhova}\ and\ \citenamefont {Ou}(2016)}]{chekhova2016nonlinear}%
  \BibitemOpen
  \bibfield  {author} {\bibinfo {author} {\bibfnamefont {M.}~\bibnamefont {Chekhova}}\ and\ \bibinfo {author} {\bibfnamefont {Z.}~\bibnamefont {Ou}},\ }\href {https://doi.org/10.1364/AOP.8.000104} {\bibfield  {journal} {\bibinfo  {journal} {Adv. Opt. Photonics}\ }\textbf {\bibinfo {volume} {8}},\ \bibinfo {pages} {104} (\bibinfo {year} {2016})}\BibitemShut {NoStop}%
\bibitem [{\citenamefont {Thekkadath}\ \emph {et~al.}(2020)\citenamefont {Thekkadath}, \citenamefont {Mycroft}, \citenamefont {Bell}, \citenamefont {Wade}, \citenamefont {Eckstein}, \citenamefont {Phillips}, \citenamefont {Patel}, \citenamefont {Buraczewski}, \citenamefont {Lita}, \citenamefont {Gerrits} \emph {et~al.}}]{thekkadath2020quantum}%
  \BibitemOpen
  \bibfield  {author} {\bibinfo {author} {\bibfnamefont {G.}~\bibnamefont {Thekkadath}}, \bibinfo {author} {\bibfnamefont {M.}~\bibnamefont {Mycroft}}, \bibinfo {author} {\bibfnamefont {B.}~\bibnamefont {Bell}}, \bibinfo {author} {\bibfnamefont {C.}~\bibnamefont {Wade}}, \bibinfo {author} {\bibfnamefont {A.}~\bibnamefont {Eckstein}}, \bibinfo {author} {\bibfnamefont {D.}~\bibnamefont {Phillips}}, \bibinfo {author} {\bibfnamefont {R.}~\bibnamefont {Patel}}, \bibinfo {author} {\bibfnamefont {A.}~\bibnamefont {Buraczewski}}, \bibinfo {author} {\bibfnamefont {A.}~\bibnamefont {Lita}}, \bibinfo {author} {\bibfnamefont {T.}~\bibnamefont {Gerrits}}, \emph {et~al.},\ }\href {https://doi.org/10.1038/s41534-020-00320-y} {\bibfield  {journal} {\bibinfo  {journal} {NPJ Quantum Inf.}\ }\textbf {\bibinfo {volume} {6}},\ \bibinfo {pages} {89} (\bibinfo {year} {2020})}\BibitemShut {NoStop}%
\bibitem [{\citenamefont {Machado}\ \emph {et~al.}(2020)\citenamefont {Machado}, \citenamefont {Frascella}, \citenamefont {Torres},\ and\ \citenamefont {Chekhova}}]{machado2020optical}%
  \BibitemOpen
  \bibfield  {author} {\bibinfo {author} {\bibfnamefont {G.~J.}\ \bibnamefont {Machado}}, \bibinfo {author} {\bibfnamefont {G.}~\bibnamefont {Frascella}}, \bibinfo {author} {\bibfnamefont {J.~P.}\ \bibnamefont {Torres}},\ and\ \bibinfo {author} {\bibfnamefont {M.~V.}\ \bibnamefont {Chekhova}},\ }\bibfield  {journal} {\bibinfo  {journal} {App. Phys. Lett.}\ }\textbf {\bibinfo {volume} {117}},\ \href {https://doi.org/10.1063/5.0016259} {10.1063/5.0016259} (\bibinfo {year} {2020})\BibitemShut {NoStop}%
\bibitem [{\citenamefont {Qin}\ \emph {et~al.}(2023)\citenamefont {Qin}, \citenamefont {Deng}, \citenamefont {Zhong}, \citenamefont {Peng}, \citenamefont {Su}, \citenamefont {Luo}, \citenamefont {Xu}, \citenamefont {Wu}, \citenamefont {Gong}, \citenamefont {Liu} \emph {et~al.}}]{qin2023unconditional}%
  \BibitemOpen
  \bibfield  {author} {\bibinfo {author} {\bibfnamefont {J.}~\bibnamefont {Qin}}, \bibinfo {author} {\bibfnamefont {Y.-H.}\ \bibnamefont {Deng}}, \bibinfo {author} {\bibfnamefont {H.-S.}\ \bibnamefont {Zhong}}, \bibinfo {author} {\bibfnamefont {L.-C.}\ \bibnamefont {Peng}}, \bibinfo {author} {\bibfnamefont {H.}~\bibnamefont {Su}}, \bibinfo {author} {\bibfnamefont {Y.-H.}\ \bibnamefont {Luo}}, \bibinfo {author} {\bibfnamefont {J.-M.}\ \bibnamefont {Xu}}, \bibinfo {author} {\bibfnamefont {D.}~\bibnamefont {Wu}}, \bibinfo {author} {\bibfnamefont {S.-Q.}\ \bibnamefont {Gong}}, \bibinfo {author} {\bibfnamefont {H.-L.}\ \bibnamefont {Liu}}, \emph {et~al.},\ }\href {https://doi.org/10.1103/PhysRevLett.130.070801} {\bibfield  {journal} {\bibinfo  {journal} {Phys. Rev. Lett.}\ }\textbf {\bibinfo {volume} {130}},\ \bibinfo {pages} {070801} (\bibinfo {year} {2023})}\BibitemShut {NoStop}%
\bibitem [{\citenamefont {Eckstein}\ \emph {et~al.}(2011{\natexlab{b}})\citenamefont {Eckstein}, \citenamefont {Brecht},\ and\ \citenamefont {Silberhorn}}]{eckstein2011quantum}%
  \BibitemOpen
  \bibfield  {author} {\bibinfo {author} {\bibfnamefont {A.}~\bibnamefont {Eckstein}}, \bibinfo {author} {\bibfnamefont {B.}~\bibnamefont {Brecht}},\ and\ \bibinfo {author} {\bibfnamefont {C.}~\bibnamefont {Silberhorn}},\ }\href {https://doi.org/10.1364/OE.19.013770} {\bibfield  {journal} {\bibinfo  {journal} {Opt. Exp.}\ }\textbf {\bibinfo {volume} {19}},\ \bibinfo {pages} {13770} (\bibinfo {year} {2011}{\natexlab{b}})}\BibitemShut {NoStop}%
\bibitem [{\citenamefont {Manurkar}\ \emph {et~al.}(2016)\citenamefont {Manurkar}, \citenamefont {Jain}, \citenamefont {Silver}, \citenamefont {Huang}, \citenamefont {Langrock}, \citenamefont {Fejer}, \citenamefont {Kumar},\ and\ \citenamefont {Kanter}}]{manurkar2016multidimensional}%
  \BibitemOpen
  \bibfield  {author} {\bibinfo {author} {\bibfnamefont {P.}~\bibnamefont {Manurkar}}, \bibinfo {author} {\bibfnamefont {N.}~\bibnamefont {Jain}}, \bibinfo {author} {\bibfnamefont {M.}~\bibnamefont {Silver}}, \bibinfo {author} {\bibfnamefont {Y.-P.}\ \bibnamefont {Huang}}, \bibinfo {author} {\bibfnamefont {C.}~\bibnamefont {Langrock}}, \bibinfo {author} {\bibfnamefont {M.~M.}\ \bibnamefont {Fejer}}, \bibinfo {author} {\bibfnamefont {P.}~\bibnamefont {Kumar}},\ and\ \bibinfo {author} {\bibfnamefont {G.~S.}\ \bibnamefont {Kanter}},\ }\href {https://doi.org/10.1364/OPTICA.3.001300} {\bibfield  {journal} {\bibinfo  {journal} {Optica}\ }\textbf {\bibinfo {volume} {3}},\ \bibinfo {pages} {1300} (\bibinfo {year} {2016})}\BibitemShut {NoStop}%
\bibitem [{\citenamefont {Reddy}\ and\ \citenamefont {Raymer}(2017)}]{reddy2017engineering}%
  \BibitemOpen
  \bibfield  {author} {\bibinfo {author} {\bibfnamefont {D.~V.}\ \bibnamefont {Reddy}}\ and\ \bibinfo {author} {\bibfnamefont {M.~G.}\ \bibnamefont {Raymer}},\ }\href {https://doi.org/10.1364/OE.25.012952} {\bibfield  {journal} {\bibinfo  {journal} {Opt. Exp.}\ }\textbf {\bibinfo {volume} {25}},\ \bibinfo {pages} {12952} (\bibinfo {year} {2017})}\BibitemShut {NoStop}%
\bibitem [{\citenamefont {Silverstone}\ \emph {et~al.}(2014)\citenamefont {Silverstone}, \citenamefont {Bonneau}, \citenamefont {Ohira}, \citenamefont {Suzuki}, \citenamefont {Yoshida}, \citenamefont {Iizuka}, \citenamefont {Ezaki}, \citenamefont {Natarajan}, \citenamefont {Tanner}, \citenamefont {Hadfield} \emph {et~al.}}]{silverstone2014chip}%
  \BibitemOpen
  \bibfield  {author} {\bibinfo {author} {\bibfnamefont {J.~W.}\ \bibnamefont {Silverstone}}, \bibinfo {author} {\bibfnamefont {D.}~\bibnamefont {Bonneau}}, \bibinfo {author} {\bibfnamefont {K.}~\bibnamefont {Ohira}}, \bibinfo {author} {\bibfnamefont {N.}~\bibnamefont {Suzuki}}, \bibinfo {author} {\bibfnamefont {H.}~\bibnamefont {Yoshida}}, \bibinfo {author} {\bibfnamefont {N.}~\bibnamefont {Iizuka}}, \bibinfo {author} {\bibfnamefont {M.}~\bibnamefont {Ezaki}}, \bibinfo {author} {\bibfnamefont {C.~M.}\ \bibnamefont {Natarajan}}, \bibinfo {author} {\bibfnamefont {M.~G.}\ \bibnamefont {Tanner}}, \bibinfo {author} {\bibfnamefont {R.~H.}\ \bibnamefont {Hadfield}}, \emph {et~al.},\ }\href {https://doi.org/10.1038/nphoton.2013.339} {\bibfield  {journal} {\bibinfo  {journal} {Nature Photonics}\ }\textbf {\bibinfo {volume} {8}},\ \bibinfo {pages} {104} (\bibinfo {year} {2014})}\BibitemShut {NoStop}%
\bibitem [{\citenamefont {Luo}\ \emph {et~al.}(2019)\citenamefont {Luo}, \citenamefont {Brauner}, \citenamefont {Eigner}, \citenamefont {Sharapova}, \citenamefont {Ricken}, \citenamefont {Meier}, \citenamefont {Herrmann},\ and\ \citenamefont {Silberhorn}}]{luo2019nonlinear}%
  \BibitemOpen
  \bibfield  {author} {\bibinfo {author} {\bibfnamefont {K.-H.}\ \bibnamefont {Luo}}, \bibinfo {author} {\bibfnamefont {S.}~\bibnamefont {Brauner}}, \bibinfo {author} {\bibfnamefont {C.}~\bibnamefont {Eigner}}, \bibinfo {author} {\bibfnamefont {P.~R.}\ \bibnamefont {Sharapova}}, \bibinfo {author} {\bibfnamefont {R.}~\bibnamefont {Ricken}}, \bibinfo {author} {\bibfnamefont {T.}~\bibnamefont {Meier}}, \bibinfo {author} {\bibfnamefont {H.}~\bibnamefont {Herrmann}},\ and\ \bibinfo {author} {\bibfnamefont {C.}~\bibnamefont {Silberhorn}},\ }\href {https://doi.org/10.1126/sciadv.aat1451} {\bibfield  {journal} {\bibinfo  {journal} {Science advances}\ }\textbf {\bibinfo {volume} {5}},\ \bibinfo {pages} {eaat1451} (\bibinfo {year} {2019})}\BibitemShut {NoStop}%
\bibitem [{\citenamefont {Bao}\ \emph {et~al.}(2023)\citenamefont {Bao}, \citenamefont {Fu}, \citenamefont {Pramanik}, \citenamefont {Mao}, \citenamefont {Chi}, \citenamefont {Cao}, \citenamefont {Zhai}, \citenamefont {Mao}, \citenamefont {Dai}, \citenamefont {Chen} \emph {et~al.}}]{bao2023very}%
  \BibitemOpen
  \bibfield  {author} {\bibinfo {author} {\bibfnamefont {J.}~\bibnamefont {Bao}}, \bibinfo {author} {\bibfnamefont {Z.}~\bibnamefont {Fu}}, \bibinfo {author} {\bibfnamefont {T.}~\bibnamefont {Pramanik}}, \bibinfo {author} {\bibfnamefont {J.}~\bibnamefont {Mao}}, \bibinfo {author} {\bibfnamefont {Y.}~\bibnamefont {Chi}}, \bibinfo {author} {\bibfnamefont {Y.}~\bibnamefont {Cao}}, \bibinfo {author} {\bibfnamefont {C.}~\bibnamefont {Zhai}}, \bibinfo {author} {\bibfnamefont {Y.}~\bibnamefont {Mao}}, \bibinfo {author} {\bibfnamefont {T.}~\bibnamefont {Dai}}, \bibinfo {author} {\bibfnamefont {X.}~\bibnamefont {Chen}}, \emph {et~al.},\ }\href {https://doi.org/10.1038/s41566-023-01187-z} {\bibfield  {journal} {\bibinfo  {journal} {Nature Photonics}\ }\textbf {\bibinfo {volume} {17}},\ \bibinfo {pages} {573} (\bibinfo {year} {2023})}\BibitemShut {NoStop}%
\bibitem [{\citenamefont {Bulmer}\ \emph {et~al.}(2022)\citenamefont {Bulmer}, \citenamefont {Paesani}, \citenamefont {Chadwick},\ and\ \citenamefont {Quesada}}]{bulmer2022threshold}%
  \BibitemOpen
  \bibfield  {author} {\bibinfo {author} {\bibfnamefont {J.~F.~F.}\ \bibnamefont {Bulmer}}, \bibinfo {author} {\bibfnamefont {S.}~\bibnamefont {Paesani}}, \bibinfo {author} {\bibfnamefont {R.~S.}\ \bibnamefont {Chadwick}},\ and\ \bibinfo {author} {\bibfnamefont {N.}~\bibnamefont {Quesada}},\ }\href {https://doi.org/10.1103/PhysRevA.106.043712} {\bibfield  {journal} {\bibinfo  {journal} {Phys. Rev. A}\ }\textbf {\bibinfo {volume} {106}},\ \bibinfo {pages} {043712} (\bibinfo {year} {2022})}\BibitemShut {NoStop}%
\bibitem [{\citenamefont {Jeong}\ \emph {et~al.}(2000)\citenamefont {Jeong}, \citenamefont {Lee},\ and\ \citenamefont {Kim}}]{jeong2000dynamics}%
  \BibitemOpen
  \bibfield  {author} {\bibinfo {author} {\bibfnamefont {H.}~\bibnamefont {Jeong}}, \bibinfo {author} {\bibfnamefont {J.}~\bibnamefont {Lee}},\ and\ \bibinfo {author} {\bibfnamefont {M.~S.}\ \bibnamefont {Kim}},\ }\href {https://doi.org/10.1103/PhysRevA.61.052101} {\bibfield  {journal} {\bibinfo  {journal} {Phys. Rev. A}\ }\textbf {\bibinfo {volume} {61}},\ \bibinfo {pages} {052101} (\bibinfo {year} {2000})}\BibitemShut {NoStop}%
\bibitem [{\citenamefont {Quesada}\ and\ \citenamefont {Bra\ifmmode~\acute{n}\else \'{n}\fi{}czyk}(2019)}]{quesada2019broadband}%
  \BibitemOpen
  \bibfield  {author} {\bibinfo {author} {\bibfnamefont {N.}~\bibnamefont {Quesada}}\ and\ \bibinfo {author} {\bibfnamefont {A.~M.}\ \bibnamefont {Bra\ifmmode~\acute{n}\else \'{n}\fi{}czyk}},\ }\href {https://doi.org/10.1103/PhysRevA.99.013830} {\bibfield  {journal} {\bibinfo  {journal} {Phys. Rev. A}\ }\textbf {\bibinfo {volume} {99}},\ \bibinfo {pages} {013830} (\bibinfo {year} {2019})}\BibitemShut {NoStop}%
\bibitem [{\citenamefont {Agrawal}(2013)}]{agarwal2013nonlinear}%
  \BibitemOpen
  \bibfield  {author} {\bibinfo {author} {\bibfnamefont {G.}~\bibnamefont {Agrawal}},\ }in\ \href {https://doi.org/https://doi.org/10.1016/B978-0-12-397023-7.00007-3} {\emph {\bibinfo {booktitle} {Nonlinear Fiber Optics (Fifth Edition)}}},\ \bibinfo {series and number} {Optics and Photonics},\ \bibinfo {editor} {edited by\ \bibinfo {editor} {\bibfnamefont {G.}~\bibnamefont {Agrawal}}}\ (\bibinfo  {publisher} {Academic Press},\ \bibinfo {address} {Boston},\ \bibinfo {year} {2013})\ \bibinfo {edition} {fifth edition}\ ed.,\ pp.\ \bibinfo {pages} {245--293}\BibitemShut {NoStop}%
\bibitem [{\citenamefont {Killoran}\ \emph {et~al.}(2019)\citenamefont {Killoran}, \citenamefont {Izaac}, \citenamefont {Quesada}, \citenamefont {Bergholm}, \citenamefont {Amy},\ and\ \citenamefont {Weedbrook}}]{killoran2019strawberry}%
  \BibitemOpen
  \bibfield  {author} {\bibinfo {author} {\bibfnamefont {N.}~\bibnamefont {Killoran}}, \bibinfo {author} {\bibfnamefont {J.}~\bibnamefont {Izaac}}, \bibinfo {author} {\bibfnamefont {N.}~\bibnamefont {Quesada}}, \bibinfo {author} {\bibfnamefont {V.}~\bibnamefont {Bergholm}}, \bibinfo {author} {\bibfnamefont {M.}~\bibnamefont {Amy}},\ and\ \bibinfo {author} {\bibfnamefont {C.}~\bibnamefont {Weedbrook}},\ }\href {https://doi.org/10.22331/q-2019-03-11-129} {\bibfield  {journal} {\bibinfo  {journal} {Quantum}\ }\textbf {\bibinfo {volume} {3}},\ \bibinfo {pages} {129} (\bibinfo {year} {2019})}\BibitemShut {NoStop}%
\bibitem [{\citenamefont {Boyd}(2003)}]{boyd2003nonlinear}%
  \BibitemOpen
  \bibfield  {author} {\bibinfo {author} {\bibfnamefont {R.}~\bibnamefont {Boyd}},\ }\href@noop {} {\emph {\bibinfo {title} {Nonlinear Optics}}}\ (\bibinfo  {publisher} {Academic Press},\ \bibinfo {year} {2003})\BibitemShut {NoStop}%
\bibitem [{\citenamefont {Quesada}\ \emph {et~al.}(2018)\citenamefont {Quesada}, \citenamefont {Arrazola},\ and\ \citenamefont {Killoran}}]{quesada2018gaussian}%
  \BibitemOpen
  \bibfield  {author} {\bibinfo {author} {\bibfnamefont {N.}~\bibnamefont {Quesada}}, \bibinfo {author} {\bibfnamefont {J.~M.}\ \bibnamefont {Arrazola}},\ and\ \bibinfo {author} {\bibfnamefont {N.}~\bibnamefont {Killoran}},\ }\href {https://doi.org/10.1103/PhysRevA.98.062322} {\bibfield  {journal} {\bibinfo  {journal} {Phys. Rev. A}\ }\textbf {\bibinfo {volume} {98}},\ \bibinfo {pages} {062322} (\bibinfo {year} {2018})}\BibitemShut {NoStop}%
\bibitem [{\citenamefont {Helt}\ and\ \citenamefont {Quesada}(2020)}]{helt2020degenerate}%
  \BibitemOpen
  \bibfield  {author} {\bibinfo {author} {\bibfnamefont {L.}~\bibnamefont {Helt}}\ and\ \bibinfo {author} {\bibfnamefont {N.}~\bibnamefont {Quesada}},\ }\href {https://doi.org/10.1088/2515-7647/ab87fc} {\bibfield  {journal} {\bibinfo  {journal} {Journal of Physics: Photonics}\ }\textbf {\bibinfo {volume} {2}},\ \bibinfo {pages} {035001} (\bibinfo {year} {2020})}\BibitemShut {NoStop}%
\bibitem [{\citenamefont {Quesada}(2015)}]{quesada2015thesis}%
  \BibitemOpen
  \bibfield  {author} {\bibinfo {author} {\bibfnamefont {N.}~\bibnamefont {Quesada}},\ }\emph {\bibinfo {title} {Very Nonlinear Quantum Optics}},\ \href@noop {} {Ph.D. thesis},\ \bibinfo  {school} {University of Toronto}, \bibinfo {address} {Toronto, Canada} (\bibinfo {year} {2015})\BibitemShut {NoStop}%
\end{thebibliography}%

\newpage
\onecolumngrid












\onecolumngrid
\renewcommand{\theequation}{S\arabic{equation}}
\renewcommand{\thefigure}{S\arabic{figure}}

\section*{Supplemental material}

\newcommand{\dd}{\text{d}}

\newcommand{\ddt}[1]{\frac{\text{d}{#1}}{\text{d}t}}

\newcommand{\hc}{\text{h.c.}}
\newcommand{\cc}{\text{c.c.}}

\newcommand{\nico}[1]{\textcolor{violet}{Nico: #1 \\}}
\newcommand{\will}[1]{\textcolor{blue}{Will: #1 \\}}
\newcommand{\mh}[1]{\textcolor{green}{Martin: #1 \\}}
\newcommand{\gt}[1]{\textcolor{red}{Guillaume: #1 \\}}

\subsection{Hamiltonian}
\label{sec:Hamiltonian}
We start by considering propagation in a quasi-one-dimensional geometry with three beams that we label signal (s), idler (i), and pump (p). Each beam is characterized by a central frequency $\bar{\omega}_\mu$, a wavevector $\bar{k}_\mu$ ($\{s,i,p\} \ni \mu$), and a dispersion relation
\begin{align}
\underbrace{k_\mu - \bar{k}_\mu}_{\equiv \delta k_\mu} = \frac{1}{v_\mu} \underbrace{(\omega_\mu - \bar{\omega}_\mu)}_{\equiv \delta \omega_\mu} ,
\end{align}
where $v_\mu$ is the group velocity of beam $\mu$ which is assumed to be constant over the range of frequencies involved. 
We assume that higher order dispersion terms have a negligible effect given the (narrow) frequency support of the beams in considerations. 
As we are interested in type-\MakeUppercase{\romannumeral 2} SPDC, we assume that the central frequencies and wavevectors satisfy energy conservation and (quasi-)phase matching
\begin{align}\label{eq:phase-matching}
\bar{\omega}_p - \bar{\omega}_s - \bar{\omega}_i = &0, \\ \bar{k}_p - \bar{k}_s - \bar{k}_i =& 0\  (\text{or } =\pm \frac{2 \pi}{\Lambda} \text{if quasi-phase-matching}).
\end{align}
For the quasi-phase matching case, $\Lambda$ is the poling period.
The free (i.e. linear) Hamiltonian of the beams is given by
\begin{align}
\hat{\mathcal{H}}_{\text{L}} &=  \sum_{\{s,i,p\}\ni \mu} \int dk \hbar \omega_\mu(k_\mu) \hat{a}^\dagger_\mu(k_\mu) \hat{a}_\mu(k_\mu) \\
&= 
\sum_{\{s,i,p\}\ni \mu} \int dz \hbar \left[ \bar{\omega}_\mu \hat{\psi}_\mu^\dagger (z) \hat{\psi}_\mu(z) -\tfrac{i}{2} v_\mu \left\{ \hat{\psi}_\mu^\dagger(z,t) \frac{\partial \hat{\psi}_\mu(z,t)}{\partial z}  - \frac{\partial \hat{\psi}_\mu^\dagger(z,t)}{\partial z}  \hat{\psi}_\mu(z,t) \right\}  \right] \nonumber 
\end{align}
where we have introduced the bosonic creation and annihilation operators which satisfy
\begin{align}
[\hat{a}_i(k),\hat{a}^\dagger_j(k')] = \delta(k-k')\delta_{i,j},
\end{align}
and all others commutation relations are zero. 
Similarly, we expressed the Hamiltonian in terms of the field operators which are defined as
\begin{align}
\hat{\psi}_{\mu}(z) = \int \frac{dk}{\sqrt{2 \pi}} \hat{a}_\mu(k) e^{i (k - \bar{k}_\mu) z}.
\end{align}

Under the rotating-wave approximation, we keep only the resonant terms for the nonlinear portion of the Hamiltonian. 
Doing so, we find  
\begin{align}
\hat{\mathcal{H}}_{\text{NL}} = - \hbar  \int dz \Bigl\{ \zeta_{s,i,p} g(z) \hat{\psi}^\dagger_s(z) \hat{\psi}^\dagger_i(z) \hat{\psi}_p(z)	+ \frac{1}{2}\zeta_{p} h(z) \hat{\psi}^\dagger_p(z)^\dagger\hat{\psi}^\dagger_p(z)\hat{\psi}_p(z)\hat{\psi}_p(z) \nonumber  \\  + \zeta_{s}h(z)\hat{\psi}^\dagger_p(z)\hat{\psi}_p(z)\hat{\psi}^\dagger_s(z)\hat{\psi}_s(z)+\zeta_{i}h(z)\hat{\psi^\dagger}_p(z)\hat{\psi}_p(z)\hat{\psi}^\dagger_i(z)\hat{\psi}_i(z)       +\text{h.c.}  \Bigr\}.
\end{align}
In the first term, responsible for pair creation, we have $\zeta_{s,i,p} = \sqrt{\frac{\hbar\bar{\omega}_s \bar{\omega}_i \bar{\omega}_p}{2 \epsilon_0 \bar{n}_s \bar{n}_i \bar{n}_p A_{s,i,p} }}\bar{\chi}_{2}$, where $\bar{n}_j$ is the index of refraction of beam $j$ at $\bar{\omega}_j$, $\epsilon_0$ is the permittivity of vacuum, and $A_{s,i,p}$ is a characteristic area describing the region over which the interaction occurs~\cite{quesada2022beyond}. 
The function $g(z)$ is the poling function which takes on the values $g(z)=0$ where the nonlinearity is absent and either $+1$ or $-1$ depending on the orientation of the nonlinear medium. 
As an example, for quasi-phase matching, $g(z)$ would be an alternating function of $\pm 1$ over small intervals of length $\Lambda$. 
The next three terms are responsible for self- and cross-phase modulation of the beams. 
The function $h(z)$ is a square function which is $+1$ where the nonlinearity is present and $0$ otherwise. 
For the self-phase modulation of the pump, we have $\zeta_{p}=\frac{3}{4\epsilon_0}\left(\frac{\hbar \bar{\omega}_{p}^2}{\bar{n}_{p}^{2}\sqrt{A_p}}   \right)\bar{\chi}_{3}$ and for the cross-phase modulation we have $\zeta_{\mu}=\frac{3}{2\epsilon_0}\left(\frac{\hbar \bar{\omega}_{p}\bar{\omega}_{\mu}}{\bar{n}_{p}\bar{n}_{\mu}\sqrt{A_\mu}}   \right)\bar{\chi}_{3}$ where $\mu=s,i$. 
Again, $A_{j}$ for $\{s,i,p\} \ni j$ is a characteristic area describing the area over which the self- and cross-phase modulation interactions occur.  

We can then construct an effective Hamiltonian for our model which we use to obtain the Schr\"{o}dinger equation satisfied by the evolution operator
\begin{align}\label{eq:schrodingereom}
    i\hbar \frac{d}{dt}\hat{\mathcal{U}}(t,t_{0}) = \hat{\mathcal{H}}_{\text{eff}}\hat{\mathcal{U}}(t,t_{0}) = \left( \hat{\mathcal{H}}_{\text{L}}+\hat{\mathcal{H}}_{\text{NL}} \right)\hat{\mathcal{U}}(t,t_{0}),
\end{align}
where at the initial time $t_{0}$, $\hat{\mathcal{U}}(t_{0},t_{0})=\hat{\mathbb{1}}$ with $\hat{\mathbb{1}}$ being the identity operator. 
With the evolution operator in hand, we can now consider how initial states evolve by either evaluating $\ket{{\psi}(t)} = \hat{\mathcal{U}}(t,t_0) \ket{{\psi}(t_0)}$ or by solving the Heisenberg equations of motion for the Heisenberg field operators $\hat{\psi}_{j}(z,t) = \hat{\mathcal{U}}^\dagger (t,t_0) \hat{\psi}_{j}(z,t_0) \hat{\mathcal{U}}(t,t_0)$.

The state that we are interested in evolving is a product of vacuum for the signal and idler beams and a strong coherent state for the pump:
\begin{align}\label{eq:initialstate}
\ket{\Psi(t_0)} =\exp\left( \int dz f_p(z) \hat{\psi}^\dagger_p(z) -\text{h.c.} \right) \ket{\text{vac}} = \exp\left( \int dk \alpha_p(k) \hat{a}^\dagger_p(k) -\text{h.c.} \right) \ket{\text{vac}} , \  \ket{\text{vac}}=\ket{\text{vac}_p} \otimes \ket{\text{vac}_s} \otimes \ket{\text{vac}_i} 
\end{align}
where $\ket{\text{vac}}$ is the three-beam vacuum state, and $\exp\left( \int dz f_p(z) \psi_p(z)^\dagger -\text{h.c.} \right)$ is a displacement operator. 
We will assume that the nonlinear region lies in the region $-L/2 < z < L/2$ , where $L$ is the length of the crystal. 
Moreover, we assume that at time $t_0$, the pump envelope function has not entered yet entered the crystal such that $\int_{-L/2}^{L/2} dz |f_p(z)|^2 \approx 0$.

\subsection{Heisenberg dynamics}
From our effective Hamiltonian, we find that the Heisenberg equation for the pump field operator is
\begin{align}
    \left( \frac{\partial}{\partial t} +v_{p}\frac{\partial}{\partial z} +i\bar{\omega}_{p} \right)\hat{\psi}_p(z,t) = i\zeta_{p}h(z)\hat{\psi}_p(z,t)^{\dagger}\hat{\psi}_p(z,t)\hat{\psi}_p(z,t) +\text{back action terms},
\end{align}
where the ``back action terms" are interaction terms which contain the signal and idler field operators. 
We make the undepleted-classical pump approximation such that $\hat{\psi}_p(z,t)\rightarrow \langle \hat{\psi}_p(z,t)\rangle $ and assume that the number of pump photons remains unchanged. Under these assumptions, we can ignore these back action terms, and obtain a solution for the mean pump field
\begin{align}
    \langle \hat{\psi}_p(z,t)\rangle = f_{p}(z-v_{p}(t-t_{0}))e^{i\bar{\omega}_{p}(t-t_{0})+i\theta(z,t)}
\end{align}
where 
\begin{align}
    \langle \hat{\psi}_p(z,t_{0})\rangle = f_{p}(z)
\end{align}
is the spatial pump envelope function at time $t_{0}$ [Eq.~\eqref{eq:initialstate}] and the phase accumulated by self-phase modulation is
\begin{align}
    \theta(z,t)=|f_{p}(z-v_{p}(t-t_{0}))|^{2}\int_{t_{0}}^{t} dt' \zeta_{p}h(z-v_{p}(t-t')).
\end{align}

Having a solution for the mean pump field, we now consider the Heisenberg equations of motion for the signal and idler fields
\begin{align}
    \left( \frac{\partial}{\partial t} +v_{s}\frac{\partial}{\partial z} +i\bar{\omega}_{s} \right)\hat{\psi}_s(z,t) &= i\zeta_{s,i,p}g(z) \langle \hat{\psi}_p(z,t)\rangle \hat{\psi}^{\dagger}_i(z,t) +i \zeta_{s}h(z)| \langle \hat{\psi}_p(z,t_{0})\rangle|^{2}\hat{\psi}_s(z,t)\\
    \left( \frac{\partial}{\partial t} +v_{i}\frac{\partial}{\partial z} +i\bar{\omega}_{i} \right)\hat{\psi}^{\dagger}_i(z,t) &= -i\zeta^{*}_{s,i,p}g(z) \langle \hat{\psi}_p(z,t)\rangle \hat{\psi}_s(z,t) -i \zeta^{*}_{i}h(z)| \langle \hat{\psi}_p(z,t_{0})\rangle|^{2}\hat{\psi}^{\dagger}_i(z,t).
\end{align}
We solve these equations via two different, yet similar, methods which are all included in the Python package \texttt{NeedALight}. 
For the first method, we follow the steps of Ref.~\cite{triginer2020understanding,quesada2020theory} and Fourier transform into frequency space. 
Doing so gives us integro-differential equations in ($z,\omega$) space. 
The terms on the right-hand side of the equations become convolutions. 
We solve numerically by discretizing the frequencies onto a grid of $N$ points and approximating the integrals as sums which give rise to matrix equations of motion. 
For the second method, we follow the steps of Ref.~\cite{helt2020degenerate} and Fourier transform into momentum space which gives us equations of motion in ($k,t$) space. 
Again, the terms on the right-hand side become convolutions. 
Similarly, we solve numerically by discretizing the momenta on a grid of $N$ points and approximating the integrals as sums. 

Once the solutions are obtained, we can evaluate certain quantities directly in the chosen space (e.g. the JSA) or Fourier transform back into ($z,t$) to see how the energy densities evolve.

\subsection{Magnus Expansion}
\label{sec:magnus}

By numerically solving the Heisenberg equations of motion, we can study how the signal and idler energy densities evolve as a function of time [Fig.~\ref{fig:fig2}]. 
However, it is worth also considering the problem in the Schr\"odinger picture which can be solved analytically under certain simplifying assumptions (which we explicitly mention below).
In this picture, the solution to Eq.~\eqref{eq:schrodingereom} is
\begin{align}
    \hat{\mathcal{U}}(t,t_{0}) =\hat{\mathcal{T}}\exp\left[-\frac{i}{\hbar}\int_{t_{0}}^{t}dt' \left( \hat{\mathcal{H}}_{\text{L}}+\hat{\mathcal{H}}_{\text{NL}}(t')   \right)\right],
\end{align}
where $\hat{\mathcal{T}}$ denotes the time-ordering operator. 
To obtain an approximate analytic expression for $\hat{\mathcal{U}}(t,t_{0})$, one usually resorts to perturbative expansions such as the Magnus series. 
To do so, we first move into the interaction picture and define the interaction evolution operator
\begin{align}\label{eqn:intevolop}
    \hat{\mathcal{U}}_{I}(t,t_{0}) = e^{\frac{i}{\hbar}\hat{\mathcal{H}}_{\text{L}}(t-t_{0})}\hat{\mathcal{U}}(t,t_{0}). 
\end{align}
This operator obeys the new Schr\"{o}dinger equation
\begin{align}
    i\hbar\frac{d}{dt}\hat{\mathcal{U}}_{I}(t,t_{0})=\hat{\mathcal{H}}_{\text{I}}(t)\hat{\mathcal{U}}_{I}(t,t_{0}),
\end{align}
where 
\begin{align}
    \hat{\mathcal{H}}_{\text{I}}(t)= e^{\frac{i}{\hbar}\hat{\mathcal{H}}_{\text{L}}(t-t_{0})} \hat{\mathcal{H}}_{\text{NL}}(t) e^{-\frac{i}{\hbar}\hat{\mathcal{H}}_{\text{L}}(t-t_{0})}.
\end{align}
We can write the formal solution of the interaction-picture Schr\"odinger equation again in terms of time-ordered exponential
\begin{align}
\hat{\mathcal{U}}_{I}(t,t_{0}) = \hat{\mathcal{T}}\exp\left[-\frac{i}{\hbar} \int_{t_0}^t dt' \hat{\mathcal{H}}_I(t') \right].
\label{eqn:evol_operator_sm}
\end{align}
This allows us to focus solely on the nonlinear Hamiltonian which gives us the interesting dynamics.
The evolution operator in Eq.~\eqref{eqn:evol_operator_sm} can be expanded according to a Magnus series:
\begin{equation}
\hat{\mathcal{U}}_{I}(t,t_{0}) = \exp{\left[ \hat{\Omega}_1(t,t_{0}) + \hat{\Omega}_2(t,t_{0}) + \hat{\Omega}_3(t,t_{0}) + ... \right]}
\label{eqn:magnus_expansion_sm}
\end{equation}
where 
\begin{align}
 \hat{\Omega}_1(t,t_{0}) &= \frac{-i}{\hbar} \int_{t_0}^{t} dt' \hat{\mathcal{H}}_I(t') \\
 \hat{\Omega}_2(t,t_{0}) &= \frac{(-i)^2}{2\hbar^2} \int_{t_0}^{t} dt' \int_{t_0}^{t'} dt'' [\hat{\mathcal{H}}_I(t'), \hat{\mathcal{H}}_I(t'')] \\
 \hat{\Omega}_3(t,t_{0}) &= \frac{(-i)^3}{6\hbar^3} \int_{t_0}^{t} dt' \int_{t_0}^{t'} dt'' \int_{t_0}^{t''} dt''' \left( [\hat{\mathcal{H}}_I(t'),[\hat{\mathcal{H}}_I(t''), \hat{\mathcal{H}}_I(t''')]] +  [[\hat{\mathcal{H}}_I(t'),\hat{\mathcal{H}}_I(t'')], \hat{\mathcal{H}}_I(t''')]\right) 
\end{align}
with $t_0 < t'' < t' < t$.
Following the derivations in Refs.~\cite{quesada2014effects, quesada2015time}, we can arrive at analytic expressions for $\hat{\Omega}_i(t,t_{0})$ for $i\leq3$ by making a number of simplifying assumptions.

(i) Firstly, we neglect higher-order nonlinear effects such as cross-phase modulation and self-phase modulation.
As such, the interaction Hamiltonian takes the simpler form 
\begin{align}
    \hat{\mathcal{H}}_{I}(t) = -\frac{\hbar\zeta_{s,i,p}}{(2 \pi)^{3/2}} \int dk_s dk_i dk_s \left[ \int_{-L/2}^{L/2} dz g(z) e^{-i (\delta k_s + \delta k_i - \delta k_p)z} \right] \hat{a}_s^\dagger (k_s) \hat{a}_i^\dagger(k_i) \alpha_p(k_p) e^{i\left( \delta\omega_{s} +\delta\omega_{i}-\delta\omega_{p} \right) (t-t_{0})} + \text{h.c.}
\end{align}
where $\alpha_{p}(k_{p}) = \langle \hat{a}_{p}(k_{p}) \rangle$ is the momentum space pump envelope function.
It is convenient to work in terms of frequency rather than momenta, as such, we make a change of variable and recall that $\hat{a}_{j}(k_{j})\rightarrow \hat{a}_{j}(\omega_{j})\cdot\sqrt{\frac{\partial \omega_{j}}{\partial k_{j}}}=\hat{a}_{j}(\omega_{j})\sqrt{v_{j}}$ to arrive at
\begin{align}
H_I(t) &=  - \hbar C' \int d\omega_s d\omega_i d\omega_p \Phi(\Delta k )  \beta_p(\omega_p) \hat{a}_s^\dagger(\omega_s) \hat{a}_i^\dagger(\omega_i)e^{i(\omega_{s}+\omega_{i}-\omega_{p})(t-t_{0})} + \text{h.c.}\\
C' &=  \bar{\chi}_{2} \sqrt{\frac{ N \bar{\omega}_p \bar{\omega}_s \bar{\omega}_i}{2 \varepsilon_0 \bar{n}_s \bar{n}_i \bar{n}_p A (2 \pi)^3 v_p v_s  v_i}},
\end{align}
where $\beta_p(\omega_p) = \alpha_{p}(\omega_{p})/\sqrt{N} = \braket{\hat{a}_p (\omega_p)}/\sqrt{N}$ is the normalized spectral content of the pump and we have defined the phase-matching function
\begin{align}
    \Phi(\Delta k) = \int_{-L/2}^{L/2} dz g(z)e^{-i\Delta k z}.
    \label{eqn:pmf_preIntegral}
\end{align}
where $\Delta k \equiv \delta k_s(\omega_s) + \delta k_i(\omega_i) - \delta k_p(\omega_p)$ is the wavevector mismatch.

(ii) Secondly, we assume both the pump spectral mode and phase-matching functions can be approximated by Gaussians.
The pump mode is given by
\begin{align}
\beta_p(\omega_p)= \sqrt[4]{\frac{2}{\pi}} \sqrt{\tau} \exp(-\tau^2 (\omega_p - \bar{\omega}_p)^2).
\end{align}
For the phase-matching function, we perform the integral in Eq.~\eqref{eqn:pmf_preIntegral} with $g(z) \equiv 1$ to obtain
\begin{equation}
\begin{split}
\Phi(\Delta k)  &= \mathrm{sinc}\left( \frac{\Delta k L }{2} \right) \\
&\approx \exp\left[-\gamma \left( \frac{\Delta k L}{2} \right)^2  \right],
\end{split}
\label{eqn:pmf_withPhase}
\end{equation}
where $\gamma \approx 0.193$ is the optimal value to approximate the sinc function.
In practice, an apodized poling function $g(z)$ can be used to obtain a Gaussian phase matching function~\cite{branczyk2011engineered}.
We also note that the \texttt{NeedALight} numerical model can solve the dynamics for an arbitrary phase-matching function.
However, a Gaussian form makes the problem analytically solvable, as we show below.


We now begin by computing $\hat{\Omega}_1(t,t_{0})$.
We take $t_0 \to -\infty$ and $t \to \infty$ (in practice this just means $t_0$ before the pump enters the nonlinear region and $t$ after it exits it) in which case $\int_{-\infty}^\infty dt' e^{i (\omega_s+ \omega_i - \omega_p) t'} = 2 \pi \delta(\omega_s+ \omega_i - \omega_p)$.
Using this result we can write
\begin{equation}
\begin{split}
\hat{\Omega}_1(t,t_{0}) =& -\frac{i}{\hbar} \int_{t_0}^t dt' H_I(t')  \\
&= 2 \pi i C' \int d\omega_s  d\omega_i \Phi(\Delta k) \beta_p(\omega_s+ \omega_i) \hat{a}_s^\dagger(\omega_s) \hat{a}^\dagger_i(\omega_i) + \text{h.c.}
\end{split}
\label{eqn:omega1_temp}
\end{equation}
Expanding the wavevector mismatch $\Delta k$, we find 
\begin{align}
\sqrt{\gamma}\Delta k L /2 = \underbrace{\frac{\sqrt{\gamma} L}{2}\left( \frac{1}{v_s} -\frac{1}{v_p}\right)}_{\equiv \eta_s} \underbrace{(\omega_s - \bar{\omega}_s)}_{\equiv \delta \omega_s} + \underbrace{\frac{\sqrt{\gamma} L}{2}\left( \frac{1}{v_i} -\frac{1}{v_p}\right)}_{\equiv \eta_i} \underbrace{(\omega_i - \bar{\omega}_i)}_{\equiv \delta \omega_i}
\end{align}
which allows us to re-write Eq.~\eqref{eqn:omega1_temp} as
\begin{equation}
\begin{split}
\hat{\Omega}_1(t,t_{0}) &= 2\pi i C'   \sqrt[4]{\frac{2}{\pi}} \sqrt{\tau} \int d\omega_s  d\omega_i\exp\left[ -(\eta_s^2+\tau^2) [\delta \omega_s]^2 - (\eta_i^2+\tau^2) [\delta \omega_i]^2 - 2 (\tau^2 + \eta_s \eta_i) \delta \omega_s \delta \omega_i  \right] \hat{a}_s^\dagger(\omega_s) \hat{a}_i^\dagger(\omega_i) + \mathrm{h.c.}
\end{split}
\end{equation}

(iii) Thirdly, we will assume that $\tau^2 + \eta_s \eta_i = 0$, meaning that there are no correlations between signal and idler to first-order in perturbation. 
Moreover, we will take $\eta_s = -\eta_i = \pm \tau $.
These conditions can be met using group velocity matching.
We then find:
\begin{equation}
\hat{\Omega}_1(t,t_{0}) =  2\pi i C' \sqrt[4]{\frac{2}{\pi}} \sqrt{\frac{\pi}{4\tau}} \int d\omega_s d\omega_i f_0(\delta \omega_s) f_0(\delta \omega_i)  \hat{a}_s^\dagger(\omega_s) \hat{a}_i^\dagger(\omega_i) + \text{h.c.}
\end{equation}
where $f_0(\delta \omega_{s,i}) = \sqrt[4]{\frac{1}{ \pi}} \sqrt{2 \tau} \exp(-2 \tau^2 \delta \omega_{s,i}^2 )$ is a normalized function such that $\int d \omega_{s,i} |f_0(\delta \omega_{s,i})|^2 = 1$.
Defining $\varepsilon = 2\pi C' \sqrt[4]{\frac{2}{\pi}} \sqrt{\frac{\pi}{4\tau}}$, we arrive at the first-order Magnus term
\begin{equation}
\hat{\Omega}_1(t,t_{0}) = i \varepsilon \int d\omega_s d\omega_i J_1(\omega_s, \omega_i) \hat{a}_s^\dagger(\omega_s) \hat{a}_i^\dagger(\omega_i) + \text{h.c.}
\end{equation}
with a first-order joint spectral mode
\begin{equation}
J_1(\omega_s, \omega_i) = f_0(\delta \omega_s) f_0(\delta \omega_i)
\label{eqn:jsa_first_order_sm}
\end{equation}
which satisfies $\int d\omega_s d\omega_i |J_1(\omega_s, \omega_i)|^2 = 1$.

\begin{figure}
\centering
\includegraphics[width=1\columnwidth]{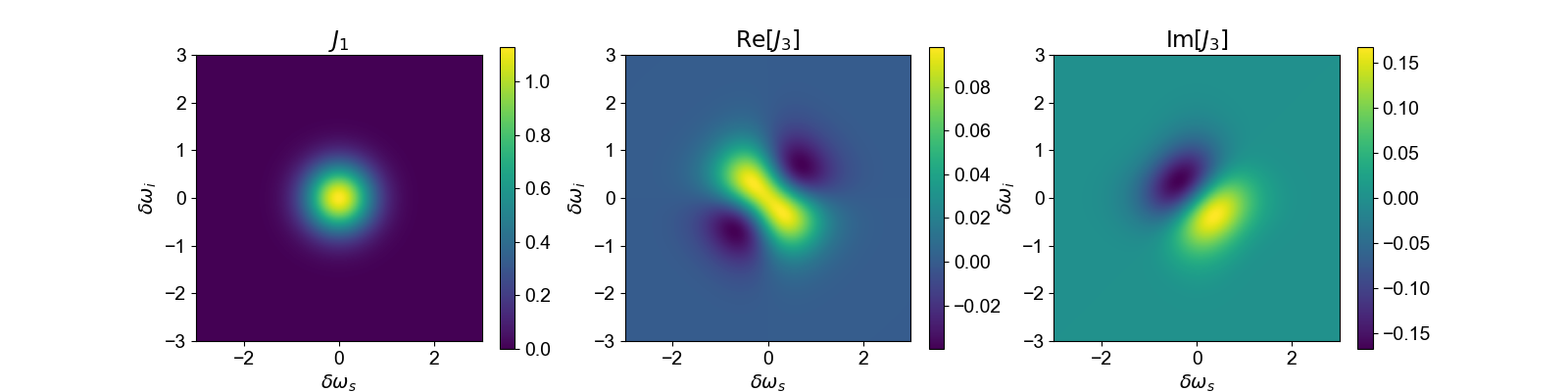}
\caption{First-order joint spectral mode $J_1(\omega_s,\omega_i)$ [Eq.~\eqref{eqn:jsa_first_order_sm}] and the next leading order mode $J_3(\omega_s,\omega_i)$ [Eq.~\eqref{eqn:jsa_third_order_sm}] using parameters $\eta_a=-\eta_b=\tau=1$.}
\label{fig:sm_JSAs}
\end{figure}

We now derive the expression for the next leading order correction to the joint spectral amplitude.
If we truncate Eq.~\eqref{eqn:magnus_expansion_sm} to third order in the Magnus expansion, we can use the Zassenhaus and Baker-Campbell-Hausdorff identities to arrive at
\begin{equation}\label{eqn:magnusapprox}
\hat{\mathcal{U}}_{I}(t,t_{0}) \approx  \exp{\left[ \hat{\Omega}_1(t,t_{0}) + \hat{\Omega}_3(t,t_{0}) - \frac{1}{2}[\hat{\Omega}_1(t,t_{0}),\hat{\Omega}_2(t,t_{0})]\right]} \exp{\left[ \hat{\Omega}_2(t,t_{0})\right]}.
\end{equation}
We note that $\exp{\left[ \hat{\Omega}_2(t,t_{0})\right]}\ket{0}_a\ket{0}_b = \ket{0}_a\ket{0}_b$.
Following a lengthy derivation which can be found in the Supplementary Materials of Ref.~\cite{quesada2015time} or in Ref.~\cite{quesada2022beyond}, one can show that:
\begin{equation}
\begin{split}
   \hat{\Omega}_3(t,t_{0}) &= i \varepsilon^3 \int d\omega_s d\omega_i L_3(\omega_s, \omega_i)  \hat{a}_s^\dagger(\omega_s) \hat{a}_i^\dagger(\omega_i) + \text{h.c.}, \\
   \frac{1}{2}[\hat{\Omega}_1(t,t_{0}),\hat{\Omega}_2(t,t_{0})] &= -\varepsilon^3 \int d\omega_s d\omega_i K_3(\omega_s, \omega_i)  \hat{a}_s^\dagger(\omega_s) \hat{a}_i^\dagger(\omega_i) + \text{h.c.}
\end{split}
\end{equation}
with 
\begin{equation}
\begin{split}
   L_3(\omega_s, \omega_i) &= \frac{1}{12}
   \left[ f_0(\delta \omega_s) f_0(\delta \omega_i) -  f_1(\delta \omega_s) f_1(\delta \omega_i) \right] \\
   K_3(\omega_s, \omega_i) &= \frac{1}{ 4\sqrt{3}} \left[  f_0(\delta \omega_s) f_1(\delta \omega_i) - f_1(\delta \omega_s) f_0(\delta \omega_i) \right]
\end{split}
\label{eqn:jsa_third_order_sm}
\end{equation}
where $f_1(\delta \omega_{s,i}) = \sqrt{3} f_0(\delta \omega_{s,i})\text{erfi}\left(\sqrt{\tfrac{4}{3}} \tau \delta \omega_{s,i}  \right)$.
We then define $J_3(\omega_s, \omega_i) =\sqrt{18}\left( L_3(\omega_s, \omega_i) - iK_3(\omega_s, \omega_i) \right) $ such that $\int d\omega_s d\omega_i |J_3(\omega_s, \omega_i)|^2 = 1$.
In Fig.~\ref{fig:sm_JSAs}, we plot $J_1(\omega_s, \omega_i)$ and both real and imaginary components of $J_3(\omega_s, \omega_i)$.
Since $\int d \omega f^*_i(\delta \omega) f_j(\delta \omega)= \delta_{i,j}$~\cite{quesada2022beyond}, the joint spectral amplitudes $J_1(\omega_s, \omega_i)$ and $J_3(\omega_s, \omega_i)$ are orthonormal functions and thus can be interpreted as different spectral-temporal modes in which the photon pairs are generated.

\subsection{Signal and idler group delay}
\label{sec:groupDelay}
In this section, we derive an analytic expression for the group delay between the signal and idler photons using the joint spectral amplitude obtained in Sec.~\ref{sec:magnus}.
Up to third order in the Magnus expansion, the joint spectral amplitude is given by 
\begin{equation}
    J(\omega_s, \omega_i) = \varepsilon J_1(\omega_s, \omega_i) + \frac{\varepsilon^3}{\sqrt{18}} J_3(\omega_s, \omega_i)
\end{equation}
where $J_1(\omega_s, \omega_i)$ is given by Eq.~\eqref{eqn:jsa_first_order_sm} and $J_3(\omega_s, \omega_i)= \sqrt{18} \left( L_3(\omega_s, \omega_i) - iK_3(\omega_s, \omega_i)\right)$ is given by Eq.~\eqref{eqn:jsa_third_order_sm}.
To obtain an expression for the group delay, we first compute the joint spectral phase which is given by:
\begin{equation}
\arg{[J(\omega_s, \omega_i)]} = \arctan{\left( \frac{\mathrm{Im}[J(\omega_s, \omega_i)]}{\mathrm{Re}[J(\omega_s, \omega_i)]} \right)} = \arctan{\left( \frac{-\varepsilon^3 K_3(\omega_s, \omega_i)}{\varepsilon J_1(\omega_s, \omega_i) + \varepsilon^3 L_3(\omega_s, \omega_i)} \right)}.
\end{equation}
Note that $f_0(\delta \omega_{s,i}) = f_0(-\delta \omega_{s,i})$ and $f_1(-\delta \omega_{s,i})=-f_1(\delta \omega_{s,i})$.
As a result, $\mathrm{Re}[J(\omega_s, \omega_i)] = \mathrm{Re}[J(\omega_i, \omega_s)]$ and $\mathrm{Im}[J(\omega_s, \omega_i)]=-\mathrm{Im}[J(\omega_i, \omega_s)]$, i.e. the real part of $J(\omega_s, \omega_i)$ is a symmetric function whereas its imaginary part is an anti-symmetric function, which can also be seen in Fig.~\ref{fig:sm_JSAs}.
Along the anti-diagonal axis $\delta \omega_s = -\delta \omega_i \equiv \delta\omega$, we find
\begin{equation}
\begin{split}
\arg{[J(\omega_s, \omega_i)]} &= \arctan{\left( \frac{  \frac{\varepsilon^2}{2\sqrt{3}} f_0(\delta \omega)f_1(\delta \omega) }{f^2_0(\delta \omega) +  \frac{\varepsilon^2}{12}\left\{ f^2_0(\delta \omega)  +  f^2_1(\delta \omega) \right\}  } \right)} \\
&= \arctan{\left( \frac{ \varepsilon^2 \mathrm{erfi}(\sqrt{4/3} \tau \delta \omega)}{2 + \frac{\varepsilon^2}{6}(1 + \mathrm{erfi}^2(\sqrt{4/3} \tau \delta \omega))} \right)}.
\end{split}
\label{eqn:jointspectralphase}
\end{equation}
Using a Taylor expansion, we can approximate that $\mathrm{erfi}(x) \approx 2x/\sqrt{\pi}$ and $\mathrm{arctan}(x) \approx x$ for $x\ll 1$ and arrive at
\begin{equation}
\begin{split}
\arg{[J(\omega_s, \omega_i)]} &\approx \left(\frac{ 24 \varepsilon^2 \tau}{\sqrt{3\pi}(12+\varepsilon^2)}\right) \delta \omega + \mathcal{O}(\delta \omega^2) \\
&= \frac{\beta}{2} \left(\delta \omega_s - \delta \omega_i \right)
\end{split}
\label{eqn:JSA_arg_tocPhase}
\end{equation}
where we defined $\beta \equiv \left(\frac{24 \varepsilon^2 \tau}{\sqrt{3\pi}(12+\varepsilon^2)}\right)$.

In our derivation of the joint spectral amplitude in Sec.~\ref{sec:magnus}, we worked in the interaction picture and thus have not accounted for the phase that the fields accumulate through linear propagation inside the crystal. 
As shown in Ref.~\cite{quesada2020theory}, this phase can be recovered by transforming the field operators according to:
\begin{equation}
\hat{a}^\dagger_\mu \rightarrow \hat{a}^\dagger_\mu e^{i\Delta k_\mu(\omega_\mu)z_0}
\end{equation}
where $z_0=-L/2$ is the start of the crystal [see Eqn.~\eqref{eqn:pmf_preIntegral}] and $\Delta k_\mu (\omega_\mu) = (v_\mu^{-1} - v_p ^{-1})\delta \omega_\mu$.
This results in an additional phase factor in the joint spectral amplitude:
\begin{equation}
J(\omega_s, \omega_s) \rightarrow J(\omega_s, \omega_s)\exp{[-i\Delta k_s(\omega_s)L/2 - i\Delta k_i(\omega_i)L/2]}.
\end{equation}
Expanding this phase factor, we find 
\begin{equation}
\begin{split}
\exp{[-i\Delta k_s(\omega_s)L/2 - i\Delta k_i(\omega_i)L/2]} &= \exp{\left[-i\frac{L}{2}\left(v_s^{-1} -v_p^{-1}\right)\delta \omega_s -i\frac{L}{2}\left(v_i^{-1}- v_p^{-1}\right)\delta \omega_i \right]} \\
&= \exp{\left[-i\frac{L}{4}\left(v_s^{-1} -v_i^{-1}\right)\left(\delta \omega_s - \delta \omega_i \right) \right]}
\end{split}
\end{equation}
where, in the second line, we used the conditions (i) $v_s^{-1} -v_p^{-1} = -(v_i^{-1}- v_p^{-1})$ and (ii) $v_s^{-1} - v_p^{-1} = \frac{1}{2}\left(v_s^{-1} - v_i^{-1} \right)$ due to symmetric group velocity matching.
Thus, Eq.~\eqref{eqn:JSA_arg_tocPhase} becomes modified to:
\begin{equation}
\arg{[J(\omega_s, \omega_i)]} = \frac{(\beta - \beta_0)}{2}(\delta \omega_s - \delta \omega_i)
\label{eqn:jsa_arg_final}
\end{equation}
where $\beta_0 = \frac{L}{2}\left(v_s^{-1} - v_i^{-1}\right)$ is a gain-independent group delay between signal and idler photons corresponding to these being generated half-way through the crystal~\cite{kwiat1995new}. 
By the Fourier shift theorem, the linear spectral phase gradient in Eq.~\eqref{eqn:jsa_arg_final} results in a group delay of $(\beta_0 - \beta)/2$ for the signal photon and $-(\beta_0 - \beta)/2$ the idler photon.
Hence, the total delay between the two photons is $T=\beta_0-\beta$.


\subsection{Spectrally-resolved coincidences}
\label{sec:coincidences}

In Eq.~\eqref{eqn:interferogram}, we derived the spectrally-resolved coincidence probability $\mathrm{pr}(\omega_1, \omega_2)$ by projecting the SPDC light after the beam splitter $\ket{\tilde{\Psi}}_{cd}$ onto pure single photons, i.e. $\mathrm{pr}(\omega_1, \omega_2) = \left|\bra{\omega_1}_c\bra{\omega_2}_d\ket{\tilde{\Psi}}_{cd}\right|^2$ where $\ket{\omega_1}_c\ket{\omega_2}_d = \hat{c}^\dagger(\omega_1)\hat{d}^\dagger(\omega_2)\ket{0}_c\ket{0}_d$.
In a low-gain regime where $\braket{n}\ll 1$, this projection measurement can be achieved using threshold (i.e. ``click") detectors since the probability to measure a coincidence is dominated by the single photon pair term of $\ket{\tilde{\Psi}}_{cd}$.
Optical losses merely reduce the rate of measuring these coincidences.
However, this approximation no longer holds in a high-gain regime because larger photon-number terms can also lead to a click event.
A complete description of the threshold detector statistics would require computing the spectrally-resolved Torontonians of the multimode lossy squeezed state~\cite{bulmer2022threshold}.
Below, we develop a simpler model which assumes that: (i) the squeezing operator acts in a single spatiotemporal mode, and (ii) the losses are sufficiently high such that the attenuated state only contains vacuum, one, or two-photon components.
This simpler model captures the main features of our observations, namely that the spectrally-resolved coincidences pattern contains an interference term which depends on the group delay and the marginals of the joint spectral amplitude. 

Using assumption (i), the joint spectral amplitude of the signal and idler modes is given by
\begin{equation}
J(\omega_s, \omega_i) = j_s(\omega_s) \times j_i(\omega_i) \times e^{-i \frac{T}{2} (\omega_s-\omega_i)}
\end{equation}
where $j_s(\omega) = \int d\omega_i |J(\omega,\omega_i)|$ and $j_i(\omega) = \int d\omega_s |J(\omega_s,\omega)| $ are their respective marginal spectral amplitude, $T$ is the group delay between the photon pairs.
Defining the following operators,
\begin{equation}
\begin{split}
\hat{A} \equiv \int d\omega j_s(\omega)e^{-i\frac{T}{2}\omega}\hat{a}(\omega) \\
\hat{B} \equiv \int d\omega j_i(\omega)e^{i\frac{T}{2}\omega}\hat{b}(\omega)
\end{split}
\end{equation}
we can re-write the SPDC output state as
\begin{equation}
\ket{\Psi}_{ab} = \exp{[\epsilon \hat{A}^\dagger \hat{B}^\dagger - \mathrm{h.c.}]} \ket{0}
\end{equation}
where the average photon number in each output mode is $\braket{n} = \sinh^2{(\epsilon)}$.
We then attenuate $\ket{\Psi}_{ab}$ down to the single-photon level in order to measure spectrally-resolved intensity correlations with threshold detectors.
The attenuation is modeled by placing a beam splitter with transmission $\eta$ in both the signal and idler modes then tracing over the reflected mode.
By working in a Gaussian formalism, one can show that the output state is given by~\cite{quesada2015thesis}:
\begin{equation}
\hat{\rho}_{ab} = \mathrm{exp}[\epsilon' \hat{A}^\dagger \hat{B}^\dagger - \mathrm{h.c.}] \left( \hat{\rho}_a^{\mathrm{th}} \otimes \hat{\rho}_b^{\mathrm{th}} \right) \mathrm{exp}[\epsilon' \hat{A} \hat{B} - \mathrm{h.c.}].
\label{eqn:squashed_state}
\end{equation}
Eq.~\eqref{eqn:squashed_state} describes a ``squashed state", which is obtained by applying a two-mode squeezing operator with gain $\epsilon'$ given by $\tanh{(2\epsilon')} = \left(2\eta \sinh{(\epsilon)}\cosh{(\epsilon)} \right)/\left(2 \sinh^2{(\epsilon)}\eta + 1\right) $ onto thermal states $\hat{\rho}_{a,b}^{\mathrm{th}}$ with mean occupation $\bar{n} = [1+2\eta\braket{n} - 4\eta^2\braket{n}^2]^{1/2} - 1$.
In the limit where $\eta \rightarrow 0$, we find that $\epsilon' \sim 0$ and $\bar{n}\sim \eta \braket{n}$, and thus the attenuated output state is two uncorrelated thermal states~\cite{jeong2000dynamics,quesada2019broadband}:
\begin{equation}
\hat{\rho}_{ab} = \hat{\rho}_a^{\mathrm{th}} \otimes \hat{\rho}_b^{\mathrm{th}} = \left(\sum_{n=0}^\infty \frac{\lambda^n}{n!} (\hat{A}^\dagger)^n \ket{0}_a\bra{0}_a \hat{A}^n \right) \otimes \left( \sum_{j=0}^\infty \frac{\lambda^j}{j!} (\hat{B}^\dagger)^j \ket{0}_b\bra{0}_b \hat{B}^j \right)
\end{equation}
where $\lambda = \eta\braket{n}/(1+\eta\braket{n})$.
We then combine the two thermal states onto a beam splitter which performs the unitary transformation $\hat{U}_{ab}^\dagger\hat{a}\hat{U}_{ab} = (\hat{a}+\hat{b})/\sqrt{2}$ and $\hat{U}_{ab}^\dagger\hat{b}\hat{U}_{ab} = (\hat{a}-\hat{b})/\sqrt{2}$.
If we measure spectrally-resolved coincidences at the output of the beam splitter, we find:
\begin{equation}
\begin{split}
\mathrm{pr}(\omega_1, \omega_2) &= \bra{0}_a\bra{0}_b \hat{a}(\omega_1) \hat{b}(\omega_2)  U_{ab} \hat{\rho}_{ab} U_{ab}^\dagger \hat{a}^\dagger(\omega_1) \hat{b}^\dagger(\omega_2) \ket{0}_a\ket{0}_b  \\
&= \frac{1}{4} \bigg( \bra{\omega_1}_a \bra{\omega_2}_a \hat{\rho}_{ab}\ket{\omega_1}_a\ket{\omega_2}_a + \bra{\omega_1}_a \bra{\omega_2}_b \hat{\rho}_{ab}\ket{\omega_1}_a\ket{\omega_2}_b \\
&\qquad + \bra{\omega_2}_a \bra{\omega_1}_b \hat{\rho}_{ab}\ket{\omega_2}_a\ket{\omega_1}_b + \bra{\omega_1}_b \bra{\omega_2}_b \hat{\rho}_{ab}\ket{\omega_1}_b\ket{\omega_2}_b \\
&\qquad - \bra{\omega_1}_a \bra{\omega_2}_b \hat{\rho}_{ab}\ket{\omega_2}_a\ket{\omega_1}_b - \bra{\omega_2}_a \bra{\omega_1}_b \hat{\rho}_{ab}\ket{\omega_1}_a\ket{\omega_2}_b \bigg) \\
&= \frac{1}{4} \bigg( 2\lambda^2 j_a^2(\omega_1)j_a^2(\omega_2)  +  \lambda^2 j_a^2(\omega_1)j_b^2(\omega_2) \\
&\qquad + \lambda^2 j_a^2(\omega_2)j_b^2(\omega_1) + 2\lambda^2 j_b^2(\omega_1)j_b^2(\omega_2) ) \\
&\qquad - \lambda^2 j_a(\omega_1)j_a(\omega_2)j_b(\omega_1)j_b(\omega_2)e^{-i\frac{T}{2}(\omega_1-\omega_2)}  \\
&\qquad -  \lambda^2 j_a(\omega_1)j_a(\omega_2)j_b(\omega_1)j_b(\omega_2)e^{-i\frac{T}{2}(\omega_2-\omega_1)} \bigg) \\
\end{split}
\label{eqn:spectrally_resolved_coincidences_full}
\end{equation}
where we used the fact that
\begin{equation}
\begin{split}
\bra{\omega_1}_a\bra{\omega_2}_a(\hat{A}^\dagger)^2\ket{0}_a &= \bra{0}_a \hat{a}(\omega_1)\hat{a}(\omega_2) \iint d\omega d\omega' j_a(\omega)j_a(\omega')e^{i\frac{T}{2}(\omega+\omega')}\hat{a}^\dagger(\omega)\hat{a}^\dagger(\omega')\ket{0}_a \\
&= 2 j_a(\omega_1)j_a(\omega_2)e^{i\frac{T}{2}(\omega_1+\omega_2)}.
\end{split}
\end{equation}
Eq.~\eqref{eqn:spectrally_resolved_coincidences_full} describes an interference pattern with a fringe along the $\omega_1=-\omega_2$ axis.
The factors of $2$ in the first two terms of Eq.~\eqref{eqn:spectrally_resolved_coincidences_full} are due to a bunching effect at the beam splitter: it is two times more likely to find two photons in one of the pulses than one photon in each pulse. 
If the signal and idler spectral amplitudes are indistinguishable, i.e. $j_a(\omega) = j_b(\omega) \equiv j(\omega)$, the equation can be simplified to:
\begin{equation}
\mathrm{pr}(\omega_1, \omega_2)  \propto [j(\omega_1)j(\omega_2)]^2 \left[3 - \cos \left(\tfrac{T}{2}[\omega_1-\omega_2]\right) \right].
\end{equation}
Effects such as multiple temporal-spectral modes, larger photon-number terms, and dark counts can all further reduce the interference visibility $\mathcal{V}$, leading to Eq.~\eqref{eqn:high_gain_coincidences} in the main text.



\subsection{Further details on experimental results}
\label{sec:exp_details}

\begin{figure}
\centering
\includegraphics[width=0.9\columnwidth]{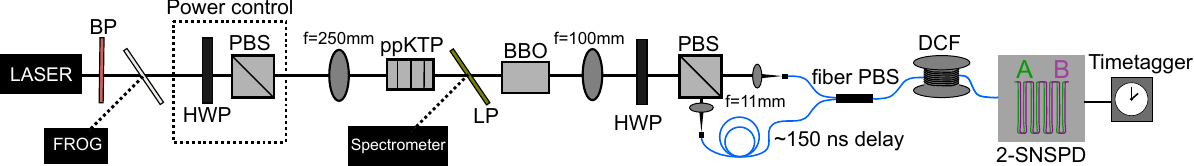}
\caption{Detailed experimental setup. Laser (LightConversion Orpheus-HP), BP: two angle-tuned bandpass filter (Semrock LD01-785/10), LP: longpass filter (Semrock LP02-808RE), HWP: half-wave plate, PBS: polarizing beam splitter, FROG: frequency-resolved optical gating (Swamp Optics), ppKTP: 2-mm-long periodically-poled potassium titanyl phosphate (Raicol), BBO: 2-mm-long $\alpha$ barium borate (NewLight Photonics), DCF: dispersion-compensating fiber (OFS SMFDK-S-60-03-10), SNSPD: two-element superconducting nanowire single photon detector (PhotonSpot), Timetagger (Swabian Instruments).}
\label{fig:detailedSetup}
\end{figure}

\begin{figure}
\centering
\includegraphics[width=1\columnwidth]{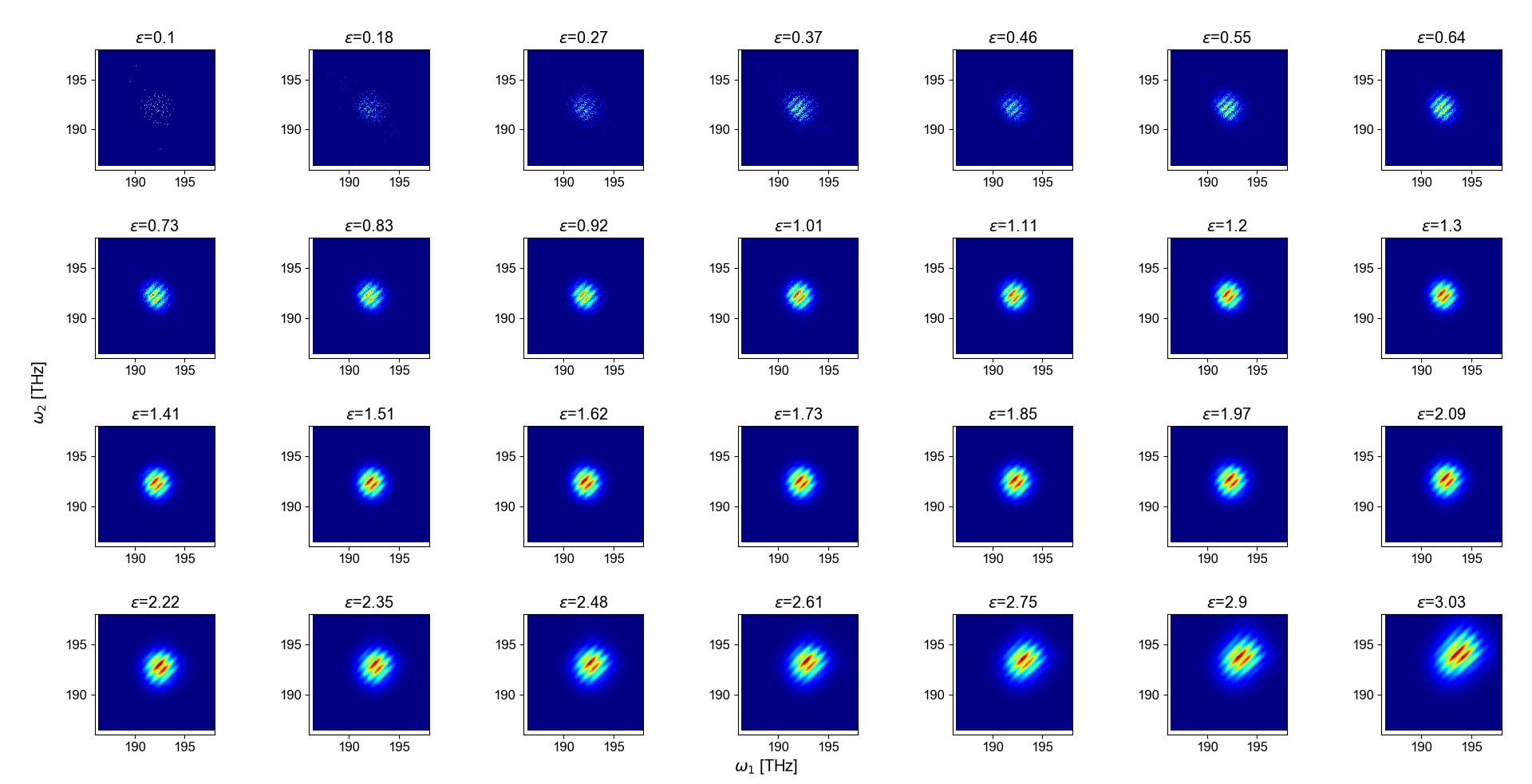}
\caption{Complete dateset for the spectrally-resolved coincidences $N(\omega_1, \omega_2)$.}
\label{fig:figS1}
\end{figure}

A detailed experimental setup is shown in Fig.~\ref{fig:detailedSetup}.
Our experiment employs a two-element superconducting nanowire detector.
We can measure up to two clicks in both the ``early" bin (transmitted mode of PBS) and the ``late" bin (reflected mode of PBS).
The histograms $N(\omega_1, \omega_2)$ are obtained by binning timetags of events where at least one photon is detected in both the ``early" and ``late" bins.
In Fig.~\ref{fig:figS1}, we plot all measured $N(\omega_1, \omega_2)$.
The center of the distributions are shifted towards higher frequencies with increasing $\varepsilon$.
This is due to a pile-up effect whereby photons that arrive at the detector earlier (i.e. with higher frequencies) are more likely to be detected than photons which arrive at the detector later due to the detector deadtime ($\sim 100~$ns).
Although this effect can be reduced by increasing the attenuation of the down-converted beams, it does not hinder our ability to resolve the fringe frequency and therefore we keep the attenuation constant to reduce the number of parameters in the model.
All parameters used in the \texttt{NeedALight} model are given in Table~\ref{table:params}.

\subsubsection{Determining $\chi^{(3)}$ parameters}
\begin{figure}
\centering
\includegraphics[width=0.3\columnwidth]{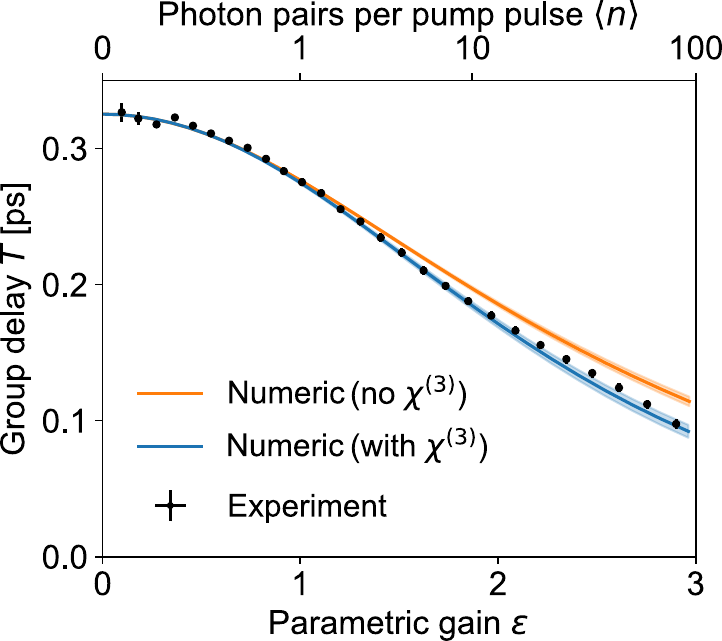}
\caption{Comparison of the numeric \texttt{NeedALight} model with and without including $\chi^{(3)}$ effects (i.e. self-phase and cross-phase modulation). Most of the observed group delay can be attributed to a high-gain $\chi^{(2)}$ effect.}
\label{fig:XPM_noXPM}
\end{figure}

To determine the $\chi^{(3)}$ nonlinear coefficients $\gamma_{s,i,p}$, we monitor the pump spectrum after the ppKTP crystal.
The self-phase modulation (SPM) coefficient, $\gamma_{p}$ is obtained by fitting a SPM model implemented in \texttt{NeedALight} [Fig~\ref{fig:spmFits}].
The cross-phase modulation (XPM) coefficients are then given by $\gamma_i=2\gamma_p$ and $\gamma_s=2\gamma_p/3$~\cite{agarwal2013nonlinear}.
These nonlinear coefficients $\gamma_\mu$ can be related to the coupling constants $\zeta_\mu$ through the equation $\gamma_\mu = \zeta_\mu / v_p v_\mu \hbar \bar{\omega}_\mu$~\cite{quesada2020theory}.
In Fig.~\ref{fig:XPM_noXPM}, we compare our experimental data to the \texttt{NeedALight} model with and without including $\chi^{(3)}$ effects.
Due to XPM, there is a small additional group delay between the signal and idler photon pulses. 
However, most of the observed delay can be attributed to high-gain $\chi^{(2)}$ interactions.

\subsubsection{Determining parametric gain $\varepsilon$}
\begin{figure}
\centering
\includegraphics[width=0.3\columnwidth]{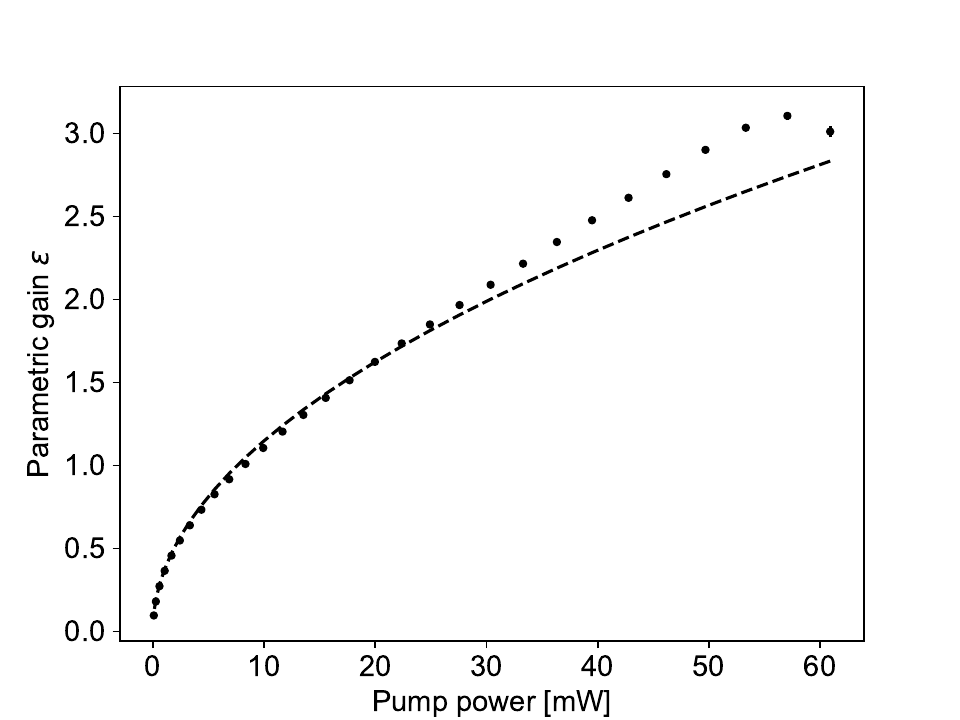}
\caption{The experimental parametric gain $\varepsilon$ values [black dots].
In the low gain regime, $\varepsilon \propto \sqrt{P}$ [dashed line] where $P$ is the pump power~\cite{triginer2020understanding}.}
\label{fig:parametricGain}
\end{figure}

To determine the parametric gain $\varepsilon$, we first rotate the HWP after the ppKTP crystal such that the signal and idler modes are separated at the PBS.
We then record the click statistics for five minutes at each pump power.
Using these statistics, we can determine $\varepsilon$ by fitting a model that uses \texttt{Strawberryfields}~\cite{killoran2019strawberry}.
In this model, the SPDC beams are described by a two-mode squeezed vacuum state of squeezing $\varepsilon$.
To model the two-element nanowire detector, both the signal and idler modes are split on a 50:50 beam splitter, and subsequently all four modes are detected by threshold detectors of efficiency $\eta$.
Note that the detector statistics are computed using Torontonians in order to accurately describe the click detectors even at higher parametric gains~\cite{bulmer2022threshold}.
The Python code for the model can be found on GitHub~\cite{SPDC_GroupDelay}.
The detection efficiency $\eta\sim 7\%$ was determined from a Klyshko measurement at low gain.
The resulting parametric gains are shown in Fig.~\ref{fig:parametricGain}.
Although our \texttt{Strawberryfields} model neglects the slight spectrally multimode nature of the down-converted light, this method of determining $\varepsilon$ provides a sufficiently accurate estimate to obtain agreement with \texttt{NeedALight} for the group delay [Fig.~\ref{fig:fig3}(d)].

\subsubsection{Validity of the single spatial mode theory}
Our theoretical treatment assumes that the nonlinear interaction occurs in a single spatial mode.
While this can be achieved in practice using waveguide structures, our experiment employs a bulk crystal which in principle can accommodate a continuum of modes.
To approximate a single mode regime, we weakly focus the pump beam (waist $w_0 = 125$~um) into the crystal. 
In this focusing regime, the Rayleigh length of the pump beam, $z_r = \pi w_0^2/\lambda \sim 63$~mm, is significantly longer than the crystal length, $L=2$~mm, and thus the pump is approximately collimated throughout the nonlinear interaction.
This regime also maximizes the heralding efficiency of the generated photon pairs~\cite{bennink2010optimal}.

\begin{figure}
\centering
\includegraphics[width=0.7\columnwidth]{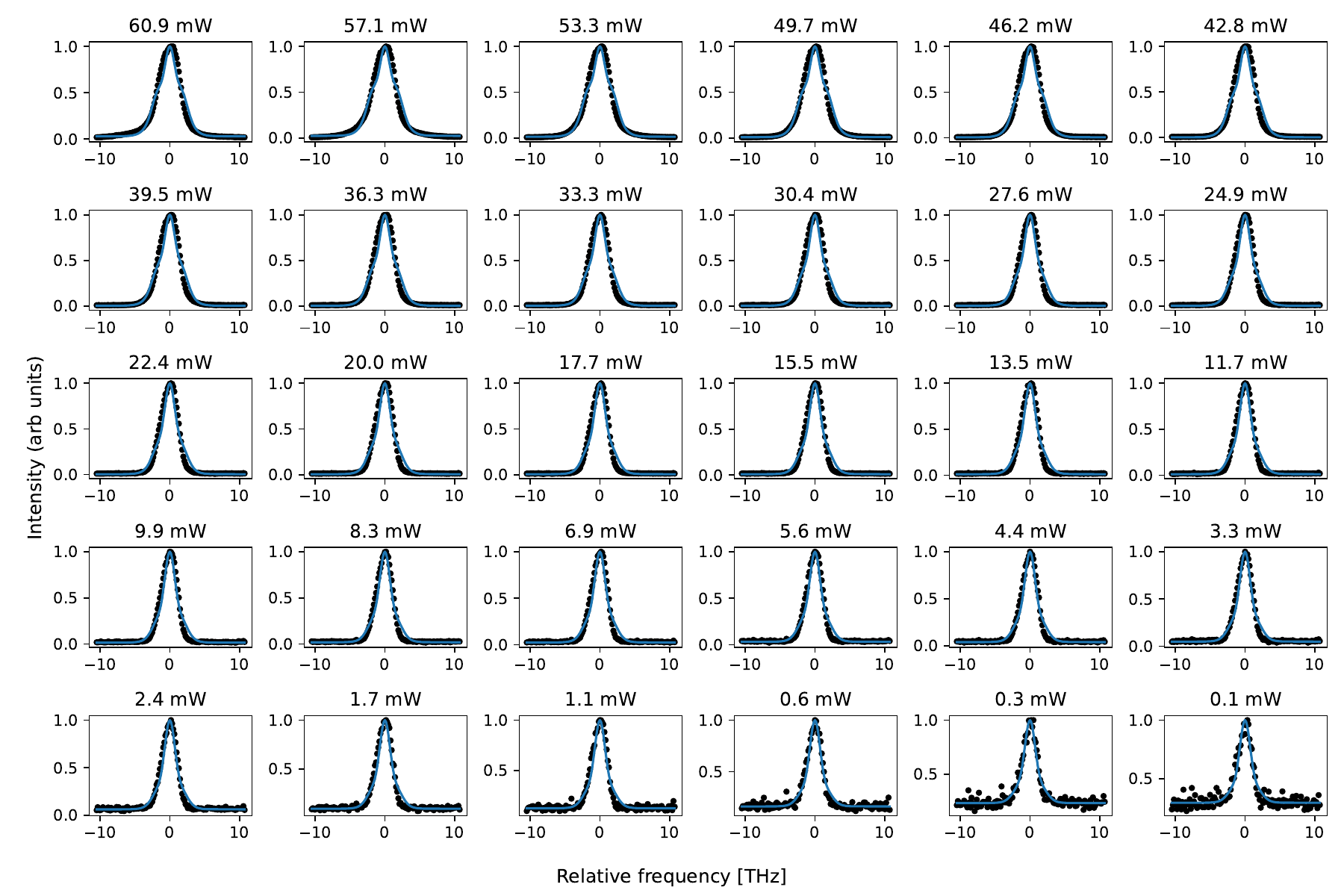}
\caption{Measured pump spectra after the ppKTP crystal.
Blue line is a fit obtained using \texttt{NeedALight}.
The spectra broaden with increasing power due to self-phase modulation.}
\label{fig:spmFits}
\end{figure}

At higher pump powers, we can expect $\chi^{(3)}$ effects to modify the spatial mode of the pump, analogously to self-phase modulation leading to the spectral broadening in Fig.~\ref{fig:spmFits}.
In particular, self-focusing can reduce the pump beam waist size inside the crystal.
We can estimate the self-focusing distance $z_{sf}$ from the crystal entrance facet as
\begin{equation}
z_{sf} = \frac{2 \bar{n}_p w_0^2}{\lambda} \frac{1}{\sqrt{P/P_{cr} -1}},
\end{equation}
where the critical power is given by $P_{cr} = 1.2 \lambda ^2 / 8 \bar{n}_p n_2$ for a Gaussian beam, and $n_2=\gamma_p \lambda w_0^2 /2$~\cite{boyd2003nonlinear}.
Using the experimentally measured values (see Table~\ref{table:params}), we find $P_{cr}\sim1.6\times10^4$~W.
At the largest pump power in our experiment (60 mW average power, or $1.6\times10^6$~W peak power), we find that $z_{sf} \sim 8~$mm.
Thus, self-focusing results in a $\sim$~20\% decrease in the pump waist at the exit of the $2$-mm-long crystal.
Since we do not re-optimize the fiber coupling at different gains, we expect that self-focusing effects will slightly reduce the coupling efficiency of the photons generated at the larger pump powers.

\begin{center}
\begin{table}
\begin{tabular}{| c | c | c |} 
 \hline
 \textbf{Parameter} & \textbf{Value} & \textbf{Method}  \\ [0.5ex] 
 \hline
 
 \multicolumn{3}{|c|}{\textit{Pump}} \\
 \hline
 Spectral amplitude bandwidth $\sigma$ & 3.22(5)\,nm & Spectrometer  \\ 
  \hline
 Pulse amplitude duration $\tau$ & 132(3)\,fs & Frequency-resolved optical gating (FROG)   \\
 \hline
 Center wavelength $\lambda$ & 779.2\,nm & Spectrometer \\
 \hline
 Beam waist $w_0$ (at focus) & 125 $\mu$m & Camera \\
 \hline
 
 \multicolumn{3}{|c|}{\textit{Down-conversion source}} \\
 \hline
  Crystal length $L$ & $2$\,mm & Manufacturer (Raicol) \\  
 \hline
 Signal efficiency $\eta_s$ & 7(1)\,\% & Klyshko measurement at low gain  \\
 \hline
 Idler efficiency $\eta_i$ & 7(1)\,\% & Klyshko measurement at low gain  \\
 \hline
 Parametric gain $\varepsilon$ & 0 to $\sim$ 3 & Fitting click statistics using $\eta_{s,i}$~\cite{quesada2018gaussian}  \\
 \hline
 Pump SPM coefficient $\gamma_p$ & $5(3)\times 10^{-4}~ \mathrm{W}^{-1} \mathrm{m} ^{-1}$ & Fitting pump spectrum after crystal [Fig.~\ref{fig:spmFits}] \\  
 \hline
 Signal XPM coefficient $\gamma_s$ & $3(2)\times 10^{-4}~ \mathrm{W}^{-1} \mathrm{m} ^{-1}$ & $\gamma_s=2\gamma_p/3$ \\ 
 \hline
 Idler XPM coefficient $\gamma_i$ & $9(5)\times 10^{-4}~ \mathrm{W}^{-1} \mathrm{m} ^{-1}$ & $\gamma_i=2\gamma_p$ \\  
 \hline
  Pump (H-pol) group index $\bar{n}_{p}$ & 1.8092 & Ref.~\cite{kato2002sellmeier} \\  
 \hline
  Signal (V-pol) group index $\bar{n}_{s}$ & 1.8514 & Ref.~\cite{kato2002sellmeier} \\  
 \hline
  Idler (H-pol) group index $\bar{n}_{i}$ & 1.7538 & Ref.~\cite{kato2002sellmeier} \\  
 \hline


\end{tabular}
\caption{A list of the parameters used in \texttt{NeedALight}. Note that the bandwidths specify one standard deviation of the amplitude functions rather than the intensity. The nonlinear coefficients $\gamma_\mu$ can be related to the coupling constants $\zeta_\mu$ through the equation $\gamma_\mu = \zeta_\mu / v_p v_\mu \hbar \bar{\omega}_\mu$~\cite{quesada2020theory}.}
\label{table:params}
\end{table}
\end{center}

\end{document}